\documentclass[12pt]{article}

\usepackage[paper=a4paper,left=24mm,right=24mm,top=24mm,bottom=24mm]{geometry}
\usepackage{amsfonts,amsmath,amssymb,amsthm}
\usepackage{bbm,booktabs,graphicx,multirow,setspace}
\usepackage{color}
\usepackage[hyphens]{url}
\usepackage[english]{babel}
\usepackage[utf8]{inputenc}
\usepackage[round]{natbib}

\newcommand{\E}{\mathbb{E}}
\newcommand{\cF}{\mathcal{F}}
\newcommand{\cL}{\mathcal{L}}
\newcommand{\cM}{\mathcal{M}}
\newcommand{\cN}{\mathcal{N}}
\newcommand{\real}{\mathbb{R}}
\newcommand{\myS}{{\rm S}}

\newcommand{\ind}{\mathbbm{1}} 
\newcommand{\dd}{{\rm d}}

\newcommand{\Ppost}{P_{\rm post}}
\newcommand{\dS}{d_{\myS}}

\newcommand{\dTV}{d_\textnormal{TV}}
\newcommand{\dW}{d_\textnormal{W}}

\newcommand{\thetam}{(\theta_i)_{i=1}^m}

\newcommand{\Xm}{(X_i)_{i=1}^m} 
\newcommand{\thetaseq}{(\theta_i)_{i=1,2,\ldots}}
\newcommand{\Xseq}{(X_i)_{i=1,2,\ldots}} 
\newcommand{\thetamk}{(\theta_i^{(k)})_{i=1}^m}
\newcommand{\Xmk}{(X_i^{(k)})_{i=1}^m} 

\newcommand{\Fc}{F_c(x \hspace{0.2mm} | \hspace{0.3mm} \theta)} 
\newcommand{\Fcd}{F_c(\cdot \hspace{0.2mm} | \hspace{0.3mm} \theta)} 
 
\newcommand{\fcd}{f_c(\cdot \hspace{0.2mm} | \hspace{0.3mm} \theta)} 
\newcommand{\Fci}{F_c(x \hspace{0.2mm} | \hspace{0.3mm} \theta_i)} 
\newcommand{\Fcdi}{F_c(\cdot \hspace{0.2mm} | \hspace{0.3mm} \theta_i)} 
\newcommand{\Fcdj}{F_c(\cdot \hspace{0.2mm} | \hspace{0.3mm} \theta_j)}

\newcommand{\FAM}{\hat{F}_m}
\newcommand{\FMP}{\hat{F}_m^{\rm MP}}
\newcommand{\fMP}{\hat{f}_m^{\rm MP}}
\newcommand{\FECDF}{\hat{F}_m^{\rm ECDF}}
\newcommand{\FGA}{\hat{F}_m^{\rm GA}}
\newcommand{\fKD}{\hat{f}_m^{\rm KD}}
\newcommand{\FKD}{\hat{F}_m^{\rm KD}}

\title{Predictive Inference Based on Markov Chain Monte Carlo Output}

\author{Fabian Kr\"uger$^{a}$ 
	\and 
	Sebastian Lerch$^{a, b}$ 
	\and 
	Thordis Thorarinsdottir$^{c}$ 
	\and 
	Tilmann Gneiting$^{b, a}$ \\
	\\
	{\small $^a$Karlsruhe Institute of Technology} \\
	{\small $^b$Heidelberg Institute for Theoretical Studies} \\  
	{\small $^c$Norwegian Computing Center}
}

\begin{document}
	
	\maketitle

\begin{abstract}
In Bayesian inference, predictive distributions are typically in the
form of samples generated via Markov chain Monte Carlo (MCMC) or
related algorithms.  In this paper, we conduct a systematic analysis
of how to make and evaluate probabilistic forecasts from such
simulation output.  Based on proper scoring rules, we develop a notion
of consistency that allows to assess the adequacy of methods for
estimating the stationary distribution underlying the simulation
output.  We then provide asymptotic results that account for the
salient features of Bayesian posterior simulators, and derive
conditions under which choices from the literature satisfy our notion
of consistency.  Importantly, these conditions depend on the scoring
rule being used, such that the choices of approximation method and
scoring rule are intertwined.  While the logarithmic rule requires
fairly stringent conditions, the continuous ranked probability score
(CRPS) yields consistent approximations under minimal assumptions.
These results are illustrated in a simulation study and an economic
data example. Overall, mixture-of-parameters approximations which
exploit the parametric structure of Bayesian models perform
particularly well. Under the CRPS, the empirical distribution function is a simple and appealing alternative option.
\end{abstract}

\onehalfspacing

\section{Introduction}

Probabilistic forecasts are predictive probability distributions
over quantities or events of interest. They implement an idea that
  was eloquently expressed already at the beginning of the 20th
  century in the context of meteorological prediction:

\begin{quote} 
\small ``It seems to me that the condition of confidence or otherwise
forms a very important part of the prediction, and ought to find
expression.''
	
\hfill \citep[pp.~23--24]{Cooke1906}
\end{quote} 

Despite this early acknowledgment of the importance of forecast
  uncertainty, constructing principled and realistic measures of the
  latter remains challenging in practice. In this context, a rapidly
growing transdisciplinary literature uses Bayesian inference to
produce posterior predictive distributions in a wide range of
applications, including economic, ecological, and meteorological
problems, among many others. Bayesian posterior predictive
  distributions naturally account for sources of uncertainty --
  such as unknown model parameters, or latent variables in state space
  models -- that are not easily captured using frequentist methods;
see, e.g., \citet{Clark2005} for an ecological perspective.

Formally, posterior predictive distributions arise as mixture
distributions with respect to the posterior distribution of the
parameter vector. In the following, we assume that the parameter
  vector contains all quantities that are subject to Bayesian
  inference, including also latent state variables, for example. For
a real-valued continuous quantity of interest, the posterior
predictive distribution, $F_0$, can be represented by its cumulative
distribution function (CDF) or the respective density.  The posterior
predictive CDF is then of the generic form
\begin{equation}  \label{eq:F0} 
F_0(x) = \int_{\Theta} \Fc \, \dd \Ppost(\theta)
\end{equation}
for $x \in \real$, where $\Ppost$ is the posterior distribution of the
parameter, $\theta$, over some parameter space, $\Theta$, and $\Fcd$
is the conditional predictive CDF when $\theta \in \Theta$ is the true
parameter.  \citet{Harris1989} argues that predictive distributions of
this form have appeal in frequentist settings as well. Often, the integral in \eqref{eq:F0} does not admit a solution
in closed form, and so the posterior predictive CDF must be
approximated or estimated in some way, typically using some form of
Markov chain Monte Carlo (MCMC); see, e.g., \citet{GelfandSmith1990}
and \citet{GilksEtAl1996book}.

Given a simulated sequence $\thetam$ of parameter values
from $\Ppost$, one approach, which we call the {\em mixture-of-parameters}\/
(MP) technique, is to approximate $F_0$ by
\begin{equation}  \label{eq:MP} 
\FMP(x) = \frac{1}{m} \sum_{i=1}^m \Fci.  
\end{equation} 
{However, this method can be used only when  the conditional
	distributions  $\Fcd$  are available in closed form.}  An
alternative route is to simulate a sequence $\Xm$ where $X_i \sim
\Fcdi$, and to approximate $F_0$ based on this sample, using either
nonparametric or parametric techniques.  The most straightforward
option is to estimate $F_0$ by the {\em empirical CDF}\/ (ECDF),
\begin{equation}  \label{eq:ECDF} 
\FECDF(x) = \frac{1}{m} \sum_{i=1}^m \ind\{ x \geq X_i \}. 
\end{equation} 
Alternatively, one might employ a {\em kernel density}\/ (KD) estimate of the posterior predictive density, namely,
\begin{equation}  \label{eq:KD}
\fKD(x) = \frac{1}{mh_m} \sum_{i=1}^m K \! \left( \frac{x - X_i}{h_m} \right) \! , 
\end{equation}
where $K$ is a kernel function, i.e., a symmetric, bounded, and
square-integrable probability density, such as the Gaussian or the
Epanechnikov kernel, and $h_m$ is a suitable bandwidth
\citep{Rosenblatt1956, Silverman1986}.  Finally, much extant work
employs a {\em Gaussian approximation}\/ (GA) to $F_0$, namely
\begin{equation}  \label{eq:GA} 
\FGA(x) = \Phi \! \left( \frac{x - \hat{\mu}_m}{\hat{\sigma}_m} \right) \! ,
\end{equation} 
where $\Phi$ is the CDF of the standard normal distribution, and
$\hat{\mu}_m$ and $\hat{\sigma}_m$ are the empirical mean and standard
deviation of the sample $\Xm$.  

Following \citet{Rubin1984} and \citet{Little2006}, it is now widely
accepted that posterior predictive inference should be evaluated using
frequentist principles, without prior information entering at the model
evaluation stage.  For the comparison and ranking of probabilistic
forecasting methods one typically uses a proper scoring rule
\citep{GneitingRaftery2007} that assigns a numerical score or penalty
based on the predictive CDF, $F$, or its density, $f$, and the
corresponding realization, $y$, such as the logarithmic score
\citep[LogS;][]{Good1952},
\begin{equation}  \label{eq:LogS} 
\text{LogS}(F, y) = -\log f(y), 
\end{equation} 
or the continuous ranked probability score
\cite[CRPS;][]{MathesonWinkler1976},
\begin{equation}  \label{eq:CRPS} 
\text{CRPS}(F,y) = \int_\real \left( F(z) - \ind\{ z \geq y \}\right)^2  \dd z.
\end{equation} 
While the LogS and CRPS are the two most popular scoring rules in
  applications, they feature interesting conceptual differences
  which we discuss in Section \ref{sec:propS}. In practice, one finds
and compares the mean score over an out-of-sample test set, and the
forecasting method with the smaller mean score is preferred.  Formal
tests of the null hypothesis of equal predictive performance can be
employed as well {\citep{DieboldMariano1995, GiacominiWhite2006,
    ClarkMcCracken2013, DelSoleTippett2014}}.

Table 1 of the Online Supplement summarizes the
use of evaluation techniques in recently published comparative studies
of probabilistic forecasting methods that use Bayesian inference via
MCMC.  As shown in the table, the mixture-of-parameters technique has
mainly been applied in concert with the logarithmic score, whereas the
empirical CDF method can be used in conjunction with the CRPS
only. However, to this date, there are few, if any, guidelines to
support choices in Table 1 of the Online
  Supplement, and it is not clear how they affect practical model
comparisons.  The present paper provides a systematic analysis of this
topic.  We focus on the following questions.  First, what defines
reasonable choices of the approximation method and scoring rule?
Second, under what conditions do extant choices from the literature
satisfy this definition?  Third, for a given scoring rule, how
accurate are alternative approximation methods in practically relevant
scenarios?

In studying these questions, our work is complementary to \cite{GneitingRaftery2007} who develop the broader theory of scoring rules and portray their rich mathematical and decision theoretic structure. While \cite{GneitingRaftery2007} mention simulated predictive distributions (see in particular their Section 4.2), the empirical literature surveyed in the Online Supplement has largely evolved after 2007, giving rise to the applied techniques that motivate the present paper. 

We emphasize that the present study --- and the use of scoring rules
in general --- concern the {\em comparative}\/ assessment of two or
more predictive models: The model with the smallest mean score is
considered the most appropriate.  Comparative assessment is essential
in order to choose among a large number of specifications typically
available in practice.  This task is different from {\em absolute}\/
assessment, which amounts to diagnosing possible misspecification,
using the probability integral transform \citep{Dawid1984,
  DieboldEtAl1998}, posterior predictive checks
(\citeauthor{GelmanEtAl1996}, \citeyear{GelmanEtAl1996};
\citeauthor{HeldEtAl2010_2}, \citeyear{HeldEtAl2010_2};
\citeauthor{GelmanEtAl2014book}, \citeyear{GelmanEtAl2014book},
Chapter 6) and related methods.

The remainder of this paper is organized as follows.  Section
\ref{sec:setup} introduces the notion of a consistent approximation to
$F_0$.  This formalizes the idea that, as the size of the
  simulated sample becomes larger and larger, and in terms of a given
  scoring rule, the approximation ought to perform as well as the
  unknown true forecast distribution.  In Section \ref{sec:theory} we
provide theoretical justifications of approximation methods
encountered in the literature.  Sections \ref{sec:simulation} and
\ref{sec:casestudy} present simulation and empirical evidence on the
performance of these methods, and Section \ref{sec:discussion}
concludes with a discussion.  Overall, our findings support the
  use of the mixture-of-parameters estimator at (\ref{eq:MP}) in order to approximate
  the posterior predictive distribution of interest. If this estimator is unavailable, the ECDF estimator at (\ref{eq:ECDF}) is a simple and appealing alternative. Technical material and supplementary analyses are deferred to Appendices
  \ref{sec:CRPS} to \ref{app:implementation}. The Online Supplement
  contains a bibliography of the pertinent applied literature and
  additional figures.

\section{Formal Setting} \label{sec:setup}

In this section, we discuss the posterior predictive distribution in
Bayesian forecasting, give a brief review of proper scoring rules and
score divergences, and introduce the concept of a consistent
approximation method based on MCMC output.

As discussed earlier, the posterior predictive cumulative distribution
function (CDF) of a Bayesian forecasting model is given by
\[
F_0(x) = \int_{\Theta} \Fc \, \dd \Ppost(\theta)
\]
where $\theta \in \Theta$ is the parameter, $\Ppost$ is the posterior
distribution of the parameter, and $\Fcd$ is the predictive
distribution {\em conditional}\/ on a parameter value $\theta$; see,
e.g., \citet[][p.~33]{Greenberg2013} or
\citet[p.~7]{GelmanEtAl2014book}.  A generic Markov chain Monte Carlo
(MCMC) algorithm designed to sample from $F_0$ can be sketched as
follows.

\begin{itemize}
	
	\item Fix $\theta_0 \in \Theta$ at some arbitrary value.
	
	\item For $i = 1, 2, \ldots$ iterate as follows:
	
	\begin{itemize}
		
		\item Draw $\theta_i \sim {\cal
                  K}(\theta_i \hspace{0.2mm} |
		\hspace{0.2mm} \theta_{i-1})$, where $\cal K$ is a transition kernel
		that specifies the conditional distribution of $\theta_i$ given
		$\theta_{i-1}$.  
		
		\item Draw $X_i \sim \Fcdi$.
		
	\end{itemize}
	
\end{itemize} 

We assume throughout that the transition kernel $\mathcal{K}$ is such
that the sequence $\thetaseq$ is stationary and ergodic {in the sense
  of \citet[][Definition 4.5.5]{Geweke2005}} with invariant
distribution $\Ppost$, as holds widely in practice
\citep{CraiuRosenthal2014}.  Importantly, stationarity and ergodicity
of $\thetaseq$ with invariant distribution $\Ppost$ imply that $\Xseq$
is stationary and ergodic with invariant distribution $F_0$
\citep[Proposition 3.1]{GenonEtAl2000}.

This generic MCMC algorithm allows for two general options for
estimating the posterior predictive distribution $F_0$ in
\eqref{eq:F0}, namely,

\begin{itemize}
	
	\item Option A: Based on parameter draws $\thetam$, 
	
	\item Option B: Based on a sample $\Xm$,
	
\end{itemize}

\noindent
where $m$ typically is on the order of a few thousands or ten
thousands.  Alternatively, some authors, such as
\citet{KruegerEtAl2015}, generate, for each $i = 1, \ldots, m$,
independent draws $X_{ij} \sim \Fcdi$, where $j = 1, \dots, J$; see
also \citet[Section III. B]{WaggonerZha1999}. The considerations below
apply in this more general setting as well.

\subsection{Approximation methods}  \label{setup-sample-generation}

In the case of Option A, the sequence $\thetam$ of parameter draws is
used to approximate the posterior predictive distribution, $F_0$, by the
mixture-of-parameters estimator $\FMP$ in \eqref{eq:MP}.  Under the
assumption of ergodicity,
\[
\FMP(x) 
= \frac{1}{m} \sum_{i=1}^m \Fci   
\longrightarrow \int_{\Theta} \Fc \, \dd P_{post}(\theta)
= F_0(x)
\]
for $x \in \real$.  This estimator was popularized by \citet[Section
2.2]{GelfandSmith1990}, based on earlier work by
\citet{TannerWong1987}, and is often called a {\em conditional}\/ or
{\em Rao-Blackwellized}\/ estimator. The latter term hints at variance reduction that may result from conditioning on the parameter draws (see Theorem 4 below). We refer to $\FMP$ as the
{\em mixture-of-parameters} (MP) estimator.

In the case of Option B, the sample $\Xm$ is employed to approximate
the posterior predictive distribution $F_0$.  Various methods for
doing this have been proposed and used, including the {\em empirical
  CDF}\/ of the sample, denoted $\FECDF$ in \eqref{eq:ECDF}, the {\em
  kernel density}\/ estimator $\fKD$ in \eqref{eq:KD}, and the {\em
  Gaussian}\/ approximation $\FGA$ in \eqref{eq:GA}.  Approaches of
this type incur `more randomness than necessary', in that the
simulation step to draw $\Xm$ can be avoided if Option A is used.
That said, Option A requires full knowledge of the model
specification, as the conditional distributions must be known in
closed form in order to compute $\FMP$.  There are situations where
this is restrictive, e.g., when the task is to predict a nonlinear
transformation of the original, possibly vector-valued predictand (see the setup in  \citealt[Section 6d]{FeldmannEtAl2015} for an example from meteorology). We emphasize, however, that the mixture-of-parameters estimator is readily available in the clear majority of applied examples that we encounter in our work.\\
The implementation of the approximation methods (based on either Option A or B) is typically straightforward, except for the case of kernel density estimation, for which we discuss implementation choices in Section \ref{sec:KD}.

\subsection{Proper scoring rules and score divergences} \label{sec:propS}

Let $\Omega \subseteq \real$ denote the set of possible values of the
quantity of interest, and let $\cF$ denote a convex class of probability
distributions on $\Omega$.  A {\em scoring rule}\/ is a function
\[ 
\myS : \cF \times \Omega \longrightarrow \real \cup \{ \infty \} 
\] 
that assigns numerical values to pairs of forecasts $F \in \cF$ and
observations $y \in \Omega$.  We typically set $\Omega = \real$, but will
occasionally restrict attention to compact subsets.

Throughout this paper, we define scoring rules to be negatively
oriented, i.e., a lower score indicates a better forecast.  A scoring
rule is {\em proper}\/ relative to $\cF$ if the expected score    
\[
\myS(F,G) = \int_{\Omega}\myS(F,y)\, \dd G(y)
\]
is minimized for $F = G$, i.e., if
\[
\myS(G,G) \leq \myS(F,G) 
\]
for all probability distributions $F, G \in \cF$. It is {\em strictly
	proper}\/ relative to the class $\cF$ if, furthermore, equality
implies that $F = G$.  The {\em score
	divergence}\/ associated with the scoring rule $S$ is given by
\[
\dS(F,G) = \myS(F,G) - \myS(G,G).
\]
Clearly, $\dS(F,G) \geq 0$ for all $F, G \in \cF$ is equivalent to
propriety of the scoring rule $\myS$, which is a critically important
property in practice.\footnote{See \cite{Brier1950} and
    \cite{ShufordEtAl1966} for early references arguing that scoring
    rules should be proper, and \cite{GneitingRaftery2007} for a
    review of the statistical implications.}

\begin{table}

	\caption{Examples of proper scoring rules, along with the associated
		score divergence and natural domain, $\cF$.  For a probability
		distribution with CDF $F$, we write $\mu_F$ for its mean, $\sigma_F$
		for its standard deviation, and $f$ for its density.
		\label{tab:scores}}
	
	\centering
	
	\bigskip
	
	\footnotesize 
	
	\begin{tabular}{lccc}
		\toprule 
		Scoring rule & $\myS(F,y)$ & $\dS(F,G)$ & $\cF$ \\
		\midrule 
		Logarithmic score \rule{0mm}{4mm} & $- \log f(y)$ & $\textstyle \int g(z) \log \frac{g(z)}{f(z)} \, \dd z$ & $\cL_1$ 
		\\ [2.5mm]
		CRPS & 
		$\int \, (F(z) - \ind\{ z \geq y \})^2 \, \dd z$ & $\int \, (F(z) - G(z))^2 \, \dd z$ & $\cM_1$ \\ [2.5mm]
		Dawid--Sebastiani score & $\log \sigma^2_F + \frac{(y-\mu_F)^2}{\sigma^2_F}$ & 
		$\frac{\sigma^2_G}{\sigma^2_F} - \log \frac{\sigma^2_G}{\sigma^2_F} + \frac{(\mu_F-\mu_G)^2}{\sigma^2_F} - 1$ & $\cM_2$ 
		\\ [2.5mm]
		\bottomrule
	\end{tabular}
	
	\bigskip
	
\end{table}

Table \ref{tab:scores} shows frequently used proper scoring rules,
along with the associated score divergences and the natural
domain. For any given scoring rule $\myS$, the associated {\em natural
  domain}\/ is the largest convex class of probability distributions
$F$ such that $\myS(F,y)$ is well-defined and finite almost surely
under $F$.  Specifically, the natural domain for the popular {\em
  logarithmic score}\/ (LogS; eq.~\eqref{eq:LogS}) is the class
$\cL_1$ of the probability distribution with densities, and the
respective score divergence is the Kullback-Leibler
divergence. The logarithmic score is local \citep{Bernardo1979},
  that is, it evaluates a predictive model based only on the density
  value at the realizing outcome. Conceptually, this means that the
  logarithmic score ignores the model's predicted probabilities of
  events that could have happened but did not.  For the {\em
  continuous ranked probability score}\/ (CRPS; eq.~\eqref{eq:CRPS}),
the natural domain is the class $\cM_1$ of the probability
distributions with finite mean. The LogS and CRPS are both
strictly proper relative to their respective natural domains. In
  contrast to the logarithmic score, the CRPS rewards predictive
  distributions that place mass close to the realizing outcome, a
  feature that is often called `sensitivity to distance'
  \citep[e.g.][Section 2]{MathesonWinkler1976}. While various authors
  have argued in favor of either locality or sensitivity to distance,
  the choice between these two contrasting features appears ultimately
  subjective.  Finally, the natural domain for the {\em
  Dawid--Sebastiani score}\/ {\citep[DSS;][]{DawidSebastiani1999}} is
the class $\cM_2$ of the probability distributions with strictly
positive, finite variance.  This score is proper, but not strictly
proper, relative to $\cM_2$.

\subsection{Consistent approximations}

To study the combined effects of the choices of approximation method
and scoring rule in the evaluation of Bayesian predictive
distributions, we introduce the notion of a {\em consistent}\/
approximation procedure.

Specifically, let $\thetaseq$ or $\Xseq$, where $X_i \sim \Fcdi$, be
output from a generic MCMC algorithm with the following property. 
\begin{itemize}
	\item[(A)] The process $\thetaseq$ is stationary and ergodic with
	invariant distribution $\Ppost$.
\end{itemize}
As noted, assumption (A) implies that $\Xseq$ is stationary and
ergodic with invariant distribution $F_0$.  Consider an approximation
method that produces, for all sufficiently large positive integers
$m$, an estimate $\FAM$ that is based on $\thetam$ or $\Xm$,
respectively.  Let $\myS$ be a proper scoring rule, and let $\cF$ be
the associated natural domain.  Then the approximation method is {\em
	consistent relative to the scoring rule\/ $\myS$ at the
	distribution\/ $F_0 \in \cF$} if $\FAM \in \cF$ for all
sufficiently large $m$, and
\[
\dS(\FAM, F_0) \longrightarrow 0 
\]
or, equivalently, $\myS(\FAM, F_0) \to \myS(F_0, F_0)$ almost surely as
$m \to \infty$.  This formalizes the idea that under continued MCMC
sampling, the approximation ought to perform as well as the unknown
true posterior predictive distribution.  We contend that this is a
highly desirable property in practical work.

Note that $\FAM$ is a random quantity that depends on the sample
$\thetam$ or $\Xm$.  The specific form of the divergence stems from
the scoring rule, which mandates convergence of a certain functional
of the estimator or approximation, $\FAM$, and the theoretical
posterior predictive distribution, $F_0$.  As we will argue, this
aspect has important implications for the choice of scoring rule and
approximation method.

Our concept of a consistent approximation procedure is independent of
the question of how well a forecast model represents the `true'
uncertainty.  The definition thus allows to separate the problem of
interest, namely, to find a good approximation $\FAM$ to $F_0$, from
the distinct task of finding a good probabilistic forecast
$F_0$.\footnote{It is possible for an inconsistent approximation
  to a {\em misspecified}\/ posterior predictive distribution
  $F_0$ to yield better forecasts than a consistent approximation
  that approaches the misguided $F_0$.  However, the
  misspecification can be detected by diagnostic tools such as
  probability integral transform histograms; see
  \citet{Dawid1984} and \citet{DieboldEtAl1998}.  The
  appropriate remedy thus is to improve the model
  specification.  Once a well-specified model has been found, the
    use of a consistent approximation improves the predictive
    performance further.}  We further emphasize that we study
convergence in the sample size, $m$, of MCMC output, given a fixed
number of observations, say, $T$, used to fit the model.  Our analysis
is thus distinct from traditional Bayesian asymptotic analyses that
study convergence of the posterior distribution as $T$ becomes larger
and larger \citep[see, e.g.,][Section 4]{GelmanEtAl2014book}, thereby
calling for markedly different technical tools.

\subsection{Relation to total variation and Wasserstein distances}\label{sec:distances}

Our focus on score divergences (in particular, on $d_{\text{LogS}}$
and $d_{\text{CRPS}}$) is motivated by their natural relation to
scoring rules, which in turn are popular tools in the applied
literature on probabilistic forecasting. As reviewed by
\cite{GibbsSu2002}, many other distance metrics for comparing two
probability distributions have been proposed in the literature. Among
these metrics, the total variation distance ($\dTV$) has received much
attention in theoretical work on MCMC
\citep[e.g.][]{Tierney1994,Rosenthal1995}, and is thus particularly
relevant in our context.  The total variation distance between two
absolutely continuous probability measures with densities $f$ and $g$
is defined as
\begin{equation*}
\dTV(F, G) = \frac{1}{2}\int_{-\infty}^\infty |f(z)-g(z)|\, \dd z.
\end{equation*}
As $2 \hspace{0.2mm} \dTV(F,G)^2 \leq d_{\text{LogS}}(F,G)$
\citep[e.g.,][]{BarronEtAl1992}, convergence in terms of
  $d_{\text{LogS}}$ implies convergence in terms of $\dTV$.

The Wasserstein distance is a divergence function motivated by
  optimal transport problems \citep{Villani2008} and has received
much attention in statistics and machine learning
  \citep{PanaretosZemel2019}. Here, we limit our discussion to the
Wasserstein distance of order 1, which is most common in
practice, and denote the corresponding metric by
\[
\dW(F,G) 
= \int_{0}^{1} |F^{-1}(\alpha) - G^{-1}(\alpha)|\, \dd \alpha 
= \int_{-\infty}^{\infty} |F(z) - G(z)|\, \dd z,
\]
where $F^{-1}$ and $G^{-1}$ are the quantile functions of $F$ and
  $G$ respectively. \citet{BellemareEtAl2017} discuss shortcomings of
Wasserstein distances in estimation with stochastic gradient
descent methods and suggest $d_{\text{CRPS}}$ as a superior
alternative. This recommendation relates to the observation that
there is no proper scoring rule with $\dW$ as score divergence
\citep[][Theorem 2]{ThorarinsdottirEtAl2013}.

As $d_{\text{CRPS}}(F,G) \leq \dW(F,G)$, convergence in terms
  of $\dW$ implies convergence in terms of $d_{\text{CRPS}}$.  If $F$
  and $G$ have densities with support in a common interval of length
  $l$, $\dW(F,G) \leq l \cdot \dTV(F,G) \leq l \cdot
  \sqrt{d_{\text{LogS}}(F,G)/2}$, so in this case consistency
relative to the logarithmic score implies consistency relative to
the CRPS.  For further relations to the Kolmororov, L\'evy,
  Prohorov and bounded Lipschitz distances see Section 2.4 of
  \citet{HuberRonchetti2009}.

\section{Consistency results and computational complexity}  \label{sec:theory}

\begin{table}
	
	\caption{Upper bounds on the computational complexity of
		approximation methods in terms of the size $m$ of the MCMC sample
		$\thetam$ or $\Xm$, respectively, for pre-processing, and for the
		exact computation of the {CRPS, Dawid--Sebastiani score (DSS)
			and logarithmic score (LogS)}. \label{tab:cost}}
	
	\bigskip
	
	\centering
	
	\begin{tabular}{lllll}
		\toprule
		Approximation method & Pre-processing  & CRPS &  DSS & LogS \\ 
		\midrule
		MP          & ${\cal O}(1)$ & ${\cal O}(m^2)$        &  ${\cal O}(m^2)$ & ${\cal O}(m)$ \\           
		ECDF      & ${\cal O}(1)$ & ${\cal O}(m \log m)$ &  ${\cal O}(m)$     & \\           
		KD          & ${\cal O}(m)$ & ${\cal O}(m^2)$        & ${\cal O}(m)$      & ${\cal O}(m)$ \\           
		Gaussian & ${\cal O}(m)$ & ${\cal O}(1)$            & ${\cal O}(1)$       & ${\cal O}(1)$ \\           
		\bottomrule
	\end{tabular}
	
	\bigskip
	
\end{table}

We now investigate sufficient conditions for consistency of the
aforementioned approximation methods, namely, the
mixture-of-parameters (MP) estimator $\FMP$ in \eqref{eq:MP}, the
empirical CDF (ECDF) method $\FECDF$ in \eqref{eq:ECDF}, the kernel
density (KD) estimate $\fKD$ in \eqref{eq:KD}, and the Gaussian
approximation (GA) $\FGA$ in \eqref{eq:GA}.  Table \ref{tab:cost}
summarizes upper bounds on the computational cost of
pre-processing and computing the CRPS, Dawid--Sebastiani score
	(DSS) and logarithmic score (LogS) under these methods in terms of
the size $m$ of the MCMC sample $\thetam$ or $\Xm$, respectively.

Consistency requires the convergence of some functional of the
approximation, $\FAM$, and the true posterior predictive distribution,
$F_0$. {The conditions} to be placed on the
Bayesian model {$F_0$}, the estimator $\FAM$, and the dependence
structure of the MCMC output {depend on} the scoring rule at hand.

\subsection{Mixture-of-parameters estimator}  \label{sec:MP}

We now establish consistency of the mixture-of-parameters estimator
$\FMP$ in \eqref{eq:MP} relative to the CRPS, DSS and
	logarithmic score. The proofs are deferred to Appendix
\ref{app:thm12}.

\medskip 
\noindent 
{\bf Theorem 1 (consistency of mixture-of-parameters approximations
	relative to the CRPS and DSS).}  {\em Under assumption (A),
	the mixture-of-parameters approximation is consistent relative to
	the CRPS at every distribution\/ $F_0$ with finite mean, {and
		consistent relative to the DSS at every distribution\/ $F_0$ with
		strictly positive, finite variance.}}

\medskip
Theorem 1 is the best possible result of its kind: It applies to every
distribution in the {natural domain} and does not invoke any
assumptions on the Bayesian model.  In contrast, Theorem 2 hinges on
rather stringent further conditions on the distribution $F_0$ and the
Bayesian model \eqref{eq:F0}, as follows.

\begin{itemize}
	\item[(B)] The distribution $F_0$ is supported on some bounded
	interval $\Omega$.  It admits a density, $f_0$, that is continuous
	and strictly positive on $\Omega$.  Furthermore, the density $\fcd$
	is continuous for every $\theta \in \Theta$.
\end{itemize}

\noindent 
{\bf Theorem 2 (consistency of mixture-of-parameters approximations
	relative to the logarithmic score).}  {\em Under assumptions (A) and
	(B), the mixture-of-parameters approximation is consistent relative
	to the logarithmic score at the distribution\/ $F_0$.}

\medskip
While we believe that the mixture-of-parameters technique is
consistent under weaker assumptions, this is the strongest result that
we have been able to prove. In particular, condition (B) does not
allow for the case $\Omega = \real$.  However, practical applications
often involve a truncation of the support for numerical reasons, as
exemplified in Section \ref{sec:simulation}, and in this sense the
assumption may not be overly restrictive.

Computing the logarithmic score and the
DSS for a predictive distribution $\FMP$ of the form \eqref{eq:MP} is straightforward. To compute the CRPS, we
note from eq.~(21) of \citet{GneitingRaftery2007} that
\begin{equation}  \label{eq:CRPS_mixture}
\text{CRPS} \left( \FMP, y \right) = 
\frac{1}{m} \sum_{i=1}^m \E |Z_i - y| - \frac{1}{2 m^2} \sum_{i=1}^m \sum_{j=1}^m \E|Z_i - Z_j|,
\end{equation}  
where $Z_i$ and $Z_j$ are independent random variables with
distribution $\Fcdi$ and $\Fcdj$, respectively.  The expectations on
the right-hand side of \eqref{eq:CRPS_mixture} often admit closed form
expressions that can be derived with techniques described by
\citet{Jordan2016} and \citet{Taillardat&2016}, including but not
limited to the {ubiquitous} case of Gaussian variables. Then the
evaluation requires ${\cal O}(m^2)$ operations, as reported in Table
\ref{tab:cost}. In Appendix \ref{sec:CRPS}, we provide details and
investigate the use of numerical integration in \eqref{eq:CRPS}, which
provides an attractive, computationally efficient alternative.

\subsection{Empirical CDF-based approximation}  \label{sec:ECDF}

The empirical CDF-based approximation $\FECDF$ in \eqref{eq:ECDF},
which builds on a simulated sample $\Xm$, is consistent relative to
the CRPS {and DSS} under minimal assumptions.  We prove the
following result in Appendix \ref{app:thm3}, which is the best
possible of its kind, as it applies to every distribution in the
{natural domain} and does not invoke any assumptions on the
Bayesian model.

\medskip 
\noindent 
{\bf Theorem {3} (consistency of empirical CDF-based
	approximations {relative to the CRPS {and DSS)}}.}  {\em
	Under assumption (A), the empirical CDF technique is consistent
	relative to the CRPS at every distribution\/ $F_0$ with finite mean,
	{and consistent relative to the DSS at every distribution\/
		$F_0$ with strictly positive, finite variance.}}

\medskip
{ As stated in Table \ref{tab:cost}}, the computation of the CRPS under $\FECDF$ requires
$\mathcal{O}(m\log{}m)$ operations only.
Specifically, let $X_{(1)} \leq \cdots \leq X_{(m)}$ denote the order
statistics of $X_1, \ldots, X_m$, which can be obtained in
$\mathcal{O}(m\log{}m)$ operations. Then 
\begin{equation}  \label{eq:Hersbach}
\text{CRPS} \left( \FECDF, y \right) = 
\frac{2}{m^2} \sum_{i = 1}^m (X_{(i)} - y)\left(m \ind\{y < X_{(i)}\} - i + \frac{1}{2}\right)\! ; 
\end{equation}
see \citet[Section 6]{Jordan2016} for details. A special case of eq.~\eqref{eq:CRPS_mixture} suggests another way of
computing the CRPS, in that
\begin{equation}  \label{eq:kernelrepr}
\text{CRPS} \left( \FECDF, y \right) = 
\frac{1}{m} \sum_{i=1}^m |X_i - y| - \frac{1}{2 m^2} \sum_{i=1}^m \sum_{j=1}^m |X_i - X_j|. 
\end{equation}
The representations in \eqref{eq:Hersbach} and \eqref{eq:kernelrepr} are 
algebraically equivalent, but the latter requires $\mathcal{O}(m^2)$ operations and thus is inefficient.


While the consistency results support the use of both $\FMP$ and $\FECDF$, Rao-Blackwellization arguments indicate superiority of $\FMP$.

\medskip 
\noindent 
{\bf  Theorem 4  (comparison of $\FMP$ and $\FECDF$).}  
{\em  Under assumption (A), $\E\, \FMP(z) = \E\, \FECDF(z)$ and $\textnormal{Var}\, \FMP(z) \leq \textnormal{Var}\, \FECDF(z)$ for any $z\in\Omega$ and $m \in \mathbb{N}$. If furthermore $F_0$ has finite mean, then $\E\, d_{\textnormal{CRPS}}\left(\FMP, F_0\right) \leq \E\, d_{\textnormal{CRPS}}\left(\FECDF, F_0\right)$ for any $m\in\mathbb{N}$.
}

\medskip 

Theorem 4 demonstrates that $\FMP$ outperforms $\FECDF$ in terms of expected divergence, for every given sample size $m$. Proposition 5 of \citet{BolinWallin2016} shows that if $F_0$ is a normal location-scale mixture then the CRPS under the mixture-of-parameters estimator additionally has smaller variance than under the
empirical CDF-based approximation. 

Despite the theoretical superiority of $\FMP$, $\FECDF$ may be attractive in practice, especially if the conditional distributions $F_c(\cdot|\theta)$ underlying $\FMP$ are difficult to compute analytically. For example, this may occur if the predictand $Y$ is modeled only indirectly (such as when $Y$ is the maximal element of a vector valued random variable). 

\subsection{Kernel density estimator}  \label{sec:KD}

We now discuss conditions for the consistency of the kernel density
estimator $\fKD$.  In the present case of dependent samples
$\Xm$, judicious choices of the bandwidth $h_m$ in \eqref{eq:KD}
require knowledge of dependence properties of the sample, and the
respective conditions are difficult to verify in practice.

{ The score divergence associated with the
  logarithmic score is the Kullback-Leibler divergence, which is
  highly sensitive to tail behavior.  Therefore, consistency of $\fKD$
  requires that the tail properties of the kernel $K$ in \eqref{eq:KD}
  and the true posterior predictive density $f_0$ be carefully
  matched, and any results tend to be technical
  (cf.~\citet{Hall1987}).  \citet{Roussas1988},
    \citet{GyoerfiEtAl1989}, \citet{Yu1993} and \citet{Liebscher1996},
    among others, establish almost sure strong uniform consistency of
  $\fKD$ under $\alpha$- or $\beta$-mixing and other conditions.  As
  noted in Appendix \ref{app:thm12}, almost sure strong uniform
  consistency then implies consistency relative to the logarithmic
  score under assumption (B).  Based on \citet{Hansen2008} who proves
  general results we give conditions for consistency of the kernel
  density estimator $\fKD$, and summarize the relevant assumptions
  in the following condition.

\begin{itemize}

	\item[(H)] For the kernel function $K$, the bandwidth $h_m$,
          and the dependence properties of $(X_i)_{i=1,2,\dots}$
          assumptions 1--3 and the conditions of Theorem 7 of
          \citet{Hansen2008} are satisfied.

\end{itemize}

\medskip 
\noindent 
{\bf  Theorem 5  (consistency of kernel density estimator-based approximations
	relative to the LogS).}  {\em  Under assumptions (A), (B), and (H),
	the kernel density estimator-based approximation technique 
	is consistent relative to the logarithmic score at the distribution $F_0$.}

\medskip 
The result is a direct consequence of \citet[][Theorem 7]{Hansen2008}
who further provides optimal convergence rates.  However, the
respective conditions are stringent and difficult to check in
practice. Indeed, \citet[p.~57]{Wasserman2006} opines that ``Despite
the natural role of Kullback-Leibler distance in parametric
statistics, it is usually not an appropriate loss function in
smoothing problems''.

Under the conditions of Theorem 5, consistency of $\FKD$ relative to
the CRPS follows directly, see Section \ref{sec:distances}.  Kernel
density estimation approximations are generally not consistent
relative to the DSS due to the variance inflation induced by typical
choices of the bandwidth. However, adaptations based on rescaling or
weighting allow for kernel density estimation under moment
constraints, see, e.g., \citet{Jones1991} and
\citet{HallPresnell1999}.

As this brief review suggests, the theoretical properties of kernel
density estimators depend on the specifics of both the MCMC sample and
the estimator. However, under the CRPS and DSS, a natural alternative
is readily available: The empirical CDF approach is simpler and
computationally cheaper than kernel density estimation, and is
consistent under weak assumptions (Theorem 3).

In our simulation and data examples, we use a simple implementation of
kernel density estimator-based approximations based on the
Gaussian kernel and the \cite{Silverman1986} plug-in
rule for bandwidth selection. This leads to the specific form
\begin{equation}  \label{eq:KD_GA} 
\FKD(x) = \frac{1}{m} \sum_{i=1}^m \Phi \! \left( \frac{x-X_i}{h_m} \right) \! ,
\end{equation} 
where $\Phi$ denotes the CDF of the standard normal distribution,
and
\begin{equation}  \label{eq:Silverman}
h_m = 1.06~\hat{A}_m \, m^{-1/5},
\end{equation} 
where $\hat A_m = \text{min}(\hat \sigma_m, \frac{\text{IQR}_m}{1.34})$ is the minimum of the standard deviation and the (scaled) interquartile range $\text{IQR}_m$ of $\Xm$. The 
pre-processing costs of the procedure are ${\cal O}(m)$, as shown in Table
\ref{tab:cost}. This choice of $h_m$, which is implemented in the \verb|R| function \verb|bw.nrd| \citep{R2016}, is motivated by simulation evidence in
\citet{HallEtAl1995}. Using the \citet{SheatherJones1991} rule or
cross-validation based methods yields slightly inferior results in our
experience.\footnote{\citet{SkoeldRoberts2003} and
  \cite{KimEtAl2016} discuss bandwidth selection rules that are
  motivated by density estimation in MCMC samples. However, both
  studies rely on mean integrated squared error criteria that are
  different from the performance measures we consider here.}  }

\subsection{Gaussian approximation}  \label{sec:Gaussian}

A parametric approximation method fits a member of a fixed
parametric family, say $\cF_\Gamma$, of probability distributions to
the MCMC sample $\Xm$.  The problem of estimating the unknown
distribution $F_0$ is thus reduced to a finite-dimensional parameter
estimation problem.  The most important case is the {\em quadratic}\/
or {\em Gaussian approximation}, which takes $\cF_\Gamma$ to be the
Gaussian family, so that
\[
\FGA(x) = \Phi \! \left( \frac{x - \hat{\mu}_m}{\hat{\sigma}_m} \right) \! ,
\]
where $\hat{\mu}_m$ and $\hat{\sigma}_m$ are the empirical mean and
standard deviation of $\Xm$.  If $F_0$ has a density $f_0$ that is
unimodal and symmetric, the approximation can be motivated by a Taylor
series expansion of the log predictive density at the mode, similar to
Gaussian approximations of posterior distributions in large-sample
Bayesian inference \citep[e.g.][Chapter 4]{KassRaftery1995, GelmanEtAl2014book}.

If $F_0$ is not Gaussian, $\FGA$ fails to be consistent relative
  to the logarithmic score and CRPS. However, the Ergodic Theorem
implies that the Gaussian approximation is consistent relative to the
Dawid--Sebastiani score under minimal conditions.

\medskip 
\noindent 
{\bf  Theorem {6}  (consistency of Gaussian approximations
	relative to the DSS).}  {\em  Under assumption (A), the Gaussian
	approximation technique is consistent relative to the DSS at every
	distribution\/ $F_0$ with  strictly positive,  finite
	variance.}

\medskip 
We also note that the logarithmic score for the Gaussian
approximation $\FGA$ corresponds to the Dawid--Sebastiani score for
the empirical CDF-based approximation $\FECDF$, in that
\[
\text{LogS} \! \left( \FGA, y \right) = 
\frac{1}{2} \left( \log 2\pi + \text{DSS} \! \left( \FECDF, y \right) \right)
\]
for $y \in \real$.  Therefore, the Gaussian approximation under the
logarithmic score yields model rankings that are identical to those
for the empirical CDF technique under the Dawid--Sebastiani score.
From an applied perspective, this equivalence suggests that the
  {inconsistency of the} Gaussian approximation may not be overly
  {problematic} when {the approximation is} used in concert with the
  logarithmic score, an assessment that is in line with empirical
  findings by \citet{WarneEtAl2016}. However, researchers should be
  aware of the fact that they are effectively using a proper, but not
  strictly proper, scoring rule (namely, the Dawid--Sebastiani score)
  that focuses on the first two moments of the predictive distribution
  only.

\section{Simulation study}  \label{sec:simulation}

We now investigate the various approximation methods in a simulation
study that is designed to emulate MCMC behavior with
dependent samples. Here, the posterior predictive distribution $F_0$
is known by construction, and so we can compare the different
approximations to the true forecast distribution. For simplicity, our choice of $F_0$ is fixed and does not correspond to a particular Bayesian model.\footnote{In Section S4 of the Online Supplement, we consider another simulation design that is based on a concrete Bayesian model (analysis of the normal model, using normal and inverse Gamma priors), yielding a posterior predictive distribution $F_0$ that depends on the data but is otherwise similar to the one considered here. While the design in the Online Supplement is necessarily more complex, all results remain qualitatively the same.}

In order to judge the quality of an approximation $\FAM$ of $F_0$ we
consider the score divergence $\dS(\FAM, F_0)$.  Note that $\dS(\FAM,
F_0)$ is a random variable, since $\FAM$ depends on the particular
MCMC sample $\thetam$ or $\Xm$.  In our results below, we therefore
consider the distribution of $\dS(\FAM, F_0)$ across repeated
simulation runs.  For a generic approximation method producing an
estimate $\FAM$, we proceed as follows:

\begin{itemize} 
	
	\item For simulation run $k = 1, \dots, K$:
	
	\begin{itemize} 
		
		\item Draw MCMC samples $\thetamk$ and $\Xmk$ 
		
		\item Compute the approximation $\FAM^{(k)}$ and the divergence
		$\dS(\FAM^{(k)}, F_0)$ for the approximation methods and scoring
		rules under consideration.
		
	\end{itemize} 
	
	\item For each approximation method and scoring rule, summarize the
	distribution of \newline $\dS(\FAM^{(1)}, F_0), \ldots, \dS(\FAM^{(K)},
	F_0)$.
	
\end{itemize}

\noindent
In order to simplify notation, we typically suppress the superscript
that identifies the Monte Carlo iteration.  The results below are
based on $K = 1\,000$ replicates.

\subsection{Data generating process}  \label{sec:dgp}

We generate sequences $\thetam$ and $\Xm$ in such a way that the 
invariant distribution,
\[
F_0(x) =  \int_{(0,\infty)} \Phi \! \left( \frac{x}{\theta} \right) \dd H_0(\theta^2),
\]
where $\Phi$ denotes the standard normal CDF, is a compound Gaussian
distribution or normal scale mixture.  Depending on the measure $H_0$,
which assumes the role of the posterior distribution $\Ppost$ in the
general Bayesian model \eqref{eq:F0}, $F_0$ can be modeled flexibly,
including many well-known parametric distributions
\citep{Gneiting1997}. As detailed below, our choice of $H_0$ implies
that
\begin{equation}  \label{eq:sim.F0}
F_0(x) = \textsf{T} \left( x \, \bigg| \, 0, \frac{ns}{n+2}, n + 2 \right)  \; 
\end{equation}
where $\textsf{T}(\cdot | a, b, c )$ denotes the CDF of a variable $Z$
with the property that $(Z-a)/\sqrt{b}$ is standard Student $t$
distributed with $c$ degrees of freedom.  To mimic a realistic MCMC
scenario with dependent draws, we proceed as proposed by
\citet{FoxWest2011}. Given parameter values $n > 0$, $s > 0$ and
$\alpha \in (-1, 1)$, let
\begin{align}
\psi_i                        & \sim \text{IG} \! \left( \frac{1}{2} \, (n+3), \frac{1}{2} \, ns(1-\alpha^2) \right) \! , \label{eq:sim.begin} \\
\upsilon_i \, | \, \psi_i & \sim \cN \! \left( \alpha, \frac{\psi_i}{ns} \right) \! , \\
\theta_i^2                  & = \psi_i + \upsilon_i^2 \, \theta_{i-1}^2, \rule{0mm}{6mm} \\
X_i \, | \, \theta_i^2     & \sim \cN(0,\theta_i^2) \rule{0mm}{7mm} \label{eq:sim.end}
\end{align}
where IG is the Inverse Gamma distribution, parametrized such that $Z
\sim \text{IG}(a, b)$ when $1/Z \sim \text{G}(a, b)$, with G being the
Gamma distribution with shape $a \geq 0$ and rate $b > 0$.

\begin{table}

	\caption{Hyper-parameters for the data generating process in the
		simulation setting of equations \eqref{eq:sim.begin} to
		\eqref{eq:sim.end}.  \label{tab:sim}}
	
	\bigskip
	
	\centering
	
	\begin{tabular}{lll}
		\toprule
		Parameter & Main role                              & Value(s) considered \\ 
		\midrule
		$\alpha$  & Persistence of $\theta_i^2$            & \{0.1, 0.5, 0.9\} \\
		$s$       & Unconditional mean of $\theta_i^2$     & 2 \\
		$n$       & Unconditional variance of $\theta_i^2$ & \{12, 20\} \\ 
		\bottomrule
	\end{tabular}
	
	\bigskip
	
\end{table}

Table \ref{tab:sim} summarizes our choices for the parameter
configurations of the data generating process.  The parameter $\alpha$
determines the persistence of the chain, in that the unconditional
mean of $\upsilon_i^2$, which can be viewed as an average autoregressive
coefficient \citep[][Section 2.3]{FoxWest2011}, is given by
$(n\alpha^2 + 1) / (n + 1)$.  We consider three values, aiming to
mimic MCMC chains with different persistence properties.  The
parameter $s$ represents a scale effect, and $n$ governs the tail thickness of the
unconditional Student $t$ distribution in \eqref{eq:sim.F0}.  We
consider values of 12 and 20 that seem realistic for macroeconomic
variables, such as the growth rate of the gross domestic product, that feature prominently in the empirical literature.

\subsection{Approximation methods}  \label{sec:approximations}

We consider the following approximation methods, which have been
discussed in detail in Section \ref{sec:theory}. The first
approximation uses a sequence $\thetam$ of parameter draws, and the
other three employ an MCMC sample $\Xm$.

\begin{itemize}
	
	\item Mixture-of-parameters estimator $\FMP$ in \eqref{eq:MP}, which here is of the form 
	\[
	\FMP(x) = \frac{1}{m} \sum_{i=1}^m  \Phi \! \left( \frac{x}{\theta_i} \right) \! ,
	\]
	where $\theta_i$ is the predictive standard deviation drawn in MCMC iteration $i$.
	
	\item Empirical CDF-based approximation $\FECDF$ in \eqref{eq:ECDF}.
	
	\item The nonparametric kernel density estimator $\fKD$ using a
	Gaussian kernel and the Silverman rule for bandwidth selection, with
	predictive CDF $\FKD$ of the form \eqref{eq:KD_GA}. 
	
	\item Gaussian approximation $\FGA$ in \eqref{eq:GA}.
	
\end{itemize}

\noindent
It is interesting to observe that here $\FMP$ is a scale mixture of
centered Gaussian distributions, and $\FKD$ is a location mixture of
normal distributions, whereas the quadratic approximation $\FGA$ is a
single Gaussian.

The conditions for consistency of the mixture-of-parameters and
empirical CDF approximations relative to the CRPS in Theorems 1 and
{3} are satisfied.  Furthermore, one might argue that
numerically the support of $F_0$ and $\FMP$ is bounded (cf.~below),
and then the assumptions of Theorem 2 also are satisfied.  Clearly,
the Gaussian approximation fails to be consistent relative to the CRPS
or the logarithmic score, as $F_0$ is not Gaussian.

For each approximation method, scoring rule $\myS$, sample size $m$
and replicate $k$, we evaluate the score divergence $\dS(\FAM^{(k)},
F_0)$.  The divergence takes the form of a univariate integral
(cf.~Table \ref{tab:scores}) that is not available in closed form.
Therefore, we compute $\dS(\FAM^{(k)}, F_0)$ by numerical integration
as implemented in the \textsf{R} function \textsf{integrate}.  This is
unproblematic if the scoring rule is the CRPS.  For the logarithmic
score, the integration is numerically challenging, as the logarithm of
the densities needs to be evaluated in their tails.  We therefore
truncate the support of the integral to the minimal and maximal values
that yield numerically finite values of the integrand.

\subsection{Main results}

\begin{figure}[t]
	
	\centering
	
	\begin{tabular}{cc}
		Logarithmic score & CRPS \\ [-1cm]
		\includegraphics[width=0.5\textwidth]{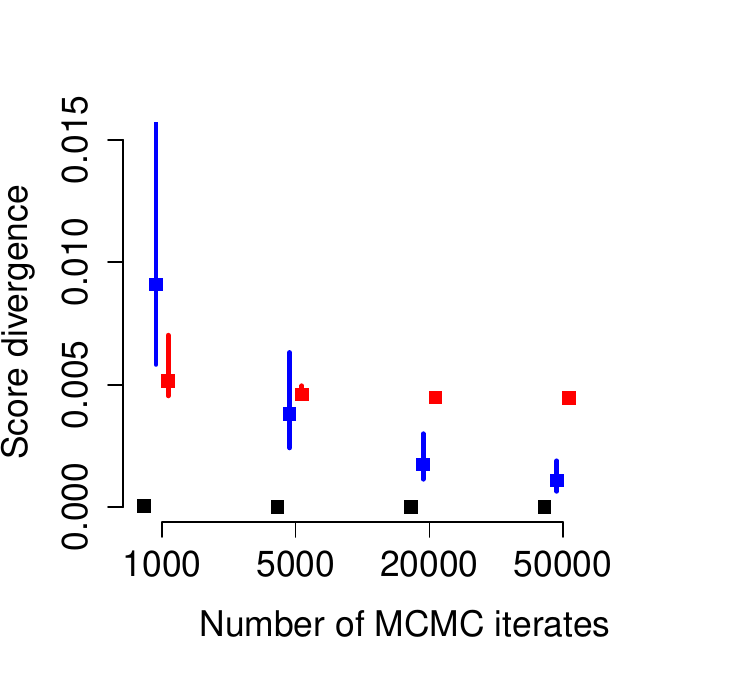} &
		\includegraphics[width=0.5\textwidth]{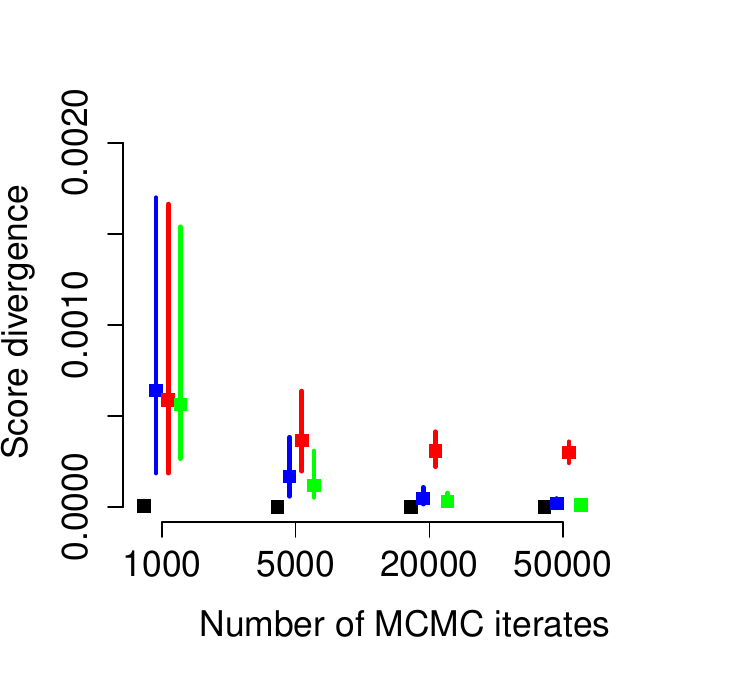} \\[-1.4cm] 
		\multicolumn{2}{c}{\includegraphics[width=\textwidth]{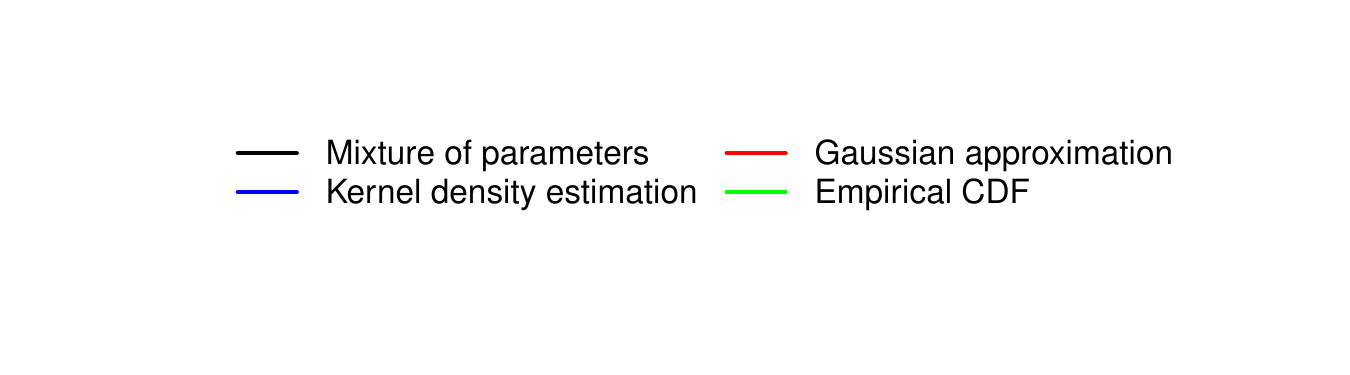}}
	\end{tabular}\vspace{-1.8cm}
	
	\caption{Score divergences in the simulation study with $(\alpha, s,
		n) = (0.5, 2, 12)$.  For a given method and MCMC sample size, the
		bars range from the 10th to the 90th percentile of the score
		divergences across 1\,000 replicates.  The squares mark the respective
		medians.  \label{fig:sim-size}}
	
\end{figure}

\begin{figure}[t]
	
	\centering
	
	\begin{tabular}{cc}
		Logarithmic score & CRPS \\ [-1cm]
		\includegraphics[width=0.5\textwidth]{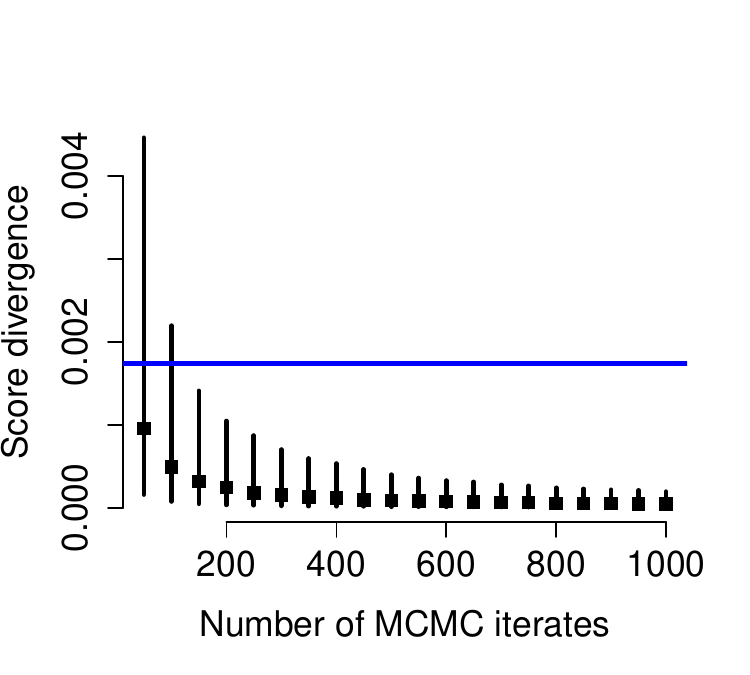} &
		\includegraphics[width=0.5\textwidth]{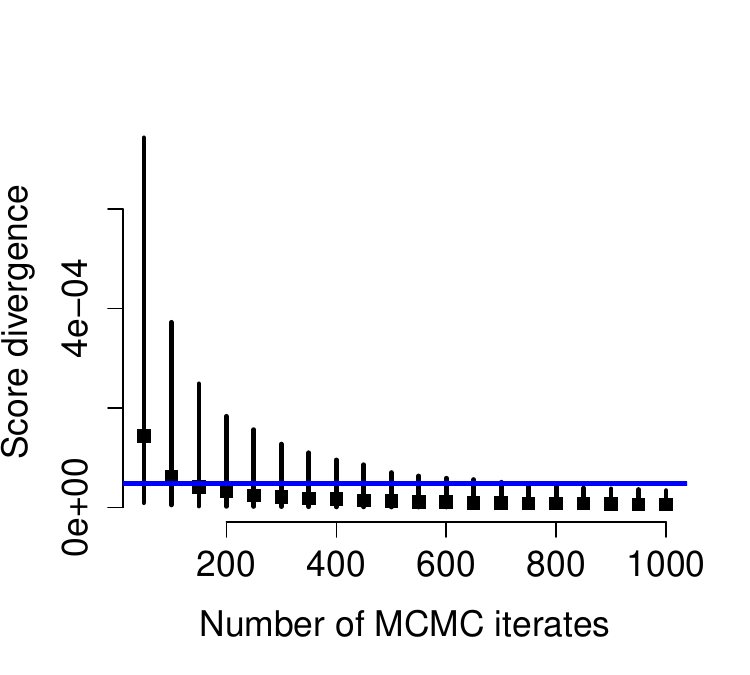}
	\end{tabular}
	
	\caption{Performance of the mixture-of-parameters estimator.  The
		design is as in Figure \ref{fig:sim-size}, but for smaller sample
		sizes.  For comparison, the blue horizontal line marks the median
		divergence of the kernel density estimator based on 20\,000
		draws.  \label{fig:sim-zoom}}
	
\end{figure}

In the interest of brevity, we restrict attention to results for a
single set of parameters of the data generating process, namely
$(\alpha, s, n) = (0.5, 2, 12)$.  This implies an unconditional
Student $t$ distribution with 14 degrees of freedom, and intermediate
autocorrelation of the MCMC draws. The results for the other
parameter constellations in Table \ref{tab:sim} are similar and
available in the Online Supplement.

Figure \ref{fig:sim-size} illustrates the performance of the
approximation methods under the logarithmic score and the CRPS, by
showing the distribution of the score divergence $\dS(\FAM, F_0)$ as
the sample size $m$ grows.  The mixture-of-parameters estimator
dominates the other methods by a wide margin: Its divergences are very
close to zero, and show little variation across replicates.  Under the
logarithmic score, the performance of the kernel density estimator is
highly variable across the replicates, even for large sample sizes.
The variability is less under the CRPS, where the kernel density
approach using the \citet{Silverman1986} rule of thumb for bandwidth
selection performs similar to the empirical CDF-based approximation.
Other bandwidth selection rules we have experimented with tend to be
inferior, as indicated by slower convergence and higher variability
across replicates.  Finally, we observe the lack of consistency of the
Gaussian approximation.

Figure \ref{fig:sim-zoom} provides insight into the performance of the
mixture-of-parameters approximation for small MCMC samples. Using as
few as 150 draws, the method attains a lower median CRPS divergence
than the kernel density estimator based on 20\,000 draws. The
superiority of the mixture-of-parameters estimator is even more
pronounced under the logarithmic score, where only 50 draws are
required to outperform the kernel density estimator based on 20\,000
draws.

\subsection{Thinning the MCMC Sample}

In Appendix \ref{sec:simulation-thinning}, we present simulation
  analyses of the effects of thinning an MCMC sample (i.e., keeping
  only every $\tau$th draw, where $\tau \in \mathbb{N}$ is the
  thinning factor), which is often done in practice with the goal of
  reducing autocorrelation in the MCMC draws. From a practical
  perspective, the analysis in Appendix \ref{sec:simulation-thinning}
  suggests that thinning is justified if, and only if, a small MCMC
  sample is desired and the mixture-of-parameter estimator is
  applied. Two arguments in favor of a small sample appear
  particularly relevant even today. First, storing large amounts of
  data on public servers (as is often done for replication purposes)
  may be costly or inconvenient.  Second, post-processing procedures
  such as score computations applied to the MCMC sample may be
  computationally demanding {(cf.~Table \ref{tab:cost})}, and
  therefore may encourage thinning.

\section{Economic data example}  \label{sec:casestudy}

In real-world uses of Bayesian forecasting methods, the posterior
predictive distribution $F_0$ is typically not available in closed
form.  Therefore, computing or estimating the object of interest for
assessing consistency, i.e., the score divergence $\dS(\FAM, F_0)$, is
not feasible.  In the subsequent data example, we thus compare the
approximation methods via their out-of-sample predictive performance,
and examine the variation of the mean scores across chains obtained by
replicates with distinct random seeds.  While studying the predictive
performance does not allow to assess consistency of the approximation
methods, it does allow us to assess the variability and applicability
of the approximations in a practical setting.

\subsection{Data} 

We consider quarterly growth rates of U.S.~real gross domestic product
(GDP), as illustrated in the Online Supplement.  The training
sample used for model estimation is recursively expanded as
forecasting moves forward in time.  We use the real-time data set
provided by the Federal Reserve Bank of
Philadelphia\footnote{\url{https://www.phil.frb.org/research-and-data/real-time-center/real-time-data/}},
which provides historical snapshots of the data vintages available at
any given date in the past, and consider forecasts for the period from
the second quarter of 1996 to the third quarter of 2014, for a total
of $T = 74$ forecast cases.  For brevity, we present results for a
prediction horizon of one quarter only.  The Online Supplement
contains results for longer horizons, which are qualitatively similar
to the ones presented here.

\subsection{Probabilistic forecasts} 

To construct density forecasts, we consider an autoregressive (AR)
model with a single lag and state-dependent error term variance, in
that
\begin{equation}   \label{eq:ms}
Y_t = \nu + \alpha Y_{t-1} + \varepsilon_t, 
\end{equation}
where $\varepsilon_t \sim \mathcal{N}(0, \eta^2_{s_t})$ and $s_{t}
\in \{1, 2\}$ is a discrete state variable that switches according to
a first-order Markov chain.  The model, which is a variant of the
Markov switching model proposed by \citet{Hamilton1989}, provides a
simple description of time-varying heteroscedasticity.  The latter is
an important stylized feature of macroeconomic time series 
\citep[see, e.g.,][]{ClarkRavazzolo2014}.

We conduct Bayesian inference via a Gibbs sampler, for which we
give details in Appendix \ref{app:implementation}.  Let $\theta_i$
denote the complete set of latent states and model parameters at
iteration $i$ of the Gibbs sampler.  Conditional on $\theta_i$, the
predictive distribution under the model in \eqref{eq:ms} is Gaussian
with mean $\mu_i = \mu(\theta_i)$ and standard deviation $\sigma_i =
\sigma(\theta_i)$, where we suppress time and forecast horizon for simplicity. At each forecast origin date $t = 1, \ldots, T = 74$, we produce 10\,000
burn-in draws, and use 40\,000 draws post burn-in.  We construct $16$
parallel chains in this way.  The (time-averaged) mean score of a given
approximation method, based on $m$ MCMC draws within chain $c = 1,
\ldots, 16$, is
\[
\bar{\myS}_{m,c} = \frac{1}{T} \sum_{t=1}^T \myS (\hat{F}_{m, c, t}, y_t),
\]
where $\hat{F}_{m, c, t}$ is the probabilistic forecast at time $t$.
The variation of $\bar{\myS}_{m, c}$ across chains $c$ is due to
differences in random seeds.  From a pragmatic perspective, a good
approximation method should be such that the values $(\bar{\myS}_{m,
	c})_{c=1}^{16}$ are small and display little variation.

\subsection{Results} 

\begin{figure}[t]
	
	\centering
	
	\begin{tabular}{cc}
		Logarithmic score & CRPS \\ 
		[-10mm]
		\includegraphics[width = 0.5\textwidth]{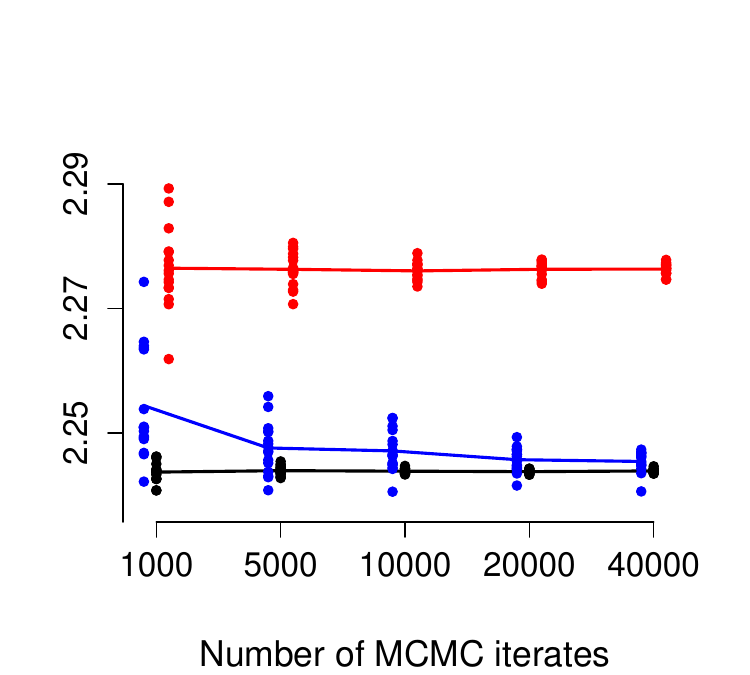} &
		\includegraphics[width = 0.5\textwidth]{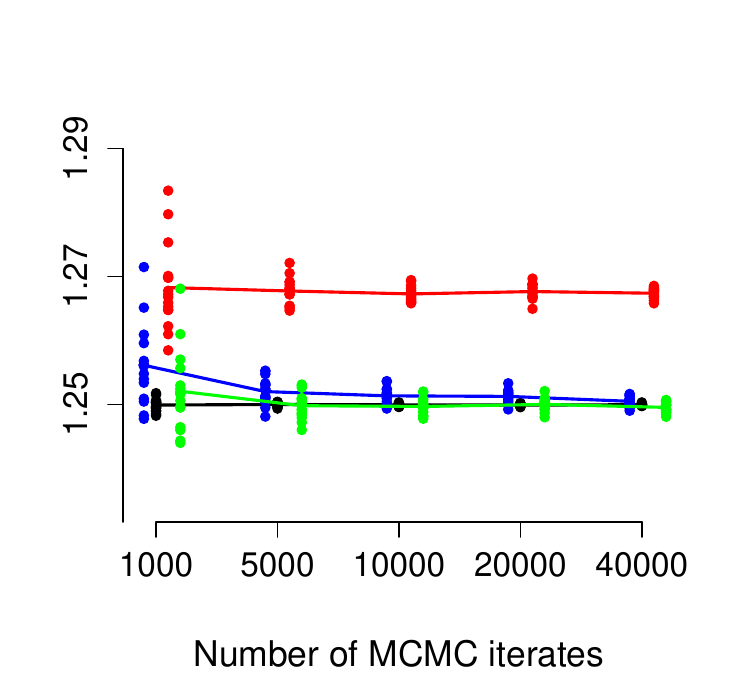} \\ [-1.4cm]
		\multicolumn{2}{c}{\includegraphics[width = \textwidth]{legend.pdf}} \\ 
	\end{tabular}\vspace{-1.8cm}
	
	\caption{Mean score in the data example against sample size.  The dots
		represent 16 parallel MCMC chains, and the lines connect averages
		across chains.
		\label{fig:empscores1}}
	
\end{figure}

In Figure \ref{fig:empscores1}, the mean score is plotted against the
size of the MCMC sample.  The mixture-of-parameters approximation
outperforms its competitors: Its scores display the smallest variation
across chains, for both the CRPS and the logarithmic score, and for
all sample sizes.  The scores of the mixture-of-parameter estimator
also tend to be lower (i.e., better) than the scores for the other
methods.  The kernel density estimator performs poorly for small
sample sizes, with the scores varying substantially across chains.
Under the CRPS, the kernel density estimator is dominated by the
empirical CDF technique, which can be interpreted as kernel density
estimation with a bandwidth of zero.

\section{Discussion}  \label{sec:discussion}

We have investigated how to make and evaluate probabilistic forecasts
based on MCMC output. The formal notion of consistency allows us to
assess the appropriateness of approximation methods within the
framework of proper scoring rules.  Despite their popularity in the
literature, Gaussian approximations generally fail to be
consistent.  Conditions for consistency depend on the scoring rule of
interest, and we have demonstrated that the mixture-of-parameters and
empirical CDF-based approximations are consistent relative to the CRPS 
under minimal conditions.  Proofs of consistency relative to the
logarithmic score generally rely on stringent assumptions.

In view of these theoretical considerations as well as the practical
perspective taken in our simulation and data examples, we {generally}
recommend the use of the mixture-of-parameters estimator, which
provides an efficient approximation method and outperforms all
alternatives.  This can be explained by the fact that it efficiently
exploits the parametric structure of the Bayesian
model. The empirical
CDF-based approximation provides a good alternative if the
  conditional distributions fail to be available in closed form, or
  if for some reason the draws are to be made directly from the
posterior predictive distribution, as opposed to using parameter
draws.  The empirical CDF-based approximation is available under the
CRPS and DSS but not under the logarithmic score, where a density is
required. Under the logarithmic score, the case for the
mixture-of-parameters estimator is thus particularly strong. In
particular, the score's sensitivity to the tails of the distribution
renders kernel density estimators unattractive from both theoretical
and applied perspectives.

Our recommendations have been implemented in the \textsf{scoringRules}
package for \textsf{R} \citep{R2016}; see \citet{JordanEtAl2015} for
details.  The functions and default choices aim to provide readily
applicable and efficient approximations.  The mixture-of-parameters
estimator depends on the specific structure of the Bayesian model and
can therefore not be covered in full generality.  However, the
implemented analytical solutions of the CRPS and logarithmic score
allow for straightforward and efficient computation. The
\textsf{scoringRules} package further contains functions and data for
replicating the simulation and case study, with details provided at
\url{https://github.com/FK83/scoringRules/blob/master/KLTG2020_replication.pdf}.

\citet{Ferro2014} studies the notion of a fair scoring rule in the
context of ensemble weather forecasts.  A scoring rule is called {\em
	fair}\/ if the expected score is optimal for samples with members
that behave as though they and the verifying observation were sampled
from the same distribution.  While certainly relevant in the context
of meteorological forecast ensembles, where the sample size $m$ is
typically between 10 and 50, these considerations seem less helpful in
the context of MCMC output, where $m$ is on the order of thousands and
can be increased at low cost.  Furthermore, the proposed small sample
adjustments and the characterization of fair scores hold for
independent samples only, an assumption that is thoroughly violated in
the case of MCMC.

We are interested in evaluating probabilistic forecasts produced via
MCMC, so that the predictive performance of a model during an
out-of-sample, test or evaluation period can be used to estimate its
forecast performance on future occasions.  In contrast, information
criteria suggest a different route towards estimating forecast
performance \citep{SpiegelhalterEtAl2002, Watanabe2010,HootenHobbs2015}.  They consider a method's in-sample performance,
and account for model complexity via penalty terms. Preferred ways of
doing so have been the issue of methodological debate, and a consensus
has not been reached; see, e.g., the comments in
\citet{GelmanEtAl2014paper} and \citet{SpiegelhalterEtAl2014}.  This
present analysis does not concern in-sample comparisons, and does not
address the question of whether these are more or less effective than
out-of-sample comparisons.  However, our results and observations
indicate that out-of-sample comparisons of the type considered here
yield robust results across a range of implementation choices.

Necessarily, the scope of this paper is restricted along several
dimensions. First, our theoretical results focus on consistency but do
not cover rates of convergence. Results on the latter tend to rely on
theoretical conditions that are hard to verify empirically, and the
plausibility of which is likely to depend on the specifics of the MCMC
algorithm. { In} contrast, many of our consistency
results require only minimal conditions that hold across a wide range
of sampling algorithms in the interdisciplinary applied
literature. Second, we have focused on univariate continuous forecast
distributions. The corresponding applied literature is large and
features a rich variety of implementation variants (cf. Table 1 of the
Online Supplement). Nevertheless, there are other empirically relevant
setups, notably simple functionals of a predictive distribution, discrete univariate distributions, and continuous multivariate
distributions. We briefly discuss each setup in turn.

Functionals such as quantiles summarize a predictive distribution, thus allowing for simpler interpretation and communication \citep{Raftery2016}. If the forecast user requires only a specific quantile of the predictive distribution, it seems natural to focus on this quantile for evaluation. Interestingly, the CRPS can be represented as the integral over (twice) the asymmetric piecewise linear scoring function which is commonly used to evaluate quantile forecasts \citep[][Equations 11 to 13]{GneitingRanjan2011}. Consequently, the CRPS divergence is the integral over the quantile score divergence. In this sense, results for quantiles are covered by our results in terms of the CRPS. The same argument applies if the functional sought is the exceedance probability at any given threshold value, as an immediate consequence of the standard representation of the CRPS (see Equation \ref{eq:CRPS}). In order to illustrate the argument numerically, Section S3 of the Online Supplement applies our simulation design to quantiles at two different levels, yielding results that are qualitatively very similar to our CRPS results for full predictive distributions. 

In relevant discrete settings, such as predicting
  probabilities of a binary or categorical outcome, the estimation
  problem becomes considerably simpler than for the real-valued
  case. The more complex case of integer-valued count data can be
  handled using methods similar to the ones we discuss. Instead of
  probability density functions, the count data case involves
  probability mass functions to which both the logarithmic score and
  the CRPS transfer naturally \citep{CzadoEtAl2009}. Furthermore, all
  of the approximation methods we discuss can be used in the count
  data case. For example, the mixture-of-parameters estimator can be
  used in concert with a Poisson or Negative Binomial
  specification. Similarly, \citet[Section 4]{ShirotaGelfand2017}
  consider eq.~\eqref{eq:kernelrepr} in a count data context, and
  kernel-type smoothing methods have been proposed for count data as
  well \citep{RajagopalanLall1995}.

The multivariate case features novel
  challenges. Perhaps most fundamentally, a consensus on practically
  appropriate choices of the scoring rule is yet to be reached
  \citep{GneitingEtAl2008, ScheuererHamill2015VS}.
  \citet[Section 4.2]{HeldEtAl2017} and \citet[Section
      3.3]{WhiteEtAl2019} propose the use of the empirical CDF
    approximation in concert with the multivariate energy score.  In
    this setting, analogues of our Theorem 3 hold, assuring
    consistency under weak conditions.  For kernel density estimators
  the `curse of dimensionality' applies, and for the
  mixture-of-parameters estimator we expect numerical challenges when
  evaluating, say, a log predictive density in a
  high-dimensional space. Clearly, there is considerable scope
    and opportunity for future research in these directions.

	\bibliographystyle{myims2} 
	\setlength{\bibsep}{0mm}
	\bibliography{bibliography_all_abbrev_table-update.bib}

\begin{thebibliography}{89}
\expandafter\ifx\csname natexlab\endcsname\relax\def\natexlab#1{#1}\fi
\expandafter\ifx\csname url\endcsname\relax
  \def\url#1{\texttt{#1}}\fi
\expandafter\ifx\csname urlprefix\endcsname\relax\def\urlprefix{URL }\fi
\providecommand{\eprint}[2][]{\url{#2}}

\bibitem[{Amisano and Giacomini(2007)}]{AmisanoGiacomini2007}
{Amisano, G.} and {Giacomini, R.} (2007).
\newblock {Comparing density forecasts via weighted likelihood ratio tests}.
\newblock \textit{J. Bus. Econom. Statist.}, {25}, 177--190.

\bibitem[{Barron et~al.(1992)Barron, Gy{\"o}rfi and van~der
  Meulen}]{BarronEtAl1992}
{Barron, A.~R.}, {Gy{\"o}rfi, L.} and {van~der Meulen, E.~C.} (1992).
\newblock Distribution estimation consistent in total variation and in two
  types of information divergence.
\newblock \textit{IEEE Trans. Inform. Theory}, {38}, 1437--1454.

\bibitem[{Bellemare et~al.(2017)Bellemare, Danihelka, Dabney, Mohamed,
  Lakshminarayanan, Hoyer and Munos}]{BellemareEtAl2017}
{Bellemare, M.~G.}, {Danihelka, I.}, {Dabney, W.}, {Mohamed, S.},
  {Lakshminarayanan, B.}, {Hoyer, S.} and {Munos, R.} (2017).
\newblock {The Cramer distance as a solution to biased Wasserstein gradients}.
\newblock Preprint, available at \url{http://arxiv.org/abs/1705.10743}.

\bibitem[{Bernardo(1979)}]{Bernardo1979}
{Bernardo, J.~M.} (1979).
\newblock Expected information as expected utility.
\newblock \textit{Ann. Statist.}, {7}, 686--690.

\bibitem[{Bolin and Wallin(2020)}]{BolinWallin2016}
{Bolin, D.} and {Wallin, J.} (2020).
\newblock {Multivariate type-G Mat\'ern stochastic partial differential
  equation random fields}.
\newblock \textit{J. R. Stat. Soc. Ser. B. Stat. Methodol.}, {82}, 215--239.

\bibitem[{Brier(1950)}]{Brier1950}
{Brier, G.~W.} (1950).
\newblock Verification of forecasts expressed in terms of probability.
\newblock \textit{Mon. Weather Rev.}, {78}, 1--3.

\bibitem[{Clark(2005)}]{Clark2005}
{Clark, J.~S.} (2005).
\newblock Why environmental scientists are becoming {B}ayesians.
\newblock \textit{Ecol. Lett.}, {8}, 2--14.

\bibitem[{Clark and McCracken(2013)}]{ClarkMcCracken2013}
{Clark, T.} and {McCracken, M.} (2013).
\newblock Advances in forecast evaluation.
\newblock In \textit{Handbook of Economic Forecasting} (G.~Elliott and
  A.~Timmermann, eds.), vol.~2. Elsevier, 1107--1201.

\bibitem[{Clark and Ravazzolo(2015)}]{ClarkRavazzolo2014}
{Clark, T.~E.} and {Ravazzolo, F.} (2015).
\newblock Macroeconomic forecasting performance under alternative
  specifications of time-varying volatility.
\newblock \textit{J. Appl. Econometrics}, {30}, 551--575.

\bibitem[{Cooke(1906)}]{Cooke1906}
{Cooke, W. E.~.} (1906).
\newblock Forecasts and verifications in {W}estern {A}ustralia.
\newblock \textit{Mon. Weather Rev.}, {34}, 23–24.

\bibitem[{Craiu and Rosenthal(2014)}]{CraiuRosenthal2014}
{Craiu, R.~V.} and {Rosenthal, J.~S.} (2014).
\newblock {Bayesian computation via Markov chain Monte Carlo}.
\newblock \textit{Annu. Rev. Stat. Appl.}, {1}, 179--201.

\bibitem[{Czado et~al.(2009)Czado, Gneiting and Held}]{CzadoEtAl2009}
{Czado, C.}, {Gneiting, T.} and {Held, L.} (2009).
\newblock Predictive model assessment for count data.
\newblock \textit{Biometrics}, {65}, 1254--1261.

\bibitem[{Dawid(1984)}]{Dawid1984}
{Dawid, A.~P.} (1984).
\newblock Present position and potential developments: {S}ome personal views.
  {S}tatistical theory: {T}he prequential approach.
\newblock \textit{J. R. Stat. Soc. Ser. A. Gen.}, {147}, 278--290.

\bibitem[{Dawid and Sebastiani(1999)}]{DawidSebastiani1999}
{Dawid, A.~P.} and {Sebastiani, P.} (1999).
\newblock {Coherent dispersion criteria for optimal experimental design}.
\newblock \textit{Ann. Statist.}, {27}, 65--81.

\bibitem[{Dehling and Philipp(2002)}]{DehlingPhilipp2002}
{Dehling, H.} and {Philipp, W.} (2002).
\newblock {Empirical process techniques for dependent data}.
\newblock In \textit{{Empirical Process Techniques for Dependent Data}}
  (H.~Dehling, T.~Mikosch and M.~S\o{}rensen, eds.). {Birkh\"auser}, {Boston},
  3--113.

\bibitem[{DelSole and Tippett(2014)}]{DelSoleTippett2014}
{DelSole, T.} and {Tippett, M.~K.} (2014).
\newblock Comparing forecast skill.
\newblock \textit{Mon. Weather Rev.}, {142}, 4658--4678.

\bibitem[{Diebold et~al.(1998)Diebold, Gunther and Tay}]{DieboldEtAl1998}
{Diebold, F.~X.}, {Gunther, T.~A.} and {Tay, A.~S.} (1998).
\newblock {Evaluating density forecasts with applications to financial risk
  management}.
\newblock \textit{Internat. Econom. Rev.}, {39}, 863--883.

\bibitem[{Diebold and Mariano(1995)}]{DieboldMariano1995}
{Diebold, F.~X.} and {Mariano, R.~S.} (1995).
\newblock {Comparing predictive accuracy}.
\newblock \textit{J. Bus. Econom. Statist.}, {13}, 253--263.

\bibitem[{Feldmann et~al.(2015)Feldmann, Scheuerer and
  Thorarinsdottir}]{FeldmannEtAl2015}
{Feldmann, K.}, {Scheuerer, M.} and {Thorarinsdottir, T.~L.} (2015).
\newblock Spatial postprocessing of ensemble forecasts for temperature using
  nonhomogeneous gaussian regression.
\newblock \textit{Mon. Weather Rev.}, {143}, 955--971.

\bibitem[{Ferro(2014)}]{Ferro2014}
{Ferro, C. A.~T.} (2014).
\newblock {Fair scores for ensemble forecasts}.
\newblock \textit{{Q. J. Royal Meteorol. Soc.}}, {140}, 1917--1923.

\bibitem[{Fox and West(2011)}]{FoxWest2011}
{Fox, E.~B.} and {West, M.} (2011).
\newblock {Autoregressive models for variance matrices: Stationary inverse
  Wishart processes}.
\newblock Preprint, available at \url{http://arxiv.org/abs/1107.5239}.

\bibitem[{Gelfand and Smith(1990)}]{GelfandSmith1990}
{Gelfand, A.~E.} and {Smith, A. F.~M.} (1990).
\newblock Sampling-based approaches to calculating marginal densities.
\newblock \textit{{J. Amer. Statist. Assoc.}}, {85}, 398--409.

\bibitem[{Gelman et~al.(2014{\natexlab{a}})Gelman, Carlin, Stern, Dunson,
  Vehtari and Rubin}]{GelmanEtAl2014book}
{Gelman, A.}, {Carlin, J.~B.}, {Stern, H.~S.}, {Dunson, D.~B.}, {Vehtari, A.}
  and {Rubin, D.~B.} (2014{\natexlab{a}}).
\newblock \textit{{Bayesian Data Analysis}}.
\newblock 3rd ed. {Chapman \& Hall/CRC}, Boca Raton.

\bibitem[{Gelman et~al.(2014{\natexlab{b}})Gelman, Hwang and
  Vehtari}]{GelmanEtAl2014paper}
{Gelman, A.}, {Hwang, J.} and {Vehtari, A.} (2014{\natexlab{b}}).
\newblock {Understanding predictive information criteria for Bayesian models}.
\newblock \textit{Stat. Comput.}, {24}, 997--1016.

\bibitem[{Gelman et~al.(1996)Gelman, Meng and Stern}]{GelmanEtAl1996}
{Gelman, A.}, {Meng, X.-L.} and {Stern, H.} (1996).
\newblock Posterior predictive assessment of model fitness via realized
  discrepancies.
\newblock \textit{Statist. Sinica}, {6}, 733--760.

\bibitem[{Genon-Catalot et~al.(2000)Genon-Catalot, Jeantheau and
  Lar{\'e}do}]{GenonEtAl2000}
{Genon-Catalot, V.}, {Jeantheau, T.} and {Lar{\'e}do, C.} (2000).
\newblock Stochastic volatility models as hidden {M}arkov models and
  statistical applications.
\newblock \textit{Bernoulli}, {6}, 1051--1079.

\bibitem[{Geweke(2005)}]{Geweke2005}
{Geweke, J.} (2005).
\newblock \textit{{Contemporary Bayesian Econometrics and Statistics}}.
\newblock John Wiley \& Sons, Hoboken.

\bibitem[{Geyer(1992)}]{Geyer1992}
{Geyer, C.~J.} (1992).
\newblock {Practical Markov chain Monte Carlo}.
\newblock \textit{Statist. Sci.}, {7}, 473--483.

\bibitem[{Giacomini and White(2006)}]{GiacominiWhite2006}
{Giacomini, R.} and {White, H.} (2006).
\newblock Tests of conditional predictive ability.
\newblock \textit{Econometrica}, {74}, 1545--1578.

\bibitem[{Gibbs and Su(2002)}]{GibbsSu2002}
{Gibbs, A.~L.} and {Su, F.~E.} (2002).
\newblock On choosing and bounding probability metrics.
\newblock \textit{Int. Stat. Rev.}, {70}, 419--435.

\bibitem[{Gilks et~al.(1996)Gilks, Richardson and
  Spiegelhalter}]{GilksEtAl1996book}
{Gilks, W.~R.}, {Richardson, S.} and {Spiegelhalter, D.~J.} (eds.)  (1996).
\newblock \textit{{Markov Chain Monte Carlo in Practice}}.
\newblock {Chapman \& Hall/CRC}, Boca Raton.

\bibitem[{Gneiting(1997)}]{Gneiting1997}
{Gneiting, T.} (1997).
\newblock Normal scale mixtures and dual probability densities.
\newblock \textit{J. Stat. Comput. Simul.}, {59}, 375--384.

\bibitem[{Gneiting and Raftery(2007)}]{GneitingRaftery2007}
{Gneiting, T.} and {Raftery, A.~E.} (2007).
\newblock {Strictly proper scoring rules, prediction, and estimation}.
\newblock \textit{J. Amer. Statist. Assoc.}, {102}, 359--378.

\bibitem[{Gneiting and Ranjan(2011)}]{GneitingRanjan2011}
{Gneiting, T.} and {Ranjan, R.} (2011).
\newblock {Comparing density forecasts using threshold- and quantile-weighted
  scoring rules}.
\newblock \textit{J. Bus. Econom. Statist.}, {29}, 411--422.

\bibitem[{Gneiting et~al.(2008)Gneiting, Stanberry, Grimit, Held and
  Johnson}]{GneitingEtAl2008}
{Gneiting, T.}, {Stanberry, L.~I.}, {Grimit, E.~P.}, {Held, L.} and {Johnson,
  N.~A.} (2008).
\newblock {Assessing probabilistic forecasts of multivariate quantities, with
  an application to ensemble predictions of surface winds (with discussion and
  rejoinder)}.
\newblock \textit{Test}, {17}, 211--264.

\bibitem[{Good(1952)}]{Good1952}
{Good, I.~J.} (1952).
\newblock {Rational decisions}.
\newblock \textit{{J. R. Stat. Soc. Ser. B. Stat. Methodol.}}, {14}, 107--114.

\bibitem[{Greenberg(2013)}]{Greenberg2013}
{Greenberg, E.} (2013).
\newblock \textit{{Introduction to Bayesian Econometrics}}.
\newblock 2nd ed. Cambridge University Press, New York.

\bibitem[{Grimit et~al.(2006)Grimit, Gneiting, Berrocal and
  Johnson}]{GrimitEtAl2006}
{Grimit, E.~P.}, {Gneiting, T.}, {Berrocal, V.~J.} and {Johnson, N.~A.} (2006).
\newblock {The continuous ranked probability score for circular variables and
  its application to mesoscale forecast ensemble verification}.
\newblock \textit{Q. J. Royal Meteorol. Soc.}, {132}, 2925--2942.

\bibitem[{Gy{\"o}rfi et~al.(1989)Gy{\"o}rfi, H{\"a}rdle, Sarda and
  Vieu}]{GyoerfiEtAl1989}
{Gy{\"o}rfi, L.}, {H{\"a}rdle, W.}, {Sarda, P.} and {Vieu, P.} (1989).
\newblock \textit{{Nonparametric Curve Estimation from Time Series}}.
\newblock Springer, Berlin.

\bibitem[{Hall(1987)}]{Hall1987}
{Hall, P.} (1987).
\newblock {On Kullback-Leibler loss and density estimation}.
\newblock \textit{Ann. Statist.}, {15}, 1491--1519.

\bibitem[{Hall et~al.(1995)Hall, Lahiri and Truong}]{HallEtAl1995}
{Hall, P.}, {Lahiri, S.~N.} and {Truong, Y.~K.} (1995).
\newblock On bandwidth choice for density estimation with dependent data.
\newblock \textit{Ann. Statist.}, {23}, 2241--2263.

\bibitem[{Hall and Presnell(1999)}]{HallPresnell1999}
{Hall, P.} and {Presnell, B.} (1999).
\newblock Density estimation under constraints.
\newblock \textit{J. Comput. Graph. Statist.}, {8}, 259--277.

\bibitem[{Hamilton(1989)}]{Hamilton1989}
{Hamilton, J.~D.} (1989).
\newblock A new approach to the economic analysis of nonstationary time series
  and the business cycle.
\newblock \textit{Econometrica}, {57}, 357--384.

\bibitem[{Hansen(2008)}]{Hansen2008}
{Hansen, B.~E.} (2008).
\newblock Uniform convergence rates for kernel estimation with dependent data.
\newblock \textit{Econom. Theory}, {24}, 726--748.

\bibitem[{Harris(1989)}]{Harris1989}
{Harris, I.~R.} (1989).
\newblock Predictive fit for natural exponential families.
\newblock \textit{Biometrika}, {76}, 675--684.

\bibitem[{Held et~al.(2017)Held, Meyer and Bracher}]{HeldEtAl2017}
{Held, L.}, {Meyer, S.} and {Bracher, J.} (2017).
\newblock Probabilistic forecasting in infectious disease epidemiology: The
  13th {A}rmitage lecture.
\newblock \textit{Stat. Med.}, {36}, 3443--3460.

\bibitem[{Held et~al.(2010)Held, Schr{\"o}dle and Rue}]{HeldEtAl2010_2}
{Held, L.}, {Schr{\"o}dle, B.} and {Rue, H.} (2010).
\newblock {Posterior and cross-validatory predictive checks: A comparison of
  MCMC and INLA}.
\newblock In \textit{Statistical Modelling and Regression Structures:
  Festschrift in Honour of Ludwig Fahrmeir} (T.~Kneib and G.~Tutz, eds.).
  Physica-Verlag HD, Heidelberg, 91--110.

\bibitem[{Hooten and Hobbs(2015)}]{HootenHobbs2015}
{Hooten, M.~B.} and {Hobbs, N.~T.} (2015).
\newblock {A guide to Bayesian model selection for ecologists}.
\newblock \textit{Ecol. Monogr.}, {85}, 3--28.

\bibitem[{Huber and Ronchetti(2009)}]{HuberRonchetti2009}
{Huber, P.~J.} and {Ronchetti, E.~M.} (2009).
\newblock \textit{{Robust Statistics}}.
\newblock 2nd ed. {Wiley}, Hoboken, New Jersey.

\bibitem[{Jones(1991)}]{Jones1991}
{Jones, M.~C.} (1991).
\newblock {On correcting for variance inflation in kernel density estimation}.
\newblock \textit{Comput. Stat. Data Anal.}, {11}, 3--15.

\bibitem[{Jordan(2016)}]{Jordan2016}
{Jordan, A.} (2016).
\newblock Facets of forecast evaluation.
\newblock Ph.D.~thesis, Karlsruhe Institute of Technology, available at
  \url{https://publikationen.bibliothek.kit.edu/1000063629}.

\bibitem[{Jordan et~al.(2019)Jordan, Kr{\"u}ger and Lerch}]{JordanEtAl2015}
{Jordan, A.}, {Kr{\"u}ger, F.} and {Lerch, S.} (2019).
\newblock Evaluating probabilistic forecasts with {scoringRules}.
\newblock \textit{J. Stat. Softw.}, {90}, 1--37.

\bibitem[{Kass and Raftery(1995)}]{KassRaftery1995}
{Kass, R.~E.} and {Raftery, A.~E.} (1995).
\newblock Bayes factors.
\newblock \textit{J. Amer. Statist. Assoc.}, {90}, 773--795.

\bibitem[{Kim et~al.(2016)Kim, MacEachern and Jung}]{KimEtAl2016}
{Kim, H.~J.}, {MacEachern, S.~N.} and {Jung, Y.} (2016).
\newblock {Bandwidth selection for kernel density estimation with a Markov
  chain Monte Carlo sample}.
\newblock Preprint, available at \url{http://arxiv.org/abs/1607.08274}.

\bibitem[{Kr\"uger et~al.(2017)Kr\"uger, Clark and Ravazzolo}]{KruegerEtAl2015}
{Kr\"uger, F.}, {Clark, T.~E.} and {Ravazzolo, F.} (2017).
\newblock {Using entropic tilting to combine BVAR forecasts with external
  nowcasts}.
\newblock \textit{J. Bus. Econom. Statist.}, {35}, 470--485.

\bibitem[{Kullback(1959)}]{Kullback1959book}
{Kullback, S.} (1959).
\newblock \textit{{Information Theory and Statistics}}.
\newblock {John Wiley \& Sons}.

\bibitem[{Liebscher(1996)}]{Liebscher1996}
{Liebscher, E.} (1996).
\newblock Strong convergence of sums of $\alpha$-mixing random variables with
  applications to density estimation.
\newblock \textit{Stochastic Process. Appl.}, {65}, 69--80.

\bibitem[{Link and Eaton(2012)}]{LinkEaton2012}
{Link, W.~A.} and {Eaton, M.~J.} (2012).
\newblock {On thinning of chains in MCMC}.
\newblock \textit{Methods Ecol. Evol.}, {3}, 112--115.

\bibitem[{Little(2006)}]{Little2006}
{Little, R.~J.} (2006).
\newblock {Calibrated Bayes: A Bayes/frequentist roadmap}.
\newblock \textit{Amer. Statist.}, {60}, 213--223.

\bibitem[{MacEachern and Berliner(1994)}]{MacEachernBerliner1994}
{MacEachern, S.~N.} and {Berliner, L.~M.} (1994).
\newblock {Subsampling the Gibbs sampler}.
\newblock \textit{Amer. Statist.}, {48}, 188--190.

\bibitem[{Matheson and Winkler(1976)}]{MathesonWinkler1976}
{Matheson, J.~E.} and {Winkler, R.~L.} (1976).
\newblock {Scoring rules for continuous probability distributions}.
\newblock \textit{Manag. Sci.}, {22}, 1087--1096.

\bibitem[{Panaretos and Zemel(2019)}]{PanaretosZemel2019}
{Panaretos, V.~M.} and {Zemel, Y.} (2019).
\newblock {Statistical aspects of {W}asserstein distances}.
\newblock \textit{Annu. Rev. Stat. Appl.}, {6}, 405--431.

\bibitem[{{R Core Team}(2019)}]{R2016}
{{R Core Team}} (2019).
\newblock \textit{R: A Language and Environment for Statistical Computing}.
\newblock R Foundation for Statistical Computing, Vienna, Austria.
\newblock \urlprefix\url{https://www.R-project.org/}.

\bibitem[{Raftery(2016)}]{Raftery2016}
{Raftery, A.~E.} (2016).
\newblock Use and communication of probabilistic forecasts.
\newblock \textit{Stat. Anal. Data. Min.}, {9}, 397--410.

\bibitem[{Rajagopalan and Lall(1995)}]{RajagopalanLall1995}
{Rajagopalan, B.} and {Lall, U.} (1995).
\newblock A kernel estimator for discrete distributions.
\newblock \textit{J. Nonparametr. Stat.}, {4}, 409--426.

\bibitem[{Rosenblatt(1956)}]{Rosenblatt1956}
{Rosenblatt, M.} (1956).
\newblock Remarks on some nonparametric estimates of a density function.
\newblock \textit{{Ann. Math. Stat.}}, {27}, 832--837.

\bibitem[{Rosenthal(1995)}]{Rosenthal1995}
{Rosenthal, J.~S.} (1995).
\newblock Minorization conditions and convergence rates for {M}arkov chain
  {M}onte {C}arlo.
\newblock \textit{J. Amer. Statist. Assoc.}, {90}, 558--566.

\bibitem[{Roussas(1988)}]{Roussas1988}
{Roussas, G.~G.} (1988).
\newblock Nonparametric estimation in mixing sequences of random variables.
\newblock \textit{J. Statist. Plann. Inference}, {18}, 135--149.

\bibitem[{Rubin(1984)}]{Rubin1984}
{Rubin, D.~B.} (1984).
\newblock Bayesianly justifiable and relevant frequency calculations for the
  applied statistician.
\newblock \textit{Ann. Statist.}, {12}, 1151--1172.

\bibitem[{Scheuerer and Hamill(2015)}]{ScheuererHamill2015VS}
{Scheuerer, M.} and {Hamill, T.~M.} (2015).
\newblock Variogram-based proper scoring rules for probabilistic forecasts of
  multivariate quantities.
\newblock \textit{Mon. Weather Rev.}, {143}, 1321--1334.

\bibitem[{Sheather and Jones(1991)}]{SheatherJones1991}
{Sheather, S.~J.} and {Jones, M.~C.} (1991).
\newblock A reliable data-based bandwidth selection method for kernel density
  estimation.
\newblock \textit{J. R. Stat. Soc. Ser. B. Stat. Methodol.}, {53}, 683--690.

\bibitem[{Shirota and Gelfand(2017)}]{ShirotaGelfand2017}
{Shirota, S.} and {Gelfand, A.~E.} (2017).
\newblock Space and circular time log {G}aussian {C}ox processes with
  application to crime event data.
\newblock \textit{The Annals of Applied Statistics}, {11}, 481--503.

\bibitem[{Shuford et~al.(1966)Shuford, Albert and Massengill}]{ShufordEtAl1966}
{Shuford, E.~H.}, {Albert, A.} and {Massengill, H.~E.} (1966).
\newblock Admissible probability measurement procedures.
\newblock \textit{Psychometrika}, {31}, 125--145.

\bibitem[{Silverman(1986)}]{Silverman1986}
{Silverman, B.~W.} (1986).
\newblock \textit{Density Estimation for Statistics and Data Analysis}.
\newblock Chapman and Hall, London.

\bibitem[{Sk{\"o}ld and Roberts(2003)}]{SkoeldRoberts2003}
{Sk{\"o}ld, M.} and {Roberts, G.~O.} (2003).
\newblock {Density estimation for the Metropolis-Hastings algorithm}.
\newblock \textit{Scand. J. Stat.}, {30}, 699--718.

\bibitem[{Spiegelhalter et~al.(2002)Spiegelhalter, Best, Carlin and van~der
  Linde}]{SpiegelhalterEtAl2002}
{Spiegelhalter, D.~J.}, {Best, N.~G.}, {Carlin, B.~P.} and {van~der Linde, A.}
  (2002).
\newblock Bayesian measures of model complexity and fit (with discussion and
  rejoinder).
\newblock \textit{J. R. Stat. Soc. Ser. B. Stat. Methodol.}, {64}, 583--639.

\bibitem[{Spiegelhalter et~al.(2014)Spiegelhalter, Best, Carlin and van~der
  Linde}]{SpiegelhalterEtAl2014}
{Spiegelhalter, D.~J.}, {Best, N.~G.}, {Carlin, B.~P.} and {van~der Linde, A.}
  (2014).
\newblock The deviance information criterion: 12 years on.
\newblock \textit{J. R. Stat. Soc. Ser. B. Stat. Methodol.}, {76}, 485--493.

\bibitem[{Taillardat et~al.(2016)Taillardat, Mestre, Zamo and
  Naveau}]{Taillardat&2016}
{Taillardat, M.}, {Mestre, O.}, {Zamo, M.} and {Naveau, P.} (2016).
\newblock Calibrated ensemble forecasts using quantile regression forests and
  ensemble model output statistics.
\newblock \textit{Mon. Weather Rev.}, {144}, 2375--2393.

\bibitem[{Tanner and Wong(1987)}]{TannerWong1987}
{Tanner, M.~A.} and {Wong, W.~H.} (1987).
\newblock The calculation of posterior distributions by data augmentation.
\newblock \textit{J. Amer. Statist. Assoc.}, {82}, 528--540.

\bibitem[{Thorarinsdottir et~al.(2013)Thorarinsdottir, Gneiting and
  Gissibl}]{ThorarinsdottirEtAl2013}
{Thorarinsdottir, T.~L.}, {Gneiting, T.} and {Gissibl, N.} (2013).
\newblock Using proper divergence functions to evaluate climate models.
\newblock \textit{SIAM/ASA J. Uncertain. Quantif.}, {1}, 522--534.

\bibitem[{Tierney(1994)}]{Tierney1994}
{Tierney, L.} (1994).
\newblock Markov chains for exploring posterior distributions.
\newblock \textit{Ann. Statist.}, {22}, 1701--1728.

\bibitem[{van~der Vaart(2000)}]{vanderVaart2000}
{van~der Vaart, A.~W.} (2000).
\newblock \textit{{Asymptotic Statistics}}.
\newblock Cambridge University Press, Cambridge.

\bibitem[{Villani(2008)}]{Villani2008}
{Villani, C.} (2008).
\newblock \textit{Optimal Transport}.
\newblock Grundlehren der mathematischen Wissenschaften, 338, Springer, Berlin,
  Heidelberg.

\bibitem[{Waggoner and Zha(1999)}]{WaggonerZha1999}
{Waggoner, D.~F.} and {Zha, T.} (1999).
\newblock Conditional forecasts in dynamic multivariate models.
\newblock \textit{Rev. Econ. Stat.}, {81}, 639--651.

\bibitem[{Warne et~al.(2016)Warne, Coenen and Christoffel}]{WarneEtAl2016}
{Warne, A.}, {Coenen, G.} and {Christoffel, K.} (2016).
\newblock {Marginalized predictive likelihood comparisons of linear Gaussian
  state-space models with applications to DSGE, DSGE-VAR and VAR models}.
\newblock \textit{J. Appl. Econometrics}, {32}, 103--119.

\bibitem[{Wasserman(2006)}]{Wasserman2006}
{Wasserman, L.} (2006).
\newblock \textit{All of Nonparametric Statistics}.
\newblock Springer, New York.

\bibitem[{Watanabe(2010)}]{Watanabe2010}
{Watanabe, S.} (2010).
\newblock {Asymptotic equivalence of Bayes cross validation and widely
  applicable information criterion in singular learning theory}.
\newblock \textit{J. Mach. Learn. Res.}, {11}, 3571--3594.

\bibitem[{White et~al.(2019)White, Gelfand, Rodrigues and
  Tzintzun}]{WhiteEtAl2019}
{White, P.~A.}, {Gelfand, A.~E.}, {Rodrigues, E.~R.} and {Tzintzun, G.} (2019).
\newblock {Pollution state modelling for {M}exico {C}ity}.
\newblock \textit{J. R. Stat. Soc. Ser. A. Stat. Soc.}, {182}, 1039--1060.

\bibitem[{Yu(1993)}]{Yu1993}
{Yu, B.} (1993).
\newblock {Density estimation in the $L^\infty$ norm for dependent data with
  applications to the Gibbs sampler}.
\newblock \textit{Ann. Statist.}, {21}, 711--735.

\end{thebibliography}


\begin{thebibliography}{78}
\expandafter\ifx\csname natexlab\endcsname\relax\def\natexlab#1{#1}\fi
\expandafter\ifx\csname url\endcsname\relax
  \def\url#1{\texttt{#1}}\fi
\expandafter\ifx\csname urlprefix\endcsname\relax\def\urlprefix{URL }\fi
\providecommand{\eprint}[2][]{\url{#2}}

\bibitem[{Aastveit et~al.(2017)Aastveit, Carriero, Clark and
  Marcellino}]{AastveitEtAl2017}
{Aastveit, K.~A.}, {Carriero, A.}, {Clark, T.~E.} and {Marcellino, M.} (2017).
\newblock {Have standard VARs remained stable since the crisis?}
\newblock \textit{J. Appl. Econom.}, {32}, 931--951.

\bibitem[{Adolfson et~al.(2007)Adolfson, Linde and Villani}]{AdolfsonEtAl2007}
{Adolfson, M.}, {Linde, J.} and {Villani, M.} (2007).
\newblock {Forecasting performance of an open economy DSGE model}.
\newblock \textit{Econometric Rev.}, {26}, 289--328.

\bibitem[{Amisano and Geweke(2017)}]{AmisanoGeweke2017}
{Amisano, G.} and {Geweke, J.} (2017).
\newblock Prediction using several macroeconomic models.
\newblock \textit{Rev. Econ. Stat.}, {99}, 912--925.

\bibitem[{Amisano and Giacomini(2007)}]{AmisanoGiacomini2007}
{Amisano, G.} and {Giacomini, R.} (2007).
\newblock {Comparing density forecasts via weighted likelihood ratio tests}.
\newblock \textit{J. Bus. Econom. Statist.}, {25}, 177--190.

\bibitem[{Ba{\c{s}}t{\"u}rk et~al.(2014)Ba{\c{s}}t{\"u}rk, Cakmakli, Ceyhan and
  Van~Dijk}]{BacsturkEtAl2014}
{Ba{\c{s}}t{\"u}rk, N.}, {Cakmakli, C.}, {Ceyhan, S.~P.} and {Van~Dijk, H.~K.}
  (2014).
\newblock {Posterior-predictive evidence on US inflation using extended new
  Keynesian Phillips curve models with non-filtered data}.
\newblock \textit{J. Appl. Econometrics}, {29}, 1164--1182.

\bibitem[{Bauwens et~al.(2015)Bauwens, Koop, Korobilis and
  Rombouts}]{BauwensEtAl2014}
{Bauwens, L.}, {Koop, G.}, {Korobilis, D.} and {Rombouts, J. V.~K.} (2015).
\newblock {The contribution of structural break models to forecasting
  macroeconomic series}.
\newblock \textit{J. Appl. Econometrics}, {30}, 596--620.

\bibitem[{Belmonte et~al.(2014)Belmonte, Koop and Korobilis}]{BelmonteEtAl2014}
{Belmonte, M. A.~G.}, {Koop, G.} and {Korobilis, D.} (2014).
\newblock Hierarchical shrinkage in time-varying parameter models.
\newblock \textit{J. Forecast.}, {33}, 80--94.

\bibitem[{Bentzien and Friederichs(2014)}]{BentzienFriederichs2014}
{Bentzien, S.} and {Friederichs, P.} (2014).
\newblock Decomposition and graphical portrayal of the quantile score.
\newblock \textit{Q. J. Royal Meteorol. Soc.}, {140}, 1924--1934.

\bibitem[{Berg(2017)}]{Berg2017}
{Berg, T.~O.} (2017).
\newblock {Forecast accuracy of a BVAR under alternative specifications of the
  zero lower bound}.
\newblock \textit{Stud. Nonlinear Dyn. Econom.}, {21}, 20150084.

\bibitem[{Berg and Henzel(2015)}]{BergHenzel2015}
{Berg, T.~O.} and {Henzel, S.~R.} (2015).
\newblock {Point and density forecasts for the Euro area using Bayesian VARs}.
\newblock \textit{Int. J. Forecast.}, {31}, 1067--1095.

\bibitem[{Berk et~al.(2018)Berk, Hoffmann and M{\"u}ller}]{BerkEtAl2018}
{Berk, K.}, {Hoffmann, A.} and {M{\"u}ller, A.} (2018).
\newblock Probabilistic forecasting of industrial electricity load with regime
  switching behavior.
\newblock \textit{Int. J. Forecast.}, {34}, 147--162.

\bibitem[{Berrocal et~al.(2014)Berrocal, Gelfand and
  Holland}]{BerrocalEtAl2014}
{Berrocal, V.~J.}, {Gelfand, A.~E.} and {Holland, D.~M.} (2014).
\newblock {Assessing exceedance of ozone standards: A space-time downscaler for
  fourth highest ozone concentrations}.
\newblock \textit{Environmetrics}, {25}, 279--291.

\bibitem[{Bitto and
  Fr{\"u}hwirth-Schnatter(2019)}]{BittoFruehwirthSchnatter2018}
{Bitto, A.} and {Fr{\"u}hwirth-Schnatter, S.} (2019).
\newblock Achieving shrinkage in a time-varying parameter model framework.
\newblock \textit{J. Econometrics}, {210}, 75--97.

\bibitem[{Brandt et~al.(2014)Brandt, Freeman and Schrodt}]{BrandtEtAl2014}
{Brandt, P.~T.}, {Freeman, J.~R.} and {Schrodt, P.~A.} (2014).
\newblock {Evaluating forecasts of political conflict dynamics}.
\newblock \textit{Int. J. Forecast.}, {30}, 944--962.

\bibitem[{Carriero et~al.(2015{\natexlab{a}})Carriero, Clark and
  Marcellino}]{CarrieroEtAl2015a}
{Carriero, A.}, {Clark, T.~E.} and {Marcellino, M.} (2015{\natexlab{a}}).
\newblock {Bayesian VARs: Specification choices and forecast accuracy}.
\newblock \textit{J. Appl. Econometrics}, {30}, 46--73.

\bibitem[{Carriero et~al.(2015{\natexlab{b}})Carriero, Clark and
  Marcellino}]{CarrieroEtAl2015c}
{Carriero, A.}, {Clark, T.~E.} and {Marcellino, M.} (2015{\natexlab{b}}).
\newblock {Realtime nowcasting with a Bayesian mixed frequency model with
  stochastic volatility}.
\newblock \textit{J. R. Stat. Soc. Ser. A. Stat. Soc.}, {178}, 837--862.

\bibitem[{Carriero et~al.(2016)Carriero, Clark and
  Marcellino}]{CarrieroEtAl2015b}
{Carriero, A.}, {Clark, T.~E.} and {Marcellino, M.} (2016).
\newblock {Common drifting volatility in large Bayesian VARs}.
\newblock \textit{J. Bus. Econom. Statist.}, {34}, 375--390.

\bibitem[{Carriero et~al.(2015{\natexlab{c}})Carriero, Mumtaz and
  Theophilopoulou}]{CarrieroEtAl2015d}
{Carriero, A.}, {Mumtaz, H.} and {Theophilopoulou, A.} (2015{\natexlab{c}}).
\newblock {Macroeconomic information, structural change, and the prediction of
  fiscal aggregates}.
\newblock \textit{Int. J. Forecast.}, {31}, 325--348.

\bibitem[{Clark(2011)}]{Clark2011}
{Clark, T.~E.} (2011).
\newblock {Real-time density forecasts from BVARs with stochastic volatility}.
\newblock \textit{J. Bus. Econom. Statist.}, {29}, 327--341.

\bibitem[{Clark et~al.(2020)Clark, McCracken and Mertens}]{ClarkEtAl2018}
{Clark, T.~E.}, {McCracken, M.~W.} and {Mertens, E.} (2020).
\newblock Modeling time-varying uncertainty of multiple-horizon forecast
  errors.
\newblock \textit{Rev. Econ. Stat.}, {102}, 17--33.

\bibitem[{Clark and Ravazzolo(2015)}]{ClarkRavazzolo2014}
{Clark, T.~E.} and {Ravazzolo, F.} (2015).
\newblock Macroeconomic forecasting performance under alternative
  specifications of time-varying volatility.
\newblock \textit{J. Appl. Econometrics}, {30}, 551--575.

\bibitem[{De~la Cruz and Branco(2009)}]{DeLaCruzBranco2009}
{De~la Cruz, R.} and {Branco, M.~D.} (2009).
\newblock {Bayesian analysis for nonlinear regression model under skewed
  errors, with application in growth curves}.
\newblock \textit{Biom. J.}, {51}, 588--609.

\bibitem[{Delatola and Griffin(2011)}]{DelatolaGriffin2011}
{Delatola, E.-I.} and {Griffin, J.~E.} (2011).
\newblock {Bayesian nonparametric modelling of the return distribution with
  stochastic volatility}.
\newblock \textit{Bayesian Anal.}, {6}, 901--926.

\bibitem[{Diks et~al.(2011)Diks, Panchenko and van Dijk}]{DiksEtAl2011}
{Diks, C.}, {Panchenko, V.} and {van Dijk, D.} (2011).
\newblock {Likelihood-based scoring rules for comparing density forecasts in
  tails}.
\newblock \textit{J. Econometrics}, {163}, 215--230.

\bibitem[{Fan et~al.(2018)Fan, Paul, Lee and Matsuo}]{FanEtAl2018}
{Fan, M.}, {Paul, D.}, {Lee, T.~C.} and {Matsuo, T.} (2018).
\newblock {A multi-resolution model for non-Gaussian random fields on a sphere
  with application to ionospheric electrostatic potentials}.
\newblock \textit{Ann. Appl. Stat.}, {12}, 459--489.

\bibitem[{Friederichs and
  Thorarinsdottir(2012)}]{FriederichsThorarinsdottir2012}
{Friederichs, P.} and {Thorarinsdottir, T.~L.} (2012).
\newblock Forecast verification for extreme value distributions with an
  application to probabilistic peak wind prediction.
\newblock \textit{Environmetrics}, {23}, 579--594.

\bibitem[{Geweke and Amisano(2010)}]{GewekeAmisano2010}
{Geweke, J.} and {Amisano, G.} (2010).
\newblock {Comparing and evaluating Bayesian predictive distributions of asset
  returns}.
\newblock \textit{Int. J. Forecast.}, {26}, 216--230.

\bibitem[{Geweke and Amisano(2011)}]{GewekeAmisano2011}
{Geweke, J.} and {Amisano, G.} (2011).
\newblock {Hierarchical Markov normal mixture models with applications to
  financial asset returns}.
\newblock \textit{J. Appl. Econometrics}, {26}, 1--29.

\bibitem[{Giannone et~al.(2015)Giannone, Lenza and
  Primiceri}]{GiannoneEtAl2015}
{Giannone, D.}, {Lenza, M.} and {Primiceri, G.~E.} (2015).
\newblock Prior selection for vector autoregressions.
\newblock \textit{Rev. Econ. Stat.}, {97}, 436--451.

\bibitem[{Gneiting and Ranjan(2011)}]{GneitingRanjan2011}
{Gneiting, T.} and {Ranjan, R.} (2011).
\newblock {Comparing density forecasts using threshold- and quantile-weighted
  scoring rules}.
\newblock \textit{J. Bus. Econom. Statist.}, {29}, 411--422.

\bibitem[{Groen et~al.(2013)Groen, Paap and Ravazzolo}]{GroenEtAl2013}
{Groen, J. J.~J.}, {Paap, R.} and {Ravazzolo, F.} (2013).
\newblock {Real-time inflation forecasting in a changing world}.
\newblock \textit{J. Bus. Econom. Statist.}, {31}, 29--44.

\bibitem[{Gschl{\"o}{\ss}l and Czado(2007)}]{GschloesslCzado2007}
{Gschl{\"o}{\ss}l, S.} and {Czado, C.} (2007).
\newblock Spatial modelling of claim frequency and claim size in non-life
  insurance.
\newblock \textit{{Scand. Actuar. J.}}, {2007}, 202--225.

\bibitem[{Held et~al.(2017)Held, Meyer and Bracher}]{HeldEtAl2017}
{Held, L.}, {Meyer, S.} and {Bracher, J.} (2017).
\newblock Probabilistic forecasting in infectious disease epidemiology: The
  13th {A}rmitage lecture.
\newblock \textit{Stat. Med.}, {36}, 3443--3460.

\bibitem[{Hoff(2009)}]{Hoff2009}
{Hoff, P.~D.} (2009).
\newblock \textit{A First Course in Bayesian Statistical Methods}.
\newblock Springer.

\bibitem[{Jensen and Maheu(2010)}]{Jensen2010}
{Jensen, M.~J.} and {Maheu, J.~M.} (2010).
\newblock Bayesian semiparametric stochastic volatility modeling.
\newblock \textit{J. Econometrics}, {157}, 306--316.

\bibitem[{Jensen and Maheu(2013)}]{Jensen2013}
{Jensen, M.~J.} and {Maheu, J.~M.} (2013).
\newblock {Bayesian semiparametric multivariate GARCH modeling}.
\newblock \textit{J. Econometrics}, {176}, 3--17.

\bibitem[{Jensen and Maheu(2014)}]{Jensen2014}
{Jensen, M.~J.} and {Maheu, J.~M.} (2014).
\newblock {Estimating a semiparametric asymmetric stochastic volatility model
  with a Dirichlet process mixture}.
\newblock \textit{J. Econometrics}, {178}, 523--538.

\bibitem[{Jin and Maheu(2013)}]{Jin2013}
{Jin, X.} and {Maheu, J.~M.} (2013).
\newblock Modeling realized covariances and returns.
\newblock \textit{J. Financ. Economet.}, {11}, 335--369.

\bibitem[{Jin and Maheu(2016)}]{Jin2016}
{Jin, X.} and {Maheu, J.~M.} (2016).
\newblock Bayesian semiparametric modeling of realized covariance matrices.
\newblock \textit{J. Econometrics}, {192}, 19--39.

\bibitem[{Jochmann et~al.(2010)Jochmann, Koop and Strachan}]{JochmannEtAl2010}
{Jochmann, M.}, {Koop, G.} and {Strachan, R.~W.} (2010).
\newblock {Bayesian forecasting using stochastic search variable selection in a
  VAR subject to breaks}.
\newblock \textit{Int. J. Forecast.}, {26}, 326--347.

\bibitem[{Kallache et~al.(2010)Kallache, Maksimovich, Michelangeli and
  Naveau}]{KallacheEtAl2010}
{Kallache, M.}, {Maksimovich, E.}, {Michelangeli, P.-A.} and {Naveau, P.}
  (2010).
\newblock {Multimodel combination by a Bayesian hierarchical model: Assessment
  of ice accumulation over the oceanic Arctic region}.
\newblock \textit{J. Clim.}, {23}, 5421--5436.

\bibitem[{Kastner(2016)}]{Kastner2016}
{Kastner, G.} (2016).
\newblock Dealing with stochastic volatility in time series using the {R}
  package stochvol.
\newblock \textit{J. Stat. Softw.}, {69}, 1--30.

\bibitem[{Kr\"uger et~al.(2017)Kr\"uger, Clark and Ravazzolo}]{KruegerEtAl2015}
{Kr\"uger, F.}, {Clark, T.~E.} and {Ravazzolo, F.} (2017).
\newblock {Using entropic tilting to combine BVAR forecasts with external
  nowcasts}.
\newblock \textit{J. Bus. Econom. Statist.}, {35}, 470--485.

\bibitem[{Kr{\"u}ger and Nolte(2016)}]{KruegerNolte2015}
{Kr{\"u}ger, F.} and {Nolte, I.} (2016).
\newblock {Disagreement versus uncertainty: Evidence from distribution
  forecasts}.
\newblock \textit{J. Bank. Finance}, {72}, 172--186.

\bibitem[{Le{\~a}o et~al.(2017)Le{\~a}o, Abanto-Valle and Chen}]{LeaoEtAl2017}
{Le{\~a}o, W.~L.}, {Abanto-Valle, C.~A.} and {Chen, M.-H.} (2017).
\newblock {Bayesian analysis of stochastic volatility-in-mean model with
  leverage and asymmetrically heavy-tailed error using generalized hyperbolic
  skew Student's $t$-distribution}.
\newblock \textit{Stat. Its Interface}, {10}, 529.

\bibitem[{Leininger et~al.(2013)Leininger, Gelfand, Allen and
  Silander~Jr}]{LeiningerEtAl2013}
{Leininger, T.~J.}, {Gelfand, A.~E.}, {Allen, J.~M.} and {Silander~Jr, J.~A.}
  (2013).
\newblock {Spatial regression modeling for compositional data with many zeros}.
\newblock \textit{J. Agric. Biol. Environ. Stat.}, {18}, 314--334.

\bibitem[{Li et~al.(2010)Li, Villani and Kohn}]{LiEtAl2010}
{Li, F.}, {Villani, M.} and {Kohn, R.} (2010).
\newblock {Flexible modeling of conditional distributions using smooth mixtures
  of asymmetric Student $t$ densities}.
\newblock \textit{J. Statist. Plann. Inference}, {140}, 3638--3654.

\bibitem[{Liu and Maheu(2009)}]{Liu2009}
{Liu, C.} and {Maheu, J.~M.} (2009).
\newblock {Forecasting realized volatility: {A} {B}ayesian model-averaging
  approach}.
\newblock \textit{J. Appl. Econometrics}, {24}, 709--733.

\bibitem[{Liu and Maheu(2018)}]{LiuMaheu2018}
{Liu, J.} and {Maheu, J.~M.} (2018).
\newblock Improving markov switching models using realized variance.
\newblock \textit{J. Appl. Econometrics}, {33}, 297--318.

\bibitem[{Lopes et~al.(2008)Lopes, Salazar and Gamerman}]{LopesEtAl2008}
{Lopes, H.~F.}, {Salazar, E.} and {Gamerman, D.} (2008).
\newblock {Spatial dynamic factor analysis}.
\newblock \textit{Bayesian Anal.}, {3}, 759--792.

\bibitem[{Louzis(2019)}]{Louzis2019}
{Louzis, D.~P.} (2019).
\newblock Steady-state modeling and macroeconomic forecasting quality.
\newblock \textit{J. Appl. Econometrics}, {34}, 285--314.

\bibitem[{Maheu and Gordon(2008)}]{Maheu2008}
{Maheu, J.~M.} and {Gordon, S.} (2008).
\newblock Learning, forecasting and structural breaks.
\newblock \textit{J. Appl. Econometrics}, {23}, 553--583.

\bibitem[{Maheu and McCurdy(2009)}]{Maheu2009}
{Maheu, J.~M.} and {McCurdy, T.~H.} (2009).
\newblock How useful are historical data for forecasting the long-run equity
  return distribution?
\newblock \textit{J. Bus. Econom. Statist.}, {27}, 95--112.

\bibitem[{Maheu et~al.(2012)Maheu, McCurdy and Song}]{MaheuEtAl2012}
{Maheu, J.~M.}, {McCurdy, T.~H.} and {Song, Y.} (2012).
\newblock {Components of bull and bear markets: Bull corrections and bear
  rallies}.
\newblock \textit{J. Bus. Econom. Statist.}, {30}, 391--403.

\bibitem[{Maheu and Song(2014)}]{Maheu2014}
{Maheu, J.~M.} and {Song, Y.} (2014).
\newblock {A new structural break model, with an application to Canadian
  inflation forecasting}.
\newblock \textit{Int. J. Forecast.}, {30}, 144--160.

\bibitem[{Maheu and Song(2018)}]{MaheuSong2018}
{Maheu, J.~M.} and {Song, Y.} (2018).
\newblock An efficient {B}ayesian approach to multiple structural change in
  multivariate time series.
\newblock \textit{J. Appl. Econometrics}, {33}, 251--270.

\bibitem[{Maheu and Yang(2016)}]{MaheuYang2016}
{Maheu, J.~M.} and {Yang, Q.} (2016).
\newblock An infinite hidden {M}arkov model for short-term interest rates.
\newblock \textit{Journal of Empirical Finance}, {38}, 202--220.

\bibitem[{Maneesoonthorn et~al.(2012)Maneesoonthorn, Martin, Forbes and
  Grose}]{ManeesoonthornEtAl2012}
{Maneesoonthorn, W.}, {Martin, G.~M.}, {Forbes, C.~S.} and {Grose, S.~D.}
  (2012).
\newblock {Probabilistic forecasts of volatility and its risk premia}.
\newblock \textit{J. Econometrics}, {171}, 217--236.

\bibitem[{Metaxoglou et~al.(2019)Metaxoglou, Pettenuzzo and
  Smith}]{Metaxoglou2018}
{Metaxoglou, K.}, {Pettenuzzo, D.} and {Smith, A.} (2019).
\newblock Option-implied equity premium predictions via entropic tilting.
\newblock \textit{J. Financ. Economet.}, {17}, 559--586.

\bibitem[{Murphy(2007)}]{Murphy07}
{Murphy, K.~P.} (2007).
\newblock Conjugate {B}ayesian analysis of the {G}aussian distribution.
\newblock Manuscript, University of British Columbia (document version: October
  3, 2007).

\bibitem[{Panagiotelis and Smith(2008)}]{PanagiotelisSmith2008}
{Panagiotelis, A.} and {Smith, M.} (2008).
\newblock {Bayesian density forecasting of intraday electricity prices using
  multivariate skew $t$ distributions}.
\newblock \textit{Int. J. Forecast.}, {24}, 710--727.

\bibitem[{Riebler et~al.(2012)Riebler, Held and Rue}]{RieblerEtAl2012}
{Riebler, A.}, {Held, L.} and {Rue, H.} (2012).
\newblock {Estimation and extrapolation of time trends in registry data ---
  Borrowing strength from related populations}.
\newblock \textit{Ann. Appl. Stat.}, {6}, 304--333.

\bibitem[{Risser and Calder(2015)}]{RisserCalder2015}
{Risser, M.~D.} and {Calder, C.~A.} (2015).
\newblock {Regression-based covariance functions for nonstationary spatial
  modeling}.
\newblock \textit{Environmetrics}, {26}, 284--297.

\bibitem[{Risser et~al.(2019)Risser, Calder, Berrocal and
  Berrett}]{RisserEtAl2018}
{Risser, M.~D.}, {Calder, C.~A.}, {Berrocal, V.~J.} and {Berrett, C.} (2019).
\newblock Nonstationary spatial prediction of soil organic carbon: Implications
  for stock assessment decision making.
\newblock \textit{Ann. Appl. Stat.}, {13}, 165--188.

\bibitem[{Rodrigues et~al.(2014)Rodrigues, Garc{\'i}a-Serrano and
  Doblas-Reyes}]{RodriguesEtAl2014}
{Rodrigues, L. R.~L.}, {Garc{\'i}a-Serrano, J.} and {Doblas-Reyes, F.} (2014).
\newblock {Seasonal forecast quality of the West African monsoon rainfall
  regimes by multiple forecast systems}.
\newblock \textit{J. Geophys. Res. Atmos.}, {119}, 7908--7930.

\bibitem[{Sahu et~al.(2015)Sahu, Bakar and Awang}]{SahuEtAl2015}
{Sahu, S.~K.}, {Bakar, K.~S.} and {Awang, N.} (2015).
\newblock {Bayesian forecasting using spatio-temporal models with applications
  to ozone levels in the eastern United States}.
\newblock In \textit{Geometry Driven Statistics} (I.~L. Dryden and J.~T. Kent,
  eds.), chap.~13. John Wiley \& Sons, 260--281.

\bibitem[{Salazar et~al.(2011)Salazar, Sans{\'o}, Finley, Hammerling,
  Steinsland, Wang and Delamater}]{SalazarEtAl2011}
{Salazar, E.}, {Sans{\'o}, B.}, {Finley, A.~O.}, {Hammerling, D.}, {Steinsland,
  I.}, {Wang, X.} and {Delamater, P.} (2011).
\newblock {Comparing and blending regional climate model predictions for the
  American Southwest}.
\newblock \textit{J. Agric. Biol. Environ. Stat.}, {16}, 586--605.

\bibitem[{Sigrist et~al.(2012)Sigrist, K{\"u}nsch and Stahel}]{SigristEtAl2012}
{Sigrist, F.}, {K{\"u}nsch, H.~R.} and {Stahel, W.~A.} (2012).
\newblock {A dynamic nonstationary spatio-temporal model for short term
  prediction of precipitation}.
\newblock \textit{Ann. Appl. Stat.}, {6}, 1452--1477.

\bibitem[{Sigrist et~al.(2015)Sigrist, K{\"u}nsch and Stahel}]{SigristEtAl2015}
{Sigrist, F.}, {K{\"u}nsch, H.~R.} and {Stahel, W.~A.} (2015).
\newblock {Stochastic partial differential equation based modelling of large
  space-time data sets}.
\newblock \textit{J. R. Stat. Soc. Ser. B. Stat. Methodol.}, {77}, 3--33.

\bibitem[{Smith and Vahey(2016)}]{SmithVahey2015}
{Smith, M.~S.} and {Vahey, S.} (2016).
\newblock {Asymmetric density forecasting of U.S. macroeconomic variables}.
\newblock \textit{J. Bus. Econom. Statist.}, {34}, 416--434.

\bibitem[{Tran et~al.(2016)Tran, Pitt and Kohn}]{TranEtAl2016}
{Tran, M.-N.}, {Pitt, M.~K.} and {Kohn, R.} (2016).
\newblock {Adaptive Metropolis--Hastings sampling using reversible dependent
  mixture proposals}.
\newblock \textit{Stat. Comput.}, {26}, 361--381.

\bibitem[{Trombe et~al.(2012)Trombe, Pinson and Madsen}]{TrombeEtAl2012}
{Trombe, P.-J.}, {Pinson, P.} and {Madsen, H.} (2012).
\newblock {A general probabilistic forecasting framework for offshore wind
  power fluctuations}.
\newblock \textit{Energies}, {5}, 621--657.

\bibitem[{Warne et~al.(2016)Warne, Coenen and Christoffel}]{WarneEtAl2016}
{Warne, A.}, {Coenen, G.} and {Christoffel, K.} (2016).
\newblock {Marginalized predictive likelihood comparisons of linear Gaussian
  state-space models with applications to DSGE, DSGE-VAR and VAR models}.
\newblock \textit{J. Appl. Econometrics}, {32}, 103--119.

\bibitem[{White et~al.(2019{\natexlab{a}})White, Gelfand, Rodrigues and
  Tzintzun}]{WhiteEtAl2019}
{White, P.~A.}, {Gelfand, A.~E.}, {Rodrigues, E.~R.} and {Tzintzun, G.}
  (2019{\natexlab{a}}).
\newblock {Pollution state modelling for {M}exico {C}ity}.
\newblock \textit{J. R. Stat. Soc. Ser. A. Stat. Soc.}, {182}, 1039--1060.

\bibitem[{White and Porcu(2019)}]{WhitePorcu2018}
{White, P.~A.} and {Porcu, E.} (2019).
\newblock {Nonseparable covariance models on circles cross time: A study of
  Mexico City ozone}.
\newblock \textit{Environmetrics}, {30}, e2558.

\bibitem[{White et~al.(2019{\natexlab{b}})White, Reese, Christensen and
  Rupper}]{WhiteEtAl2019env}
{White, P.~A.}, {Reese, C.~S.}, {Christensen, W.~F.} and {Rupper, S.}
  (2019{\natexlab{b}}).
\newblock {A model for Antarctic surface mass balance and ice core site
  selection}.
\newblock \textit{Environmetrics}, {30}, e2579.

\bibitem[{Wijeyakulasuriya et~al.(2018)Wijeyakulasuriya, Hanks, Shaby and
  Cross}]{WijeyakulasuriyaEtAl2018}
{Wijeyakulasuriya, D.~A.}, {Hanks, E.~M.}, {Shaby, B.~A.} and {Cross, P.~C.}
  (2018).
\newblock Extreme value-based methods for modeling elk yearly movements.
\newblock \textit{J. Agric. Biol. Environ. Stat.}, {24}, 73--91.

\bibitem[{Zhou et~al.(2015)Zhou, Matteson, Woodard, Henderson and
  Micheas}]{ZhouEtAl2015}
{Zhou, Z.}, {Matteson, D.~S.}, {Woodard, D.~B.}, {Henderson, S.~G.} and
  {Micheas, A.~C.} (2015).
\newblock A spatio-temporal point process model for ambulance demand.
\newblock \textit{J. Amer. Statist. Assoc.}, {110}, 6--15.

\end{thebibliography}

\section*{Acknowledgements}
The work of Tilmann Gneiting and Fabian Kr\"uger {was} funded by the
European Union Seventh Framework Programme under grant agreement
290976.  Sebastian Lerch and Thordis L.~Thorarinsdottir acknowledge
support by the Volkswagen Foundation through the program ``Mesoscale
Weather Extremes --- Theory, Spatial Modelling and Prediction
(WEX-MOP)''.  Lerch further acknowledges support by Deutsche
Forschungsgemeinschaft (DFG) through RTG 1953 ``Statistical Modeling
of Complex Systems and Processes'' and SFB/TRR 165 ``Waves to
Weather''.  Gneiting, Kr\"uger and Lerch thank the Klaus Tschira
Foundation for infrastructural support at the Heidelberg Institute for
Theoretical Studies (HITS). Helpful comments by Werner Ehm, Sylvia
Fr\"uhwirth-Schnatter, Alexander Jordan, as well as seminar and
conference participants at HITS, KIT, University of Bern,
University of Bonn, University of Oslo, the Extremes 2014 symposium
(Hannover), CFE (Pisa, 2014), GPSD (Bochum, 2016), ISBA
  (Sardinia, 2016), and Deutsche Bundesbank (Workshop on Forecasting,
2017) are gratefully acknowledged. We thank Gianni
Amisano for sharing his program code for Bayesian Markov switching
models. Furthermore, we thank an anonymous referee of a previous version of the manuscript for pointing us to the Rao-Blackwellization arguments employed in Theorem 4, and another anonymous referee for thoughtful comments on the paper. 
  
\appendix

\section{Computing the CRPS for mixtures of Gaussians}   \label{sec:CRPS} 

Here we discuss the computation of the CRPS in \eqref{eq:CRPS} when
the predictive distribution is an equally weighted mixture of normal
distributions, say $F = \FMP$, where $\Fcdi$ is Gaussian with mean
$\mu_i$ and variance $\sigma_i^2$.  \cite{GrimitEtAl2006} note that in
this case \eqref{eq:CRPS_mixture} can be written as
\begin{eqnarray}  
&& \label{eq:Grimit} \\ \text{CRPS} \left( \FMP, y \right) &=&
  \frac{1}{m} \sum_{i=1}^m A(y - \mu_m, \sigma^2_m) - \frac{1}{2m^2}
  \sum_{i = 1}^m \sum_{j=1}^m A(\mu_i - \mu_j, \sigma_i^2 +
  \sigma_j^2), \nonumber
\end{eqnarray}
where $A(\mu, \sigma^2) = 2 \sigma \phi(\frac{\mu}{\sigma}) + \mu (2
\Phi(\frac{\mu}{\sigma}) - 1)$, with $\phi$ and $\Phi$ denoting the
standard normal density and CDF, respectively.  The
\textsf{scoringRules} software package \citep{JordanEtAl2015} contains
\textsf{R}/\textsf{C++} code for the evaluation of
\eqref{eq:Grimit}, which requires $\mathcal{O}(m^2)$ operations.

\begin{figure}[t]
	
	\centering
	
	\begin{tabular}{cc}
		CRPS & Computation time \\
		&  (in seconds) \\  [-1cm]
		\includegraphics[width = 0.45\textwidth]{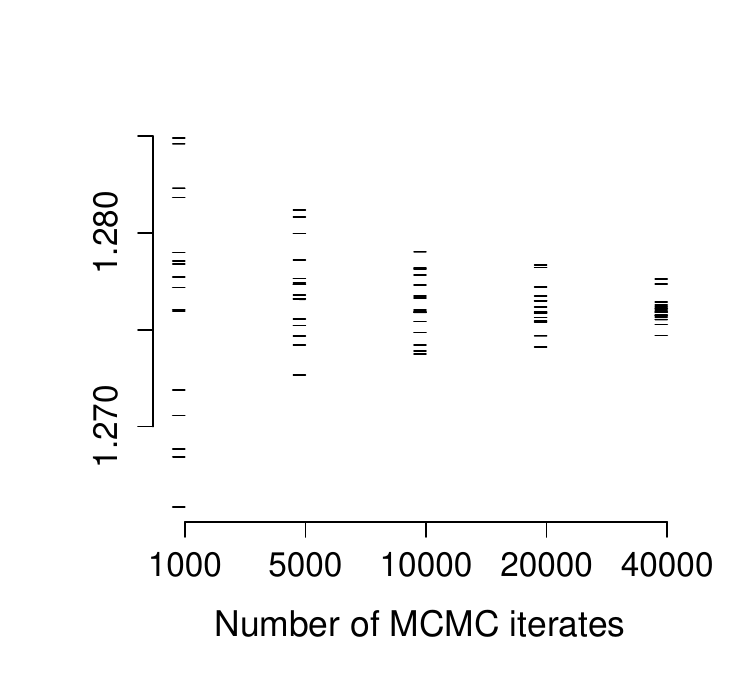} & 
		\includegraphics[width = 0.45\textwidth]{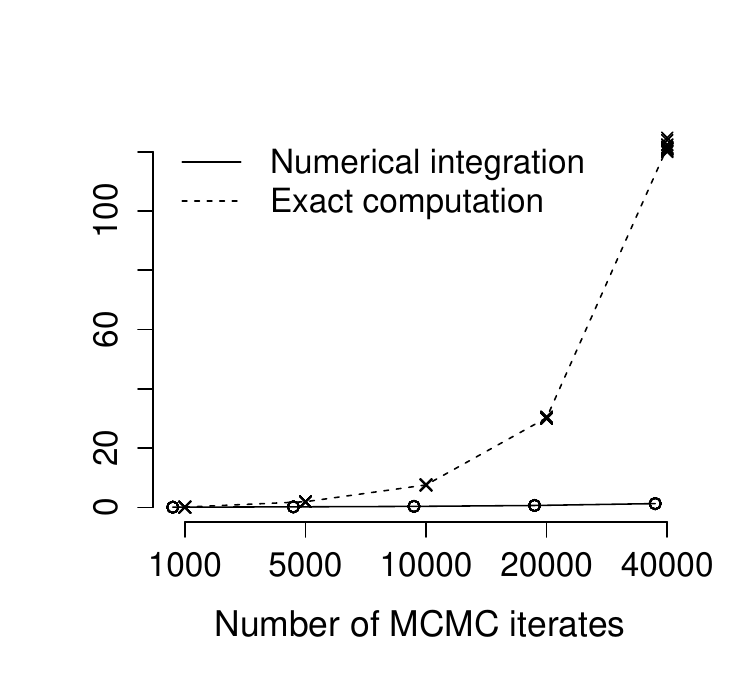}
	\end{tabular}
	
	\caption{CRPS scores for the first quarter of 2011 in the data
		example, for 16 parallel chains and various MCMC sample sizes. Left:
		The segments connect the CRPS value obtained using numerical
		integration (left node) to the score obtained using the exact
		formula (right node). Right: Computation times in seconds,
		for numerical integration (dots; solid line) and exact formula
		(crosses; dashed line).  \label{Fig:crpsExpl}}
	
\end{figure}

A potentially much faster, but not exact, alternative is to evaluate
the integral in \eqref{eq:CRPS} numerically.\footnote{Numerical
  integration could also be based on another representation of the
  CRPS that has recently been derived by \citet[p.~2390, bottom
    right]{Taillardat&2016}.}  Here we provide some evidence on the
viability of this strategy, which we implement via the \textsf{R}
function \textsf{integrate}, with arguments \textsf{rel.tol} and
\textsf{abs.tol} of \textsf{integrate} set to $10^{-6}$.  As a first
experiment, we use numerical integration to re-compute the CRPS scores
of the mixture-of-parameters estimator in our data example for the
first quarter of 2011. Figure \ref{Fig:crpsExpl} summarizes the
results for 16 parallel chains. The left panel shows that the
approximate scores are visually identical to the exact ones across all
sample sizes and chains. Indeed, the maximal absolute error incurred
by numerical integration is $8.0 \times 10^{-8}$.  The approximation
errors are dwarfed by the natural variation of the scores across MCMC
chains. The right panel compares the computation time for exact
evaluation vs.~numerical integration. The latter is much faster,
especially for large samples. For a sample of size 40,000 numerical
integration requires less than 1.5 seconds, whereas exact evaluation
requires about two minutes on an Intel i7 processor.

To obtain broad-based evidence, we next compare exact evaluation
vs.~numerical integration for all 74 forecast dates, from the second
quarter of 1996 to the third quarter of 2014, employing 16 parallel
chains for each date. We focus on the two largest MCMC sample sizes,
20\,000 and 40\,000, and find that across all 2\,368 instances (74
dates times 2 sample sizes times 16 chains), the absolute difference
of the two CRPS values never exceeds $6.3 \times 10^{-7}$.  Therefore,
we feel that numerical integration allows for the efficient evaluation
of the CRPS for mixtures of normal distributions. The differences to
the exact values are practically irrelevant and well in line with the
error bounds in \textsf{R}'s \textsf{integrate} function.

\section{Consistency of mixture-of-parameters approximations}  \label{app:thm12}

\subsection*{Proof of Theorem 1} 

In the case of the CRPS, we prove the stronger result that $\int_\real
| \FMP(z) - F_0(z) | \, \dd z \to 0$ almost surely as $m \to \infty$.
Putting $H(z) = 1 - F_0(z) + F_0(-z)$ and $\hat{H}_m(z) = 1 - \FMP(z)
+ \FMP(-z)$ for $z \in \real$, we find that, for every fixed $N > 0$,
\begin{align}
\limsup_{m \to \infty}  \int_\real |\FMP(z) - F_0(z)| \, \dd z &
\: \leq \: \limsup_{m \to \infty} \int_{-N}^N |\FMP(z) - F_0(z)| \, \dd z \nonumber \\
& \qquad + \, \int_N^\infty \! H(z) \, \dd z \: + \: \limsup_{m \to \infty} \int_N^\infty \! \hat{H}_m(z) \, \dd z. \label{eq:1.1}
\end{align}
The Ergodic Theorem implies that the first term on the right-hand
side of \eqref{eq:1.1} tends to zero, and that
\[
\int_N^\infty \! \hat{H}_m(z) \, \dd z = \int_N^\infty \frac{1}{m} \sum_{i=1}^m \left( 1 
- F_c(z \hspace{0.2mm} | \hspace{0.3mm} \theta_i)
+ F_c(-z \hspace{0.2mm} | \hspace{0.3mm} \theta_i) \right) \, \dd z 
\longrightarrow \int_N^\infty \! H(z) \, \dd z
\]
almost surely as $m \to \infty$.  In view of \eqref{eq:1.1} we
conclude that
\begin{equation}   \label{eq:1.2}
\limsup_{m \to \infty} \int_\real |\FMP(z) - F_0(z)| \, \dd z \leq 2 \, \int_N^\infty \! H(z) \, \dd z
\end{equation} 
almost surely as $m \to \infty$. As the right-hand side of
\eqref{eq:1.2} decreases to zero as $N$ grows without bounds, the
proof of the claim is complete.

In the case of the DSS, let $K(z) = 1 - F_0(z) - F_0(-z)$ and
$\hat{K}_m(z) = 1 - \FMP(z) - \FMP(-z)$ for $z \in \real$.  Due to the
finiteness of the first moments of $F_0$ and $\FMP$, $\int_\real z \,
\dd F_0(z) = \int_0^\infty K(z) \, \dd z$ and $\int_\real z \, \dd
\FMP(z) = \int_0^\infty \hat{K}_m(z) \, \dd z$.  For the second
moments, we find similarly that $\int_\real z^2 \, \dd F_0(z) = 2
\int_0^\infty z H(z) \, \dd z$ and $\int_\real z^2 \, \dd \FMP(z) = 2
\int_0^\infty z \hat{H}_m(z) \, \dd z$.  Proceeding as before, the
Ergodic Theorem implies almost sure convergence of the first and
second moments, and thereby consistency relative to the DSS.

\subsection*{Proof of Theorem 2}  

By Lemma 2.1 in Chapter 4 of \citet{Kullback1959book}, 
\[
\sup_{z \in \real} \left| 1 - \frac{\fMP(z)}{f_0(z)} \right| \longrightarrow 0
\]
almost surely as $m \to \infty$ implies the desired convergence of the
Kullback-Leibler divergence.  Let $P_m$ denote the empirical CDF of
the parameter draws $\thetam$.  Under assumption (B) almost sure
strong uniform consistency,
\[
\sup_{z \in \Omega} \left| \fMP(z) - f_0(z) \right| 
= \sup_{z \in \Omega} \left| \int_{\Theta} f_c(z|\theta) \left[\dd P_m(\theta) - \dd \Ppost(\theta) \right] \right|
\longrightarrow 0  
\]
almost surely as $m \to \infty$, yields Kullback's condition.
Finally, we establish almost sure strong uniform convergence under
assumptions (A) and (B) by applying Theorem 19.4 and Example 19.8 of
\citet{vanderVaart2000}.

\section{Consistency of empirical CDF-based approximations}  \label{app:thm3}

\subsection*{Proof of Theorem 3} 

In the case of the CRPS, we proceed in analogy to the proof of Theorem
1 and demonstrate the stronger result that $\int_\real | \FECDF(z) -
F_0(z) | \, \dd z \to 0$ almost surely as $m \to \infty$.  Putting
$H(z) = 1 - F_0(z) + F_0(-z)$ and $\hat{H}_m(z) = 1 - \FECDF(z) +
\FECDF(-z)$ for $z \in \real$, we see that, for every fixed $N > 0$,
\begin{align}
\limsup_{m \to \infty}  \int_\real |\FECDF(z) - F_0(z)| \, \dd z &
\: \leq \: \limsup_{m \to \infty} \int_{-N}^N |\FECDF(z) - F_0(z)| \, \dd z \nonumber \\
& \qquad + \, \int_N^\infty \! H(z) \, \dd z \: + \: \limsup_{m \to \infty} \int_N^\infty \! \hat{H}_m(z) \, \dd z. \label{eq:3.1}
\end{align}
The Generalized Glivenko-Cantelli Theorem \citep[][Theorem
  1.1]{DehlingPhilipp2002} implies that the first term on the
right-hand side of \eqref{eq:3.1} tends to zero almost surely as $m
\to \infty$.  If $Z_0$ has distribution $F_0$, then $\int_N^\infty \!
H(z) \, \dd z = \E (|Z_0| - N)_+$, where $(W)_+ = \max(W,0)$ denotes
the positive part of $W$.  Furthermore, by the Ergodic Theorem
\[
\int_N^\infty \! \hat{H}_m(z) \, \dd z = \frac{1}{m} \sum_{i=1}^m (|X_i| - N)_+ \longrightarrow \E (|Z_0|-N)_+
\]
almost surely as $m \to \infty$, which along with\eqref{eq:3.1}
implies that
\begin{equation}   \label{eq:3.2}
\limsup_{m \to \infty} \int_\real |\FECDF(z) - F_0(z)| \, \dd z \leq 2 \, \E (|Z_0|-N)_+
\end{equation} 
almost surely as $m \to \infty$. As the right-hand side of
\eqref{eq:3.2} gets arbitrarily close to zero as $N$ grows without
bounds, the proof {of the claim} is complete.

In the case of the DSS, it suffices to note that the moments of the
empirical CDF are the sample moments of $(X_i)_{i=1}^m$, and then to
apply the Ergodic Theorem.

\subsection*{Proof of Theorem 4}

By the law of total expectation,  $\E \FECDF(z) = \E \FMP(z)$ as
	\begin{align*}
	\E \left( \FECDF(z) | \theta_1,\dots,\theta_m \right) &= \frac{1}{m} \sum_{i=1}^{m} \mathbb{P} \left( X_i \leq z | \theta_1,\dots,\theta_m \right) \\
	&= \frac{1}{m} \sum_{i=1}^{m} \mathbb{P} \left( X_i \leq z | \theta_i \right) \\
	&= \FMP(z).
	\end{align*}
	Further, the law of total variance implies
	\begin{align*}
	\text{Var}\left(\FECDF(z)\right) &= \E \left[ \text{Var} \left( \FECDF(z)|\theta_1,\dots,\theta_m \right) \right] + \text{Var}\left[ \E \left(\FECDF(z) | \theta_1,\dots,\theta_m\right) \right]\\
	&\geq \text{Var} \left(\FMP(z)\right)
	\end{align*}
	for every $z\in\real$ and $m \in \mathbb{N}$. For a generic estimator $\hat F_m$ with finite mean, 
\begin{eqnarray*}
\E~d_{\text{CRPS}}(\hat F_m, F_0) &=& \E \int \left( \hat F_m(z) - F_0(z)\right)^2 ~dz\nonumber\\
&=& \int {\E \left( \hat F_m(z) - F_0(z)\right)^2}~dz \\
&=& \int \text{Var}~\hat F_m(z)~dz + \int \left(\E \hat F_m(z) - F_0(z) \right)^2 dz.
\end{eqnarray*}
In this light, the first part of the theorem's statement implies the second part.

\section{Simulation Study on Thinning an MCMC sample}  \label{sec:simulation-thinning}
	
\begin{figure}[t]
		
\begin{tabular}{cc}
Logarithmic score, Mixture-of-parameters & CRPS, Mixture-of-parameters \\ [-0.5cm]
\includegraphics[width = 0.45\textwidth]{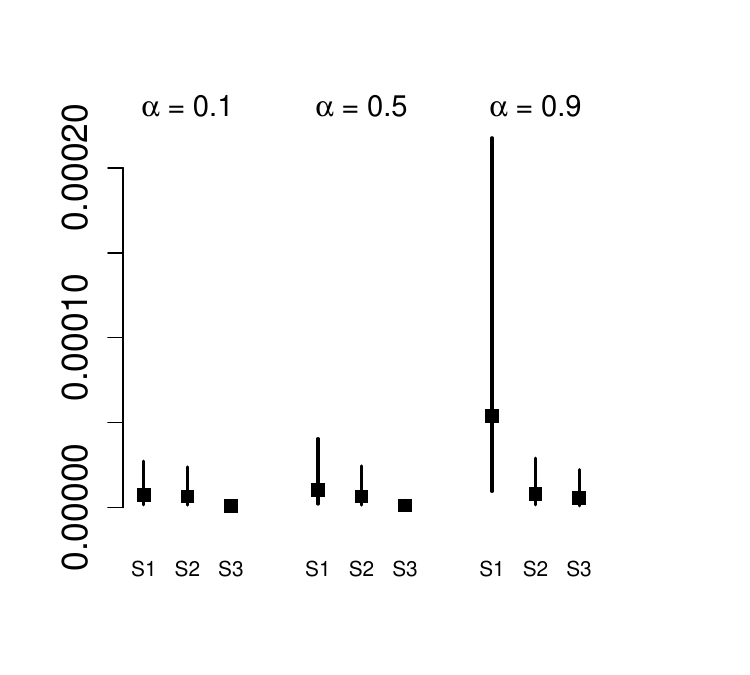} &
\includegraphics[width = 0.45\textwidth]{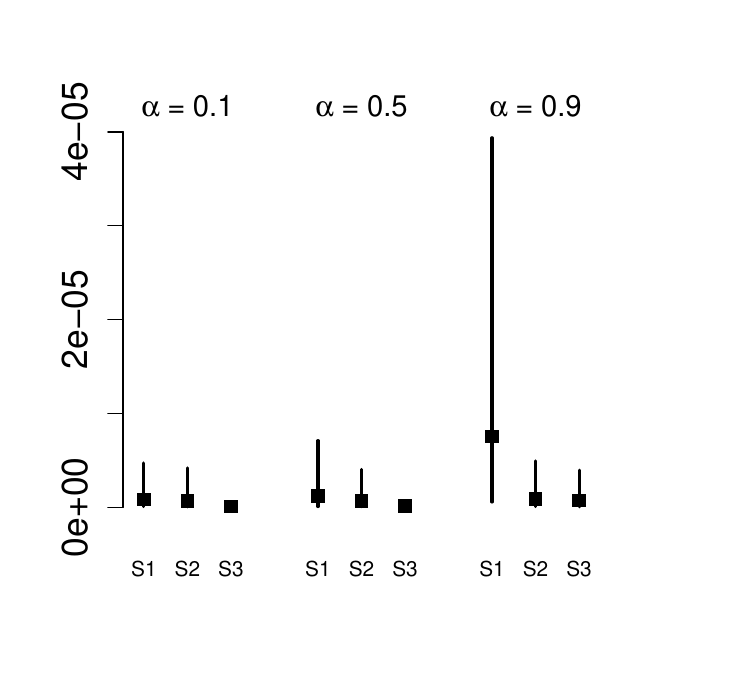} \\ [-0.35cm] 
Logarithmic score, Kernel density estimation & CRPS, Empirical CDF \\ [-0.5cm]
\includegraphics[width = 0.45\textwidth]{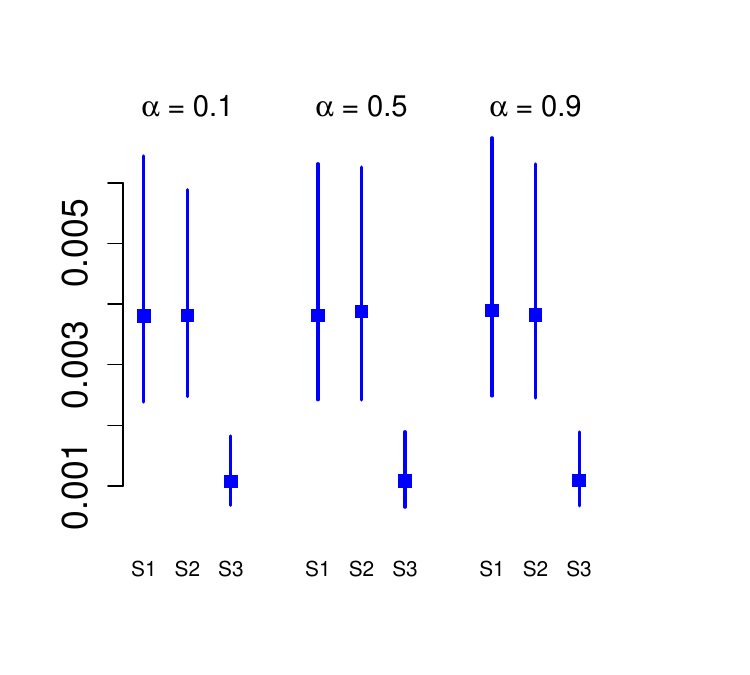} &
\includegraphics[width = 0.45\textwidth]{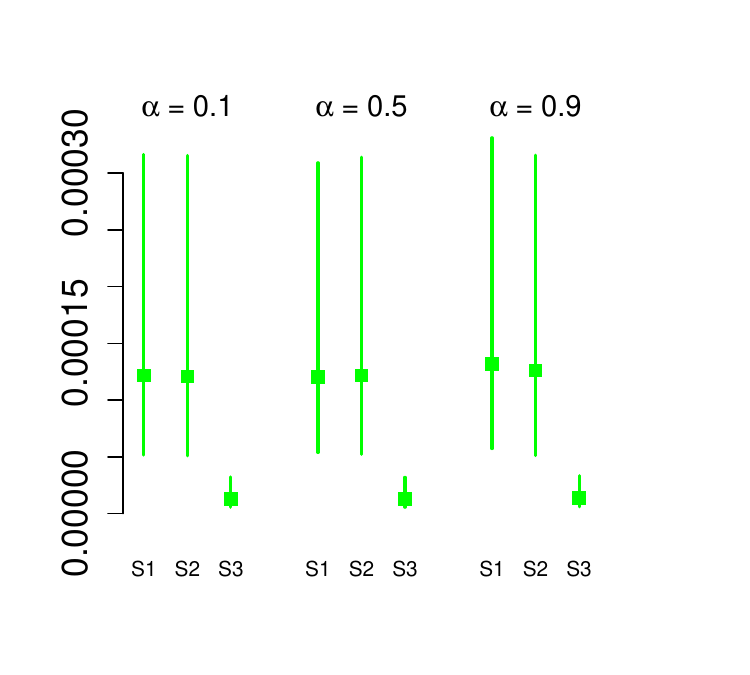} \\ [-0.75cm]
\end{tabular}
		
\caption{Performance of three sampling strategies: S1: $5\,000$ draws,
  without thinning; S2: $5\,000$ MCMC draws, retaining every $10$th
  draw from a sequence of $50\,000$ draws; S3: $50\,000$ draws,
  without thinning. Bars range from the 10th to the 90th percentile of
  the score divergences across 1\,000 replicates.  The squares mark
  the respective medians.\label{fig:thinning}}
		
\end{figure}
	
Using the same simulation setup as in Section \ref{sec:simulation}, we
further investigate the effect of thinning the Markov chains. Thinning
a chain by a factor of $\tau$ means that only every $\tau$th simulated
value is retained, and the rest is discarded. Thinning is often
applied routinely with the goal of reducing autocorrelation in the
draws.  Of the articles listed in Table 1 { of the
  Online Supplement}, about one in four explicitly reports thinning of
the simulation output, with thinning factors ranging from 2 to
100. Here we compare three sampling approaches:
\begin{itemize}
\item[(S1)] $5\,000$ MCMC draws, without thinning
\item[(S2)] $5\,000$ MCMC draws, retaining every $10$th draw from a sequence of $50\,000$ draws
\item[(S3)] $50\,000$ MCMC draws, without thinning
\end{itemize}
Note that the {samples} in S1 and S3 have the same dynamic properties,
whereas S2 will typically produce a chain with less
autocorrelation. Furthermore, S2 and S3 require the same computing
time, which exceeds that of S1 by a factor of ten. Figure
\ref{fig:thinning} summarizes the corresponding simulation results,
using parameter values $s = 2$ and $n = 12$, and varying values of the
persistence parameter $\alpha$. We report results for four popular
combinations of scoring rules and approximation methods.
	
As expected, S2 tends to outperform S1: When the sample size is held
fixed, less autocorrelation entails more precise estimators.  While
the difference in performance is modest in most cases, S2 attains
large (relative) gains over S1 when the mixture-of-parameters
estimator is applied to a very persistent sample with $\alpha = 0.9$.
This can be explained by the direct effect of the persistence
parameter $\alpha$ on the parameter draws $\thetam$, whereas the
influence is less immediate for the KDE and ECDF approximation
methods, which are based on the sequence $\Xm$ obtained in an
additional sampling step.  Furthermore, S3 outperforms S2 in all cases
covered.  While the effects of thinning have not been studied in the
context of predictive distributions before, this observation is in
line with extant reports of the greater precision of unthinned chains
\citep{Geyer1992, MacEachernBerliner1994, LinkEaton2012}. The
performance gap between S3 and S2 is modest for the
mixture-of-parameters estimator (top row of Figure
\ref{fig:thinning}), but very pronounced for the other estimators.

\section{Implementation details for the data example}  \label{app:implementation}

Here we provide additional information on the Markov switching model
for the quarterly U.S.~GDP growth rate, $Y_t$. As described in
equation~\eqref{eq:ms} in Section \ref{sec:casestudy}, the model is
given by $Y_t = \nu + \alpha Y_{t-1} + \varepsilon_t$, where
$\varepsilon_t \sim \mathcal{N}(0, \eta^2_{s_t})$, and $s_t \in \{ 1,
2 \}$ is a discrete state variable that switches according to a Markov
chain.

\begin{table}
	
	\caption{Prior parameters in the Markov switching model.  \label{tab:AG2007}}
	
	\bigskip
	
	\centering
	
	\footnotesize
	
	\begin{tabular}{lccccc}
		\toprule
		Symbol in Amisano & \multirow{2}{*}{$\boldsymbol{\underline{\mu}_{\delta}}$} & 
		\multirow{2}{*}{$\bf{\underline{H}}^{-1}_{\delta}$} & 
		\multirow{2}{*}{$\underline{s}$} & \multirow{2}{*}{$\underline{\nu}$} & 
		\multirow{2}{*}{$\mathbf{R}$} \\
		and Giacomini (2007) & & & & & \\ 
		\toprule 
		Parameter choice & $0_{[2, 1]}$ & $25 \times I_2$ & 0.3 & 3 & 
		$\begin{bmatrix} 8 & 2 \\ 2 & 8 \\ \end{bmatrix}$ \\ [3mm]
		\multirow{2}{*}{Relation to our eq.~\eqref{eq:ms}} & Prior mean & Prior variance  
		& \multicolumn{2}{c}{Prior parameters} & Dirichlet prior \\ 
		& for $(\nu, \alpha)'$ & for $(\nu, \alpha)'$ & \multicolumn{2}{c}{for $\eta^2_{s_t}$} 
		& state transitions \\ 
		\bottomrule
	\end{tabular}
	
	\bigskip 
	
\end{table}

Our implementation follows \citet[Section 6.3]{AmisanoGiacomini2007},
in that our prior distributions have the same functional forms but
possibly different parameter choices, as summarized in Table
\ref{tab:AG2007}.  However, note that we use prior parameters for the
residual variances in both latent states, whereas Amisano and
Giacomini (2007) assume the residual variance to be constant across
states.

Let $\beta = (\nu, \alpha)'$ denote the parameters for the conditional
mean equation \eqref{eq:ms}, $\overline{s}_t = (s_1, \ldots, s_t)'$
the sequence of latent states up to time $t$, $h = (\eta_1^{-2},
\eta_2^{-2})'$ the inverses of the state-dependent residual variances,
and $\mathbf{P}$ the $2 \times 2$ transition matrix for the latent
states. Our Gibbs sampler can then be sketched as follows:

\begin{itemize}
	
\item Draw $\beta \, | \, h, \overline{s}_t$ from a Gaussian
  posterior.  The mean and variance derive from a generalized least
  squares problem, with observation $t$ receiving weight
  $\eta^{-2}_{s_t}$.
	
\item Draw $h \, | \, \beta, \overline{s}_t$ from a Gamma posterior.
  The Gamma distribution parameters for $\eta_s^{-2}, s\in \{1, 2\},$
  are calculated from the observations $t$ for which $s_t = s$.  If
  necessary, permute the draws such that $\eta_1^2 > \eta_2^2$.
	
\item Draw $\overline{s}_t \, | \, \beta, h, \mathbf{P}$ using the
  algorithm described by \citet[][pp.~194--195]{Greenberg2013}.
	
\item Draw $\mathbf{P} \, | \, \overline{s}_t$ from a Dirichlet
  posterior.
	
\end{itemize}

\noindent
Gianni Amisano kindly provides implementation details and Matlab code
via his website
(\url{https://sites.google.com/site/gianniamisanowebsite/home/teaching/istanbul-2014},
last accessed: March 25, 2019). An \textsf{R} implementation of
his code is available within the \textsf{R} package
\textsf{scoringRules} \citep{JordanEtAl2015}, see
\url{https://github.com/FK83/scoringRules/blob/master/KLTG2020_replication.pdf}
for details.

\end{document}


\setcounter{page}{36} 
	
	\maketitle 

\renewcommand{\thesection}{S\arabic{section}}

\section{Literature survey}  \label{app:lit}  

\begin{table}[p]
	
	\centering
	
	\caption{Approximation methods and scoring rules in recent studies
		using probabilistic forecasts based on MCMC output. \label{tab:lit}}
	
	\medskip 
	\scriptsize

	\begin{tabular}{lll} 
		\toprule 
		Approximation method      & Logarithmic score & CRPS \\
		\midrule
		{\em Based on parameter draws} & & \\
		\midrule
		Mixture-of-parameters     & \cite{AmisanoGiacomini2007}       & \cite{KallacheEtAl2010} \\
		& \cite{GschloesslCzado2007}        & \cite{TrombeEtAl2012} \\
		& \cite{LopesEtAl2008}              & \cite{RisserCalder2015} \\
		& \cite{Maheu2008}				  &  \cite{FanEtAl2018}\\
		& \cite{Liu2009}					  & \\
		& \cite{Maheu2009}				  & \\
		& \cite{GewekeAmisano2010,GewekeAmisano2011}          & \\
		& \cite{Jensen2010,Jensen2013,Jensen2014}	  & \\
		& \cite{JochmannEtAl2010}			  & \\
		& \cite{KallacheEtAl2010}           & \\
		& \cite{LiEtAl2010}                 & \\
		& \cite{DelatolaGriffin2011}        & \\
		& \cite{MaheuEtAl2012}			  & \\
		& \cite{ManeesoonthornEtAl2012}     & \\
		& \cite{Jin2013,Jin2016} 		      & \\
		& \cite{BacsturkEtAl2014}			  & \\
		& \cite{Maheu2014}				  & \\
		& \cite{RisserCalder2015}           & \\
		& \cite{ZhouEtAl2015}               & \\
		& \cite{Kastner2016}				  & \\
		& \cite{MaheuYang2016}		  & \\
		& \cite{WarneEtAl2016}		      & \\
				& \cite{AmisanoGeweke2017} & \\
		& \cite{LeaoEtAl2017} & \\
		& \cite{BittoFruehwirthSchnatter2018} & \\
		& \cite{LiuMaheu2018} & \\
		& \cite{MaheuSong2018} & \\
		& \cite{WijeyakulasuriyaEtAl2018} & \\
		\midrule
		{\em Based on a sample} & & \\
		\midrule
		Empirical CDF     &                                   & \cite{GschloesslCzado2007} \\
		&                                   & \cite{LopesEtAl2008} \\
		&                                   & \cite{PanagiotelisSmith2008} \\
		&                                   & \cite{DeLaCruzBranco2009} \\
		&                                   & \cite{SalazarEtAl2011} \\
		&                                   & \cite{FriederichsThorarinsdottir2012} \\
		&                                   & \cite{SigristEtAl2012} \\
		&                                   & \cite{GroenEtAl2013} \\
		&                                   & \cite{LeiningerEtAl2013} \\
		&                                   & \cite{BerrocalEtAl2014} \\
		&                                   & \cite{ClarkRavazzolo2014} \\
		&                                   & \cite{KruegerEtAl2015} \\
		&                                   & \cite{SahuEtAl2015} \\
		&                                   & \cite{SigristEtAl2015} \\
		&                                   & \cite{SmithVahey2015} \\
		&                                   &  \cite{Berg2017} \\
		&									& \cite{AastveitEtAl2017} \\
		&									& \cite{BerkEtAl2018} \\
		&	& \cite{RisserEtAl2018} \\
		&									& \cite{Louzis2019} \\
		&									& \cite{WhitePorcu2018} \\
		&									& \cite{WhiteEtAl2019} \\
		&									& \cite{WhiteEtAl2019env} \\
				&	& \cite{ClarkEtAl2018} \\
		\midrule
		Kernel density estimation & \cite{BelmonteEtAl2014} & \cite{KruegerNolte2015} \\
		& \cite{BauwensEtAl2014} & \\
		& \cite{BergHenzel2015} & \\
		& \cite{CarrieroEtAl2015c, CarrieroEtAl2015d} & \\
		&  \cite{Berg2017} & \\
		& \cite{Metaxoglou2018} \\
		\midrule
		Gaussian approximation & \citet{AdolfsonEtAl2007} & \cite{BrandtEtAl2014} \\
		& \cite{Clark2011} & \cite{RodriguesEtAl2014} \\              
		& \cite{CarrieroEtAl2015a, CarrieroEtAl2015b} & \\
		& \cite{ClarkRavazzolo2014} & \\
		& \cite{GiannoneEtAl2015} & \\
		& \cite{WarneEtAl2016} & \\
		\bottomrule
	\end{tabular}
\end{table}

In Table \ref{tab:lit}, we provide a comprehensive survey of
articles that consider probabilistic forecasts based on MCMC output,
subject to the following further criteria:
\begin{itemize}

\item We include published or pre-published articles in scientific
  journals or proceedings only.  In particular, Table \ref{tab:lit}
  does not list working papers and preprints.
	
\item We focus on studies where forecasts based on Bayesian MCMC
  methods are produced, and evaluated via proper scoring rules, and we
  restricted our attention to studies of real-valued linear variables,
  ignoring articles that deal with binary or categorical observations
  only.
	
\item Furthermore, as we are interested in full probabilistic
  forecasts based on MCMC output, we disregarded articles where only
  functionals of the forecast distributions, such as means or medians,
  are evaluated, and we retained those studies only where the
  computation of the scores is documented in sufficient detail.
	
\item Finally, Table \ref{tab:lit} lists only papers that employ the
  CRPS or the logarithmic score.  Very few studies have used scoring
  rules other than {these:} \citet{RieblerEtAl2012} use the
  Dawid--Sebastiani score, and \citet{SmithVahey2015} and
  \citet{TranEtAl2016} use weighted versions of the logarithmic score
  and the CRPS, as described by \citet{DiksEtAl2011} and
  \citet{GneitingRanjan2011}, respectively.  Multivariate
    settings have been discussed in \citet{HeldEtAl2017} and
    \citet{WhiteEtAl2019}.

\end{itemize}

If a study reports several choices of approximation method and scoring
rule, each choice constitutes a separate entry.

\clearpage

\section{Additional figures for simulation study}

\begin{figure}[!htbp]

\centering

\begin{tabular}{cc}
	\multicolumn{2}{c}{$\alpha = 0.1, s = 2, n = 12$} \\ 
	Logarithmic score & CRPS \\  
	[-5mm]
	\includegraphics[width=0.4\textwidth]{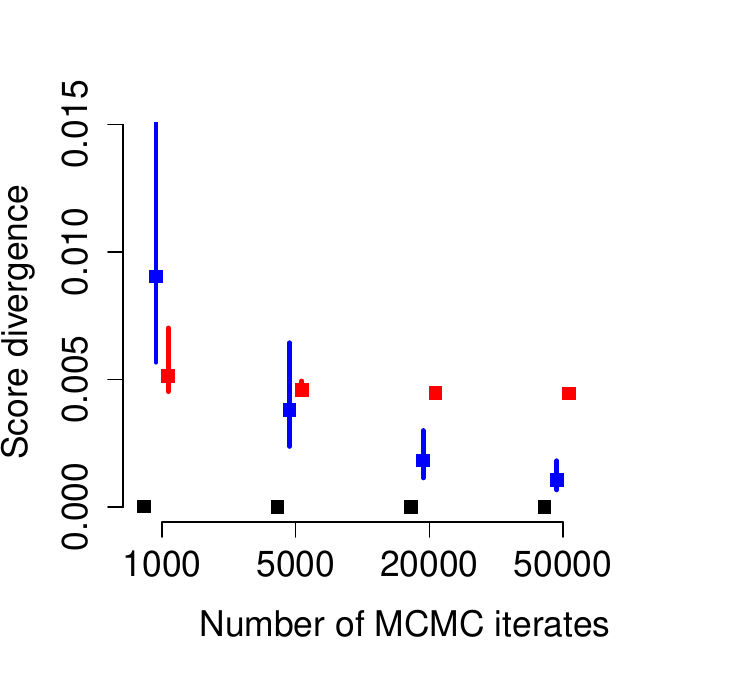} &
	\includegraphics[width=0.4\textwidth]{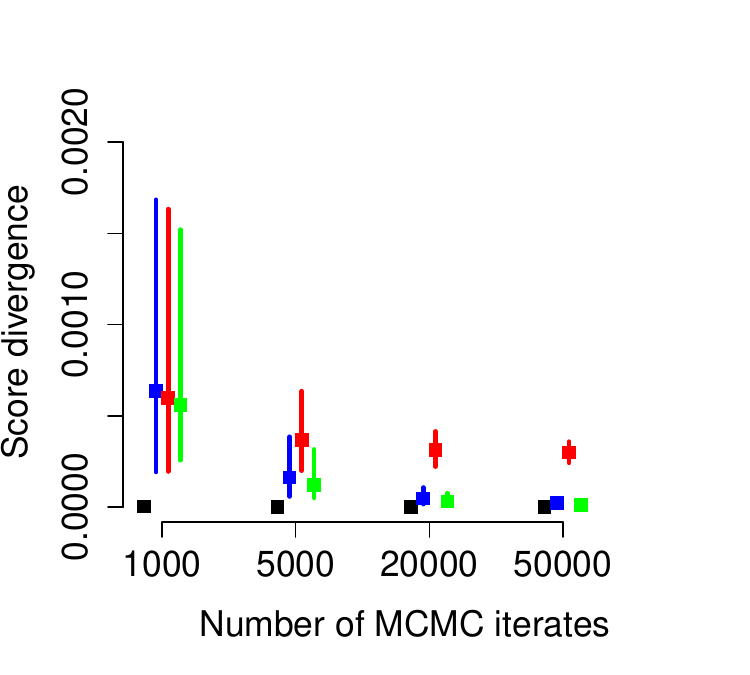} \\
	\multicolumn{2}{c}{$\alpha = 0.5, s = 2, n = 12$} \\ 
	Logarithmic score & CRPS \\  
	[-5mm]
	\includegraphics[width=0.4\textwidth]{LogS_fw_setup2.pdf} &
	\includegraphics[width=0.4\textwidth]{CRPS_fw_setup2.pdf} \\
	\multicolumn{2}{c}{$\alpha = 0.9, s = 2, n = 12$} \\ 
	Logarithmic score & CRPS \\ 
	[-5mm]
	\includegraphics[width=0.4\textwidth]{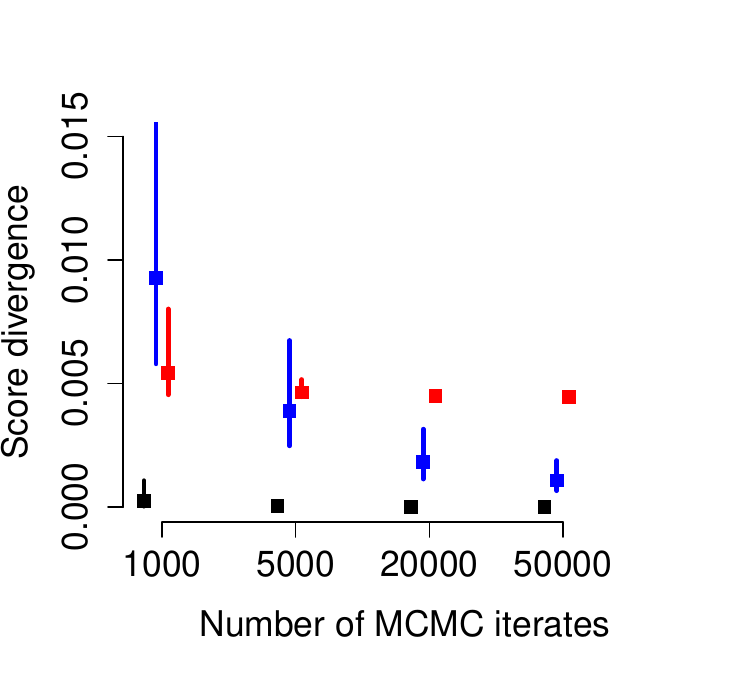} &
	\includegraphics[width=0.4\textwidth]{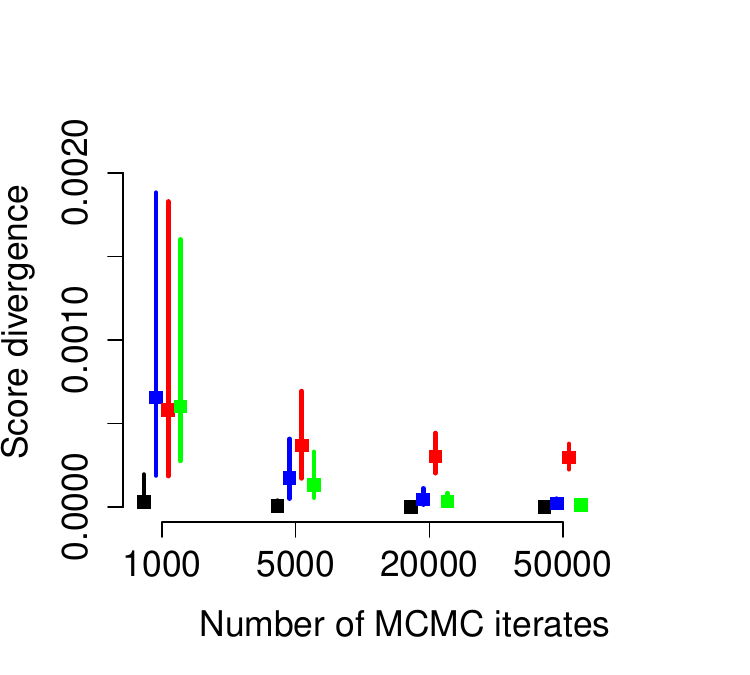} \\[-1.4cm]
	\multicolumn{2}{c}{\includegraphics[width=.8\textwidth]{legend.pdf}} \vspace{-1.6cm}
\end{tabular}

\caption{Same as Figure 1 in the paper, but for various parameter
	constellations of the data-generating process (cf.~Table 3 in
	the paper).}

\end{figure}

\begin{figure}

\centering
\begin{tabular}{cc}
	\multicolumn{2}{c}{$\alpha = 0.1, s = 2, n = 20$} \\ 
	Logarithmic score & CRPS \\  
	[-5mm]
	\includegraphics[width=0.4\textwidth]{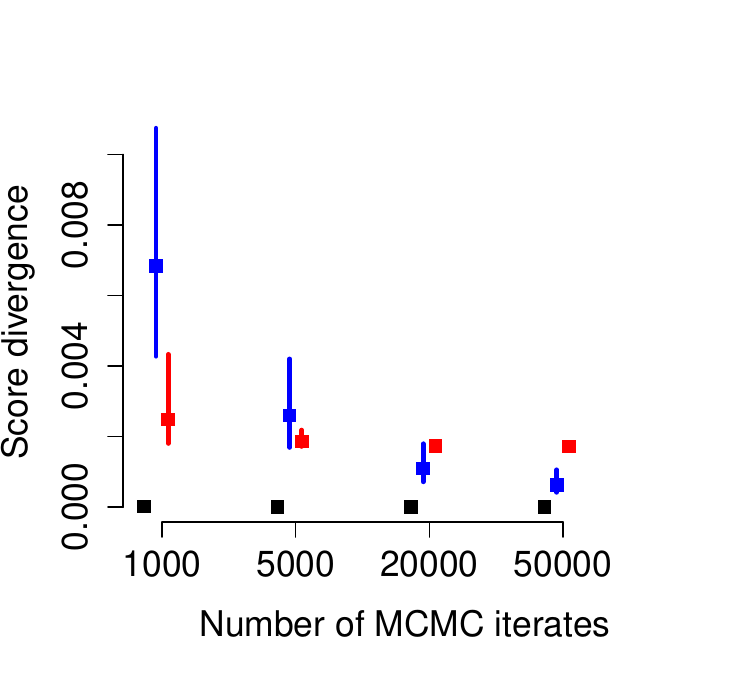} &
	\includegraphics[width=0.4\textwidth]{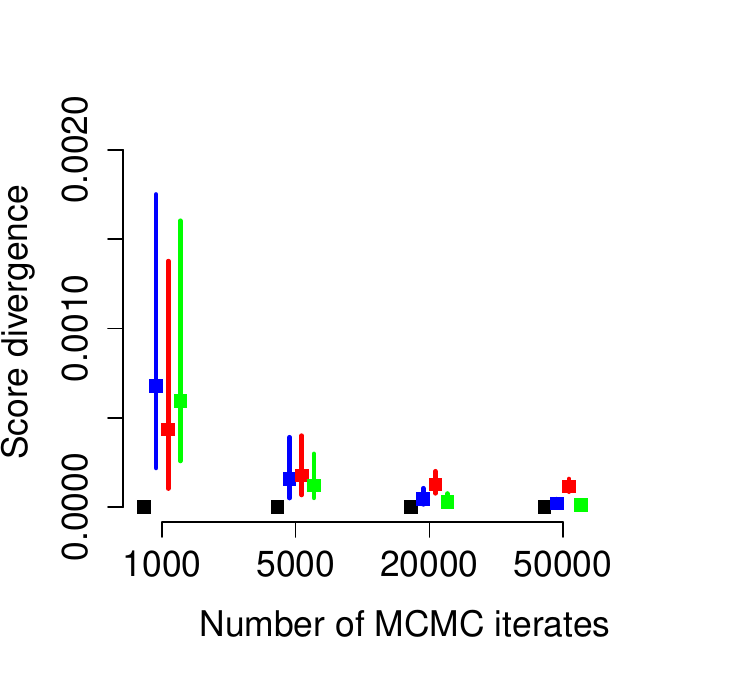} \\
	\multicolumn{2}{c}{$\alpha = 0.5, s = 2, n = 20$} \\ 
	Logarithmic score & CRPS \\  
	[-5mm]
	\includegraphics[width=0.4\textwidth]{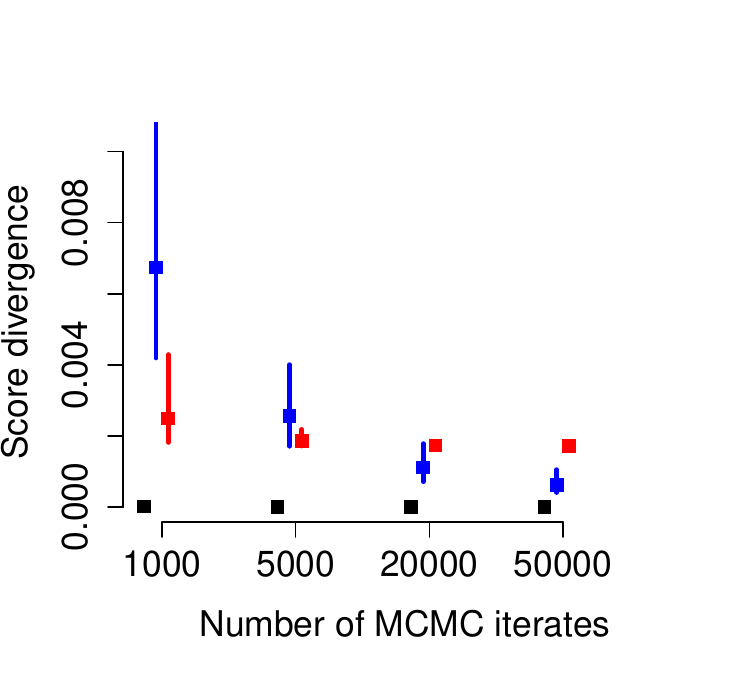} &
	\includegraphics[width=0.4\textwidth]{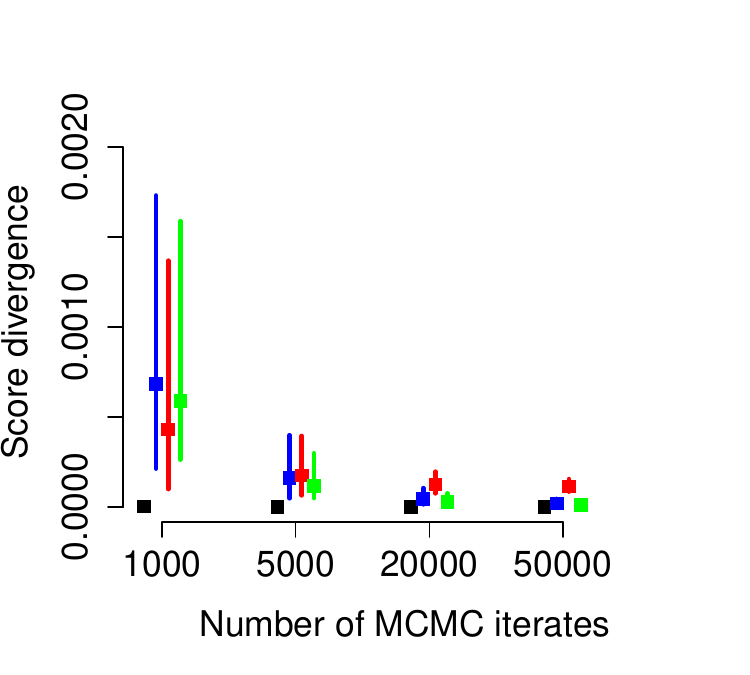} \\
	\multicolumn{2}{c}{$\alpha = 0.9, s = 2, n = 20$} \\ 
	Logarithmic score & CRPS \\ 
	[-5mm]
	\includegraphics[width=0.4\textwidth]{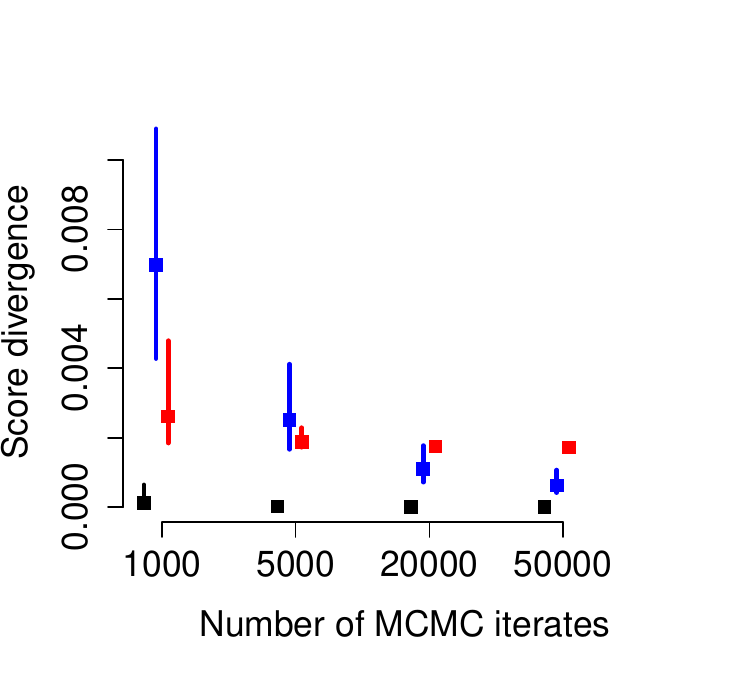} &
	\includegraphics[width=0.4\textwidth]{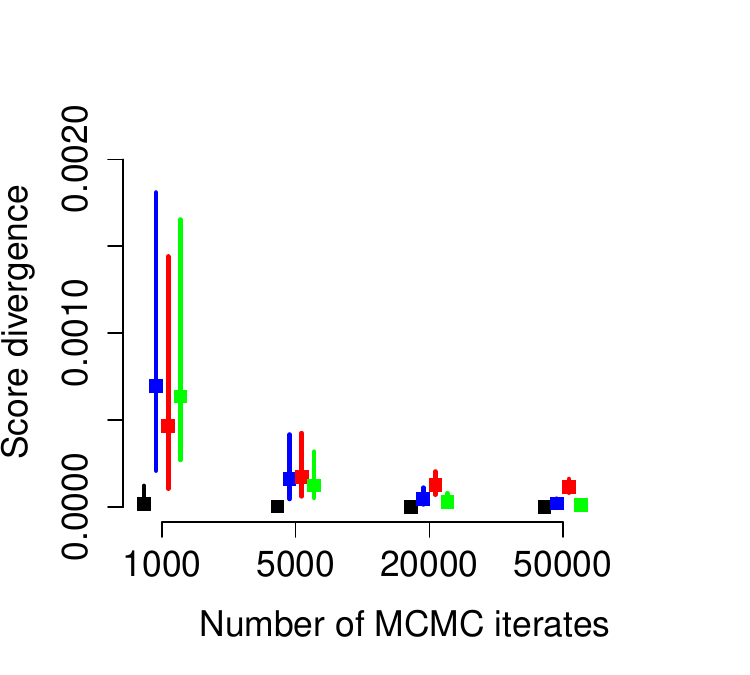} \\[-1.4cm]
	\multicolumn{2}{c}{\includegraphics[width=.8\textwidth]{legend.pdf}} \vspace{-1.6cm}
\end{tabular}

\caption{Same as Figure 1 in the paper, but for various parameter
	constellations of the data-generating process (cf.~Table 3 in
	the paper).}

\end{figure}

\section{Simulation results for quantiles}

Here we use the simulation design from Section 4 of the paper in order to compare the quantile predictions implied by various approximation methods. For a quantile prediction $x$ and outcome $y$, we consider the asymmetric piecewise linear (APL) scoring function defined as
\begin{equation}
\text{APL}(x, y) = \left(\ind\{y<x\}-\tau\right)~(x-y),\label{eq:apl}
\end{equation}
where $\tau \in (0,1)$ is the quantile level of interest. Denote the forecast distribution by $F$ and the true distribution by $G$. \citet[last formula on left column of p.~1926]{BentzienFriederichs2014} provide an expression for the divergence function of the APL score:
\begin{eqnarray}
d_S(F, G) &=& \int_\Omega\left[\text{APL}(x_\tau, y)-\text{APL}(x_\tau^*, y)\right]\dd G(y) \nonumber \\
&=& \int_{x_\tau^*}^{x_\tau} (x_\tau-y)~\dd G(y) \label{final}
\end{eqnarray}
where $x_\tau$ and $x_\tau^*$ denote the quantile forecasts implied by $F$ and $G$ respectively. Note that  $d_{\text{APL}}(F,G) \ge 0$ (i.e., the APL score is proper) but that $d_{\text{APL}}(F,G) = 0$ whenever $F$ and $G$ share the same $\tau$ percent quantile (i.e., the APL score is not strictly proper relative to typical families of distributions).\\

\begin{figure}[!htbp]
	
	\centering
	
	\begin{tabular}{cc}
	Quantile score ($\tau = .01$) & Quantile score ($\tau = .05$)\\ [-1cm]
		\includegraphics[width=0.5\textwidth]{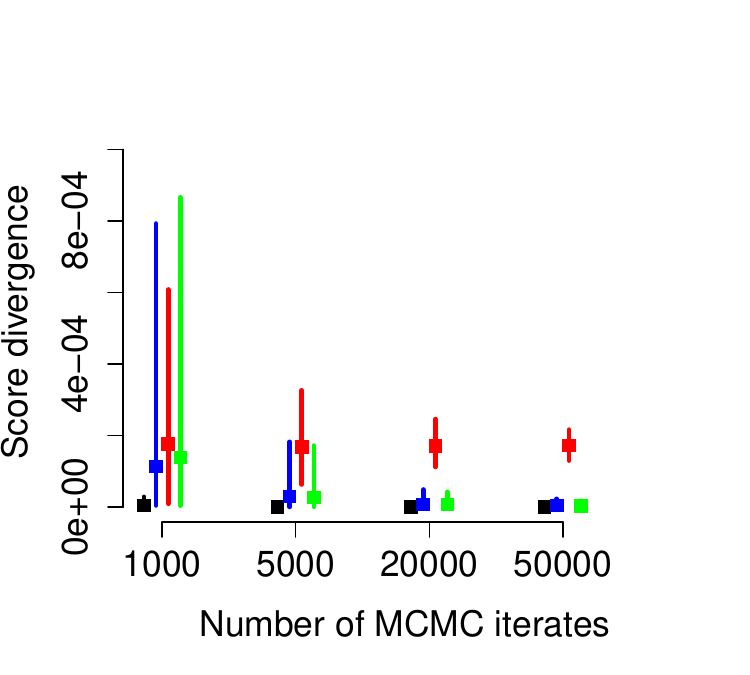} &
		\includegraphics[width=0.5\textwidth]{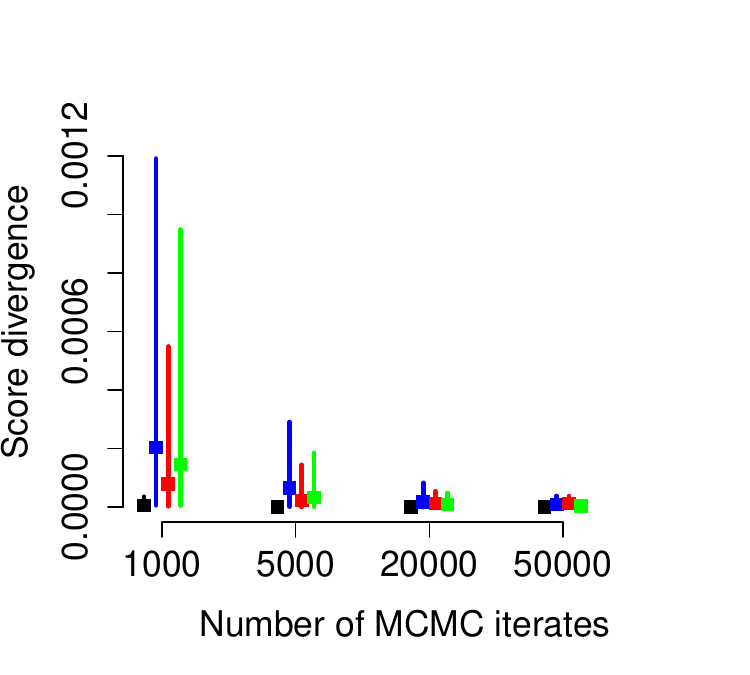} \\[-1.4cm] 
		\multicolumn{2}{c}{\includegraphics[width=\textwidth]{legend.pdf}}
	\end{tabular}\vspace{-1.8cm}
	
	\caption{Same as Figure 1 in the paper, but for the quantile scoring function in Equation (\ref{eq:apl}), together with two different quantile levels $\tau$.  \label{fig:sim-sizeq}}
	
\end{figure}

\begin{figure}[!htbp]
	
	\centering
	
	\begin{tabular}{cc}
	Quantile score ($\tau = .01$) & Quantile score ($\tau = .05$) \\ [-1cm]
		\includegraphics[width=0.5\textwidth]{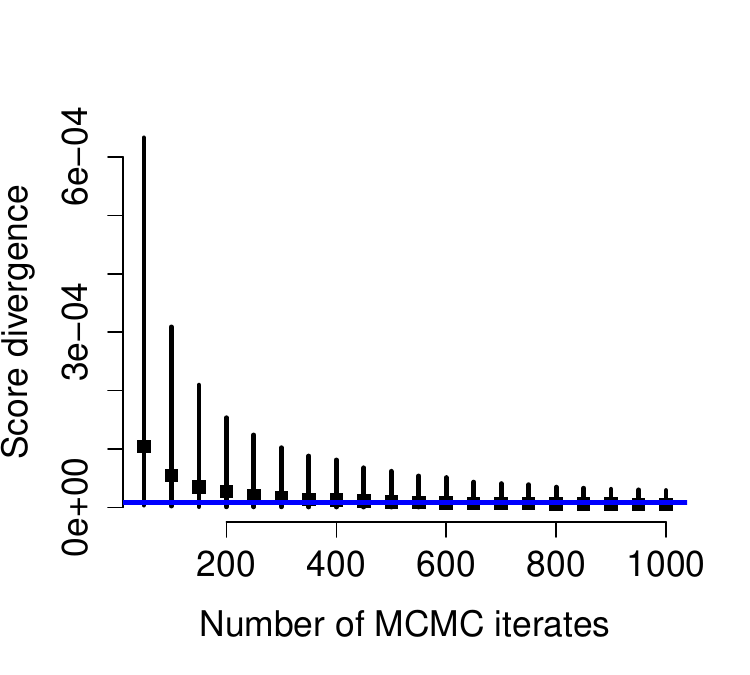} &
		\includegraphics[width=0.5\textwidth]{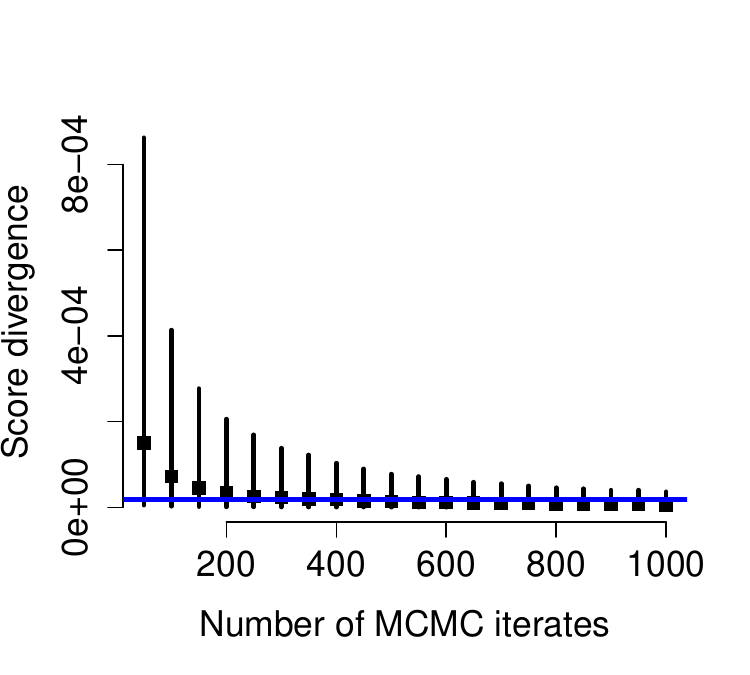}
	\end{tabular}
	
	\caption{Same as Figure 2 in the paper, but for the quantile scoring function in Equation (\ref{eq:apl}), together with two different quantile levels $\tau$.  \label{fig:sim-zoomq}}
	
\end{figure}

Figures \ref{fig:sim-sizeq} and \ref{fig:sim-zoomq} below (which are analogous to Figures 1 and 2 in the main paper) present the results for quantiles:
\begin{itemize}
	\item The mixture-of-parameters estimator attains the smallest score divergences at a given MCMC sample size (Figure \ref{fig:sim-sizeq}) and attains similar score divergences as the Kernel density estimator at a fraction of the MCMC sample size (Figure \ref{fig:sim-zoomq}). These findings are in close analogy to the ones for the LogS and CRPS. 
	\item The Gaussian approximation performs poorly for $\tau = .01$ (left panel of Figure \ref{fig:sim-sizeq}), but performs relatively well for $\tau = .05$ (right panel of the figure). These results appear to be driven by the fact that the five percent quantile of an $\mathcal{N}(\mu, \sigma^2)$ distribution is similar to the five percent quantile of a $t$ distribution with the same mean and variance, whereas the one percent quantiles of both distributions differ more markedly. We also note that the poor performance of the Gaussian approximation for some quantile levels follows from the poor performance of the Gaussian approximation in terms of the CRPS, which is an integral over quantile scores at all levels $\tau \in (0,1)$, see the discussion in Section 6 of the paper. 	
	\item The performance of the Kernel and ECDF approximations is similar, in line with the results for the CRPS.
\end{itemize}

\clearpage

\section{Alternative simulation design}
\label{sec:n_inv_g}

Here we present results for a second simulation design that is arguably closer to empirical practice than the design in Section 4 of the main paper, at the cost of increased complexity. The second design involves a posterior predictive distribution that arises from a concrete Bayesian model, as opposed to the abstract posterior predictive distribution in Section 4 of the main paper. Specifically, the second simulation design considers Bayesian analysis of the Gaussian distribution, using normal and inverse Gamma prior distributions. Our formal presentation closely follows \citet[Section 6]{Murphy07}, to which we refer for details. \citet[Section 5.3]{Hoff2009} provides a textbook treatment of the model.\\

Suppose we are given a sample from a normal population, and the prior is such that the posterior predictive distribution for a new draw $Y$ is known analytically to be a $t$ distribution. We compare this analytical distribution to various approximations based on a stylized `MCMC sampler'. (In practice, MCMC sampling is not necessary if the posterior predictive distribution is known analytically. However, comparing the true distribution to approximations based on an MCMC type sample provides guidance for more complex settings where the true distribution is not available analytically.) We mimic an MCMC sampler by producing dependent draws from either the posterior predictive distribution for $Y$ or the posterior distribution of $\theta = (\mu, \sigma^2)'$. One iteration of the simulation design consists of the following steps:
\begin{itemize}
	\item \underline{Draw a sample of data} from the predictand, using independent draws from a Gaussian distribution with parameters $\mu$ and $\sigma^2$. This sample mimics the `training' or `in-sample' data available to the statistician. 
	\item \underline{Determine posterior distribution}. Using a normal-inverse Gamma prior, the posterior distribution of $(\mu, \sigma^2)$ is again normal-inverse Gamma, taking the form $$P_{post}(\theta) = \mathcal{N}(\mu|m_n,\sigma^2V_n)IG(\sigma^2|a_n,b_n),$$
	where $IG$ denotes an inverse Gamma distribution, and $m_n,\sigma^2_n, a_n,b_n$ are posterior parameters that depend on the data from the previous step and the prior parameters $m_0, V_0, a_0, b_0$. The posterior predictive distribution for $Y$ takes the form 
	$$F_0(y) = \mathsf{T}\left(y\bigg|m_n, \frac{b_n(1+V_n)}{a_n}, 2a_n\right),$$
	where $\textsf{T}(\cdot | a, b, c )$ denotes the CDF of a variable $Z$
	with the property that $(Z-a)/\sqrt{b}$ is standard Student $t$
	distributed with $c$ degrees of freedom.
	\item \underline{Draw from posterior distribution}. In order to produce persistent `MCMC-type' draws from the posterior distribution of $\theta$ or the posterior predictive distribution of $Y$, we first generate a sequence of quantile levels $U_1, U_2, \ldots, U_m,$ where $U_i = \Phi(Z_i),$ and the sequence $(Z_i)$ is generated as 
	$$Z_i = \rho Z_{i-1} + \varepsilon_i,$$
	where $\varepsilon_i \stackrel{IID}{\sim}\mathcal{N}(0,1-\rho^2).$ Hence the unconditional distribution of $Z_i$ is standard normal, and the parameter $\rho$ determines the persistence of the stylized MCMC sampler. The $i$th draw $\sigma^2_i$ is then given by the $U_i$ quantile from the IG posterior for $\sigma^2$.\footnote{Hence the dependence of $(\sigma^2_i)_{i=1}^m$ is described by a Gaussian copula, where the $[i,j]$ entry of the correlation matrix is given by $\rho^{|i-j|}$.} We further draw $\mu_i$ from the Gaussian posterior of $\mu$ given $\sigma^2 = \sigma^2_i$. For the $i$th posterior predictive draw of $Y$, we generate a single draw from an $\mathcal{N}(\mu_i, \sigma^2_i)$ distribution.
	\item \underline{Compute the score divergence} between the approximation $\hat F_m$ and the truth $F_0$. We denote this divergence by $d_S(\hat F_m, F_0)$. We repeat this step for various approximations $\hat F_m$ and various scoring rules or scoring functions $S$.
\end{itemize}

Table \ref{tab:params} lists the parameters used in the Monte Carlo study. Most importantly, these choices imply that the posterior predictive distribution is a $t$ distribution with $2~a_n = 2~(a_0 + n/2) = 24$ degrees of freedom. These parameters are motivated by empirical practice in macroeconomic time series analysis. A sample of size $n = 20$ corresponds to five years of quarterly data, which can be a plausible choice if parameter change impedes the use of longer data samples. Similarly, a $t$ distribution with a moderately large number of $24$ degrees of freedom seems broadly plausible for macroeconomic time series.\\ 

\begin{figure}[!htbp]
	
	\centering
	
	\begin{tabular}{cc}
		Logarithmic score & CRPS \\ [-1cm]
		\includegraphics[width=0.5\textwidth]{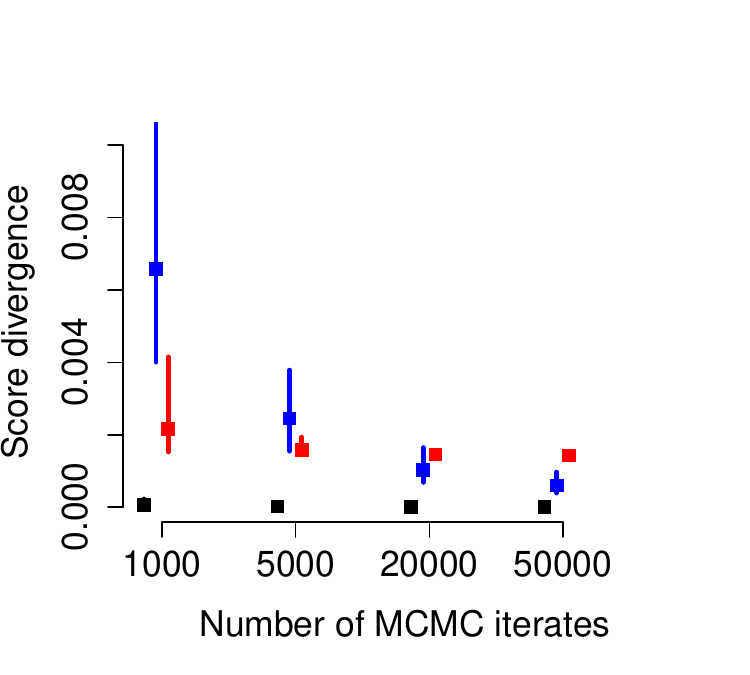} &
		\includegraphics[width=0.5\textwidth]{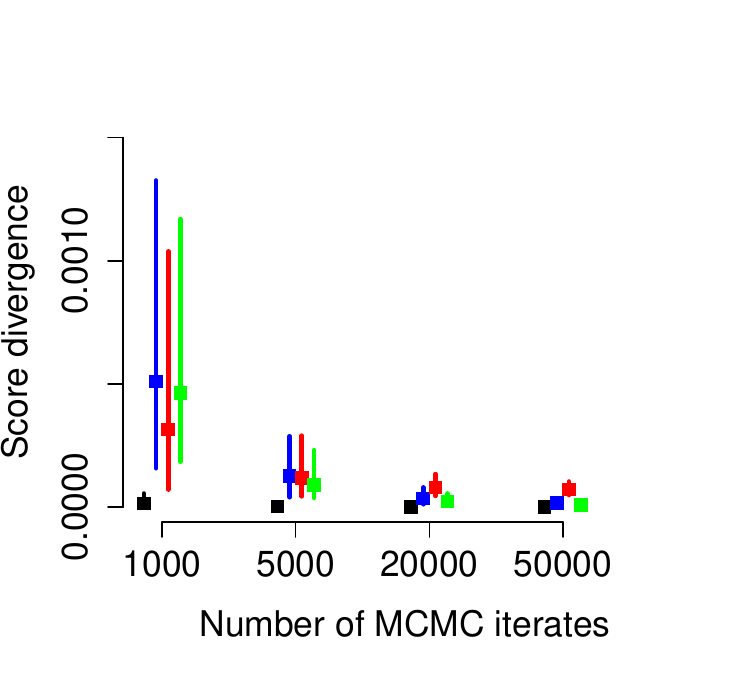} \\[-1.4cm] 
		\multicolumn{2}{c}{\includegraphics[width=\textwidth]{legend.pdf}}
	\end{tabular}\vspace{-1.8cm}
	
	\caption{Same as Figure 1 in the paper, but for the alternative simulation design described in Section \ref{sec:n_inv_g}. \label{fig:sim-sizen}}
	
\end{figure}

Figures \ref{fig:sim-sizen} to \ref{fig:sim-zoomnq} summarize the results for the alternative simulation design. The main qualitative results -- the superior performance of the mixture-of-parameters estimator, the lack of consistency of the Gaussian approximation, the similar performance of the Kernel and ECDF estimators in terms of the CRPS and the quantile score, and the poor performance of the Kernel estimator in terms of the logarithmic score -- are the same as in the simpler design. 

\begin{table}
	\centering
	\caption{Parameter choices used in the alternative simulation design. See \citet[Section 6]{Murphy07} for details on the parametrization.\label{tab:params}}\medskip
	\begin{tabular}{clr}\toprule
		Symbol & Interpretation & Value \\ \toprule
		$\mu$ & true mean parameter & 0 \\
		$\sigma^2$ & true variance parameter & 1 \\
		$n$ & length of simulated sample & 20 \\
		$m_0$ & prior mean for $\mu$ & $0$ \\
		$V_0$ & prior variance parameter for $\mu$ & $500$ \\
		$a_0$ & prior parameter for $\sigma^2$ & $2$  \\
		$b_0$ & prior parameter for $\sigma^2$ & $2$  \\
		$\rho$ & persistence parameter for toy MCMC & $0.5$ \\ \bottomrule
	\end{tabular}
\end{table}

\begin{figure}[!htbp]
	
	\centering
	
	\begin{tabular}{cc}
		Logarithmic score & CRPS \\ [-1cm]
		\includegraphics[width=0.5\textwidth]{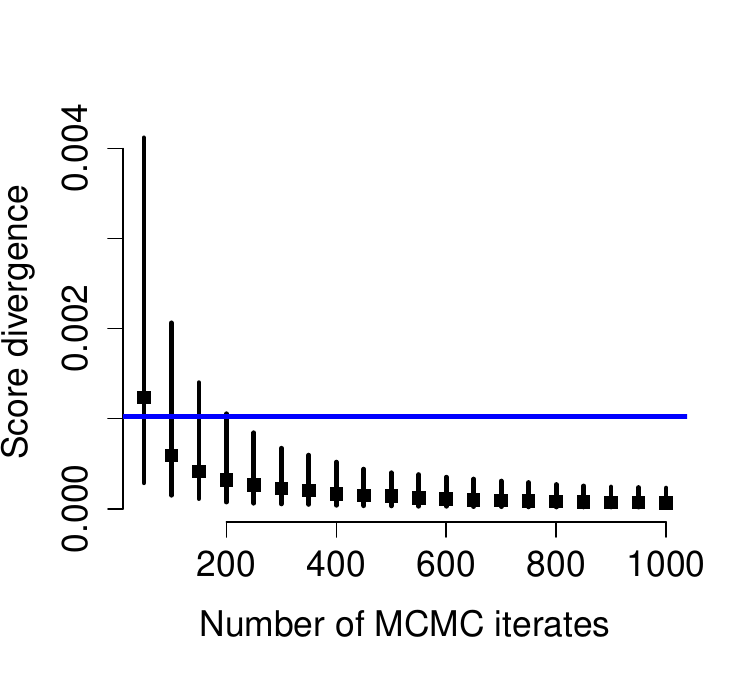} &
		\includegraphics[width=0.5\textwidth]{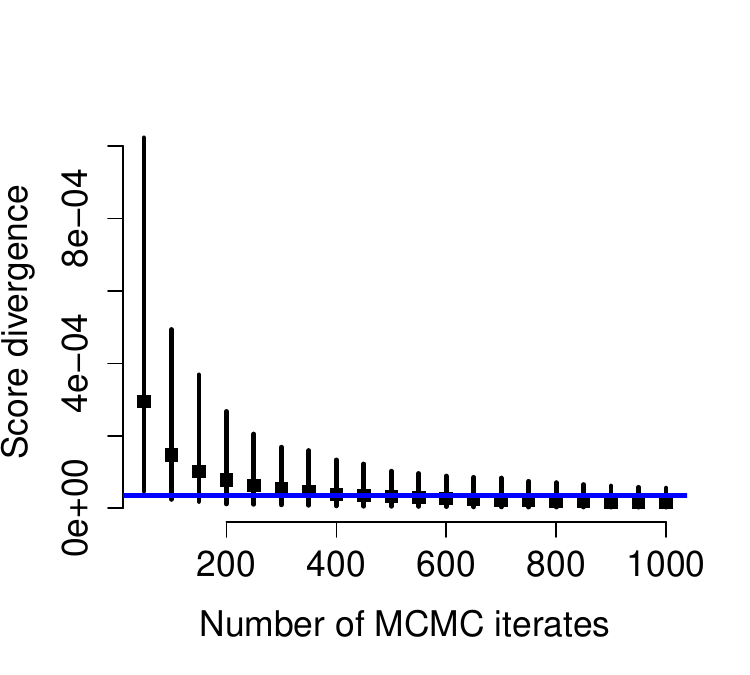} 
	\end{tabular}
	
	\caption{Same as Figure 2 in the paper, but for the alternative simulation design described in Section \ref{sec:n_inv_g}.}
	
\end{figure}

\begin{figure}[!htbp]
	
	\centering
	
	\begin{tabular}{cc}
					Quantile score ($\tau = .01$) & Quantile score ($\tau = .05$) \\ [-1cm]
				\includegraphics[width=0.5\textwidth]{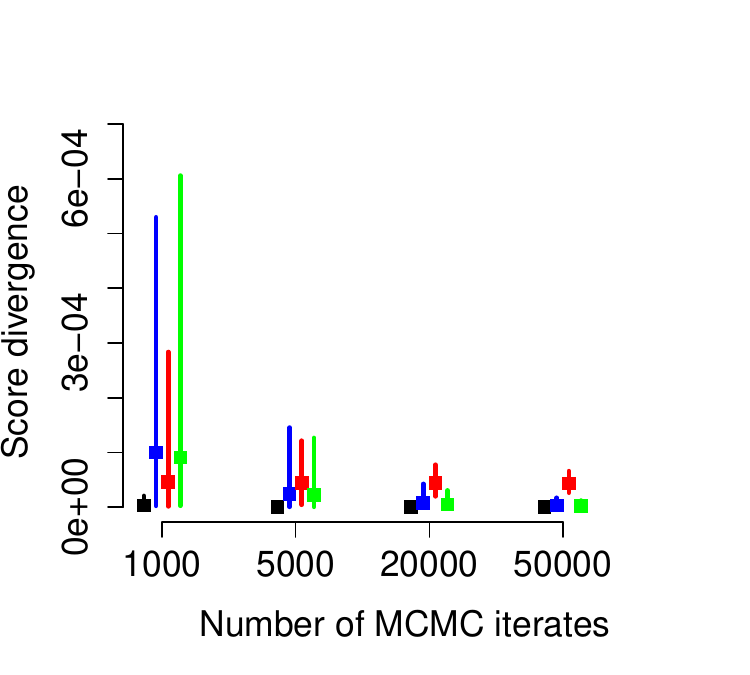} &
				\includegraphics[width=0.5\textwidth]{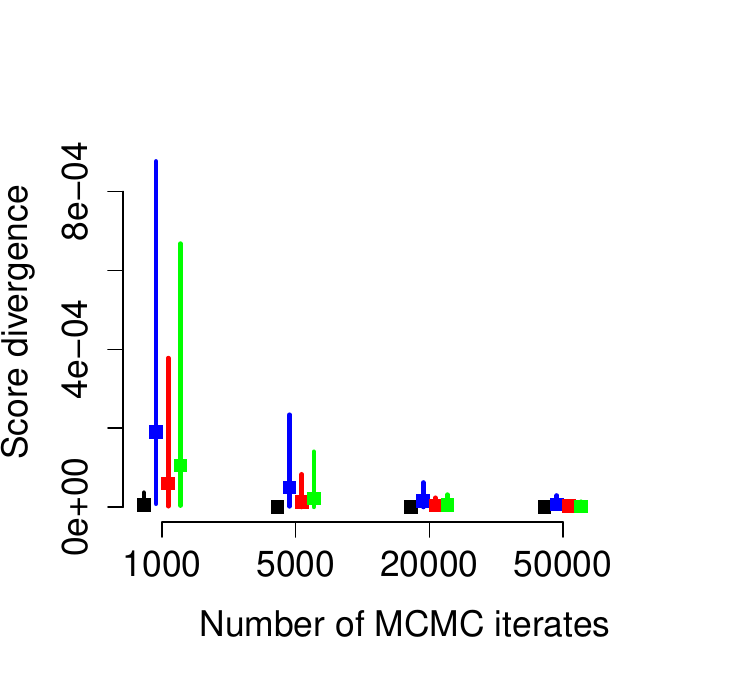} \\[-1.4cm] 
		\multicolumn{2}{c}{\includegraphics[width=\textwidth]{legend.pdf}}
	\end{tabular}\vspace{-1.8cm}
	
	\caption{Same as Figure \ref{fig:sim-sizeq} in the Online Supplement, but for the alternative simulation design described in Section \ref{sec:n_inv_g}. \label{fig:sim-sizenq}}
	
\end{figure}

\begin{figure}[!htbp]
	
	\centering
	
	\begin{tabular}{cc}
				Quantile score ($\tau = .01$) & Quantile score ($\tau = .05$) \\ [-1cm]
			\includegraphics[width=0.5\textwidth]{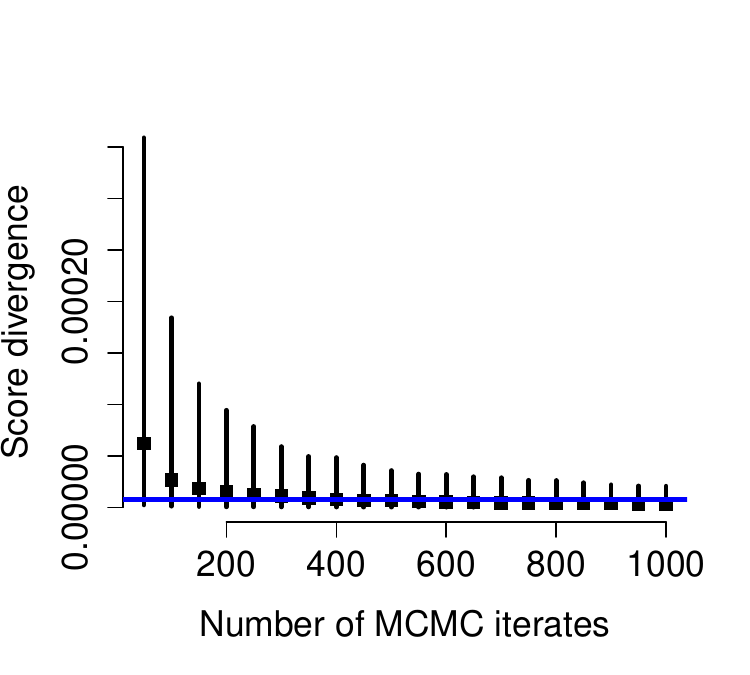} &
			\includegraphics[width=0.5\textwidth]{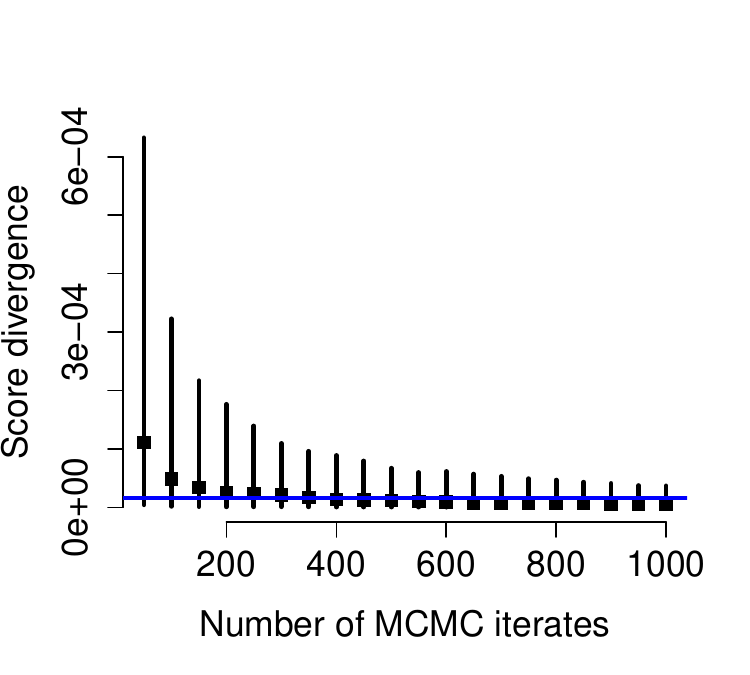}
	\end{tabular}
	
	\caption{Same as Figure \ref{fig:sim-zoomq} in the Online Supplement, but for the alternative simulation design described in Section \ref{sec:n_inv_g}.\label{fig:sim-zoomnq}}
	
\end{figure}

\clearpage

\section{Additional figures for case study}

\begin{figure}[!htbp]
	
	\centering
	
	\includegraphics[width= .8\textwidth]{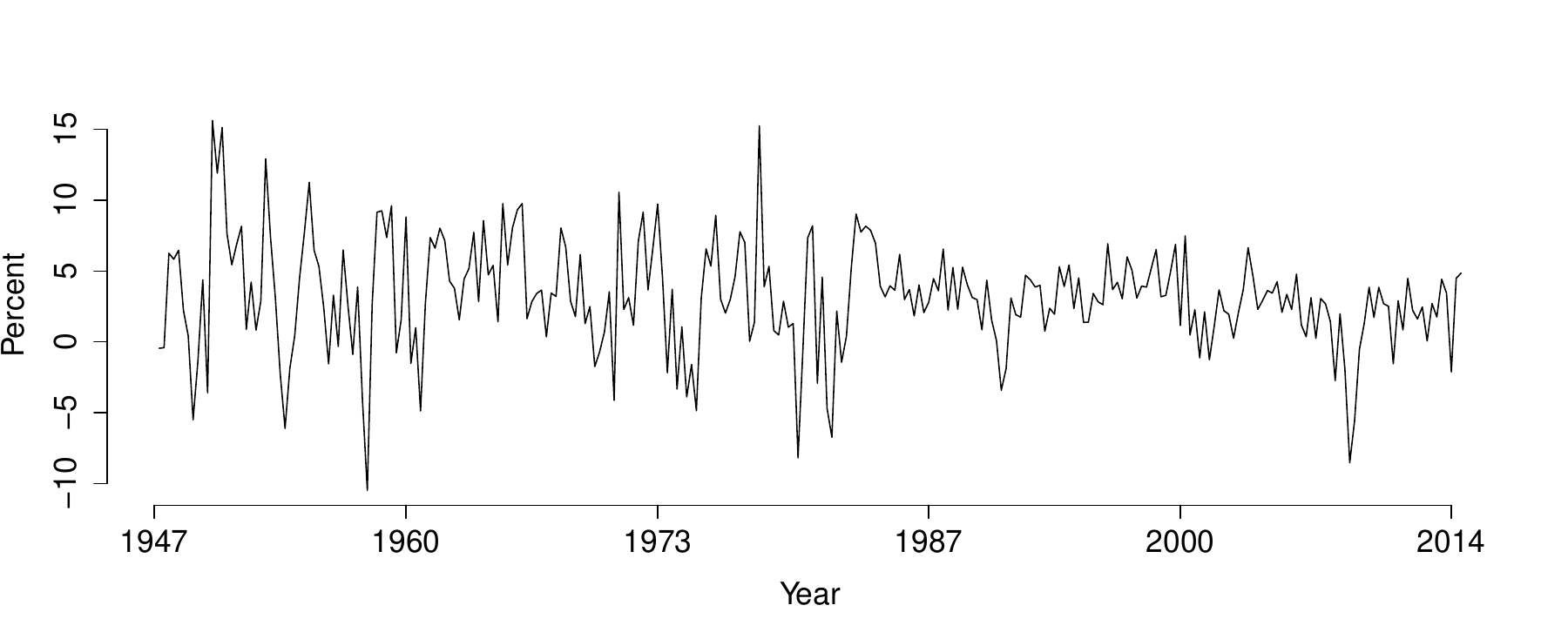}
	
	\caption{Quarterly growth rate of U.S.~real GDP from the second
		quarter of 1947 to the third quarter of 2014, using the data vintage
		from the first quarter of 2015.
		\label{fig:empirical-gdp} }
	
\end{figure}

\begin{figure}[!htbp]
	
\centering
	
\begin{tabular}{cc}
\multicolumn{2}{c}{Two quarters ahead}\\
Logarithmic score & CRPS \\ [-15mm]
\includegraphics[width = 0.45\textwidth]{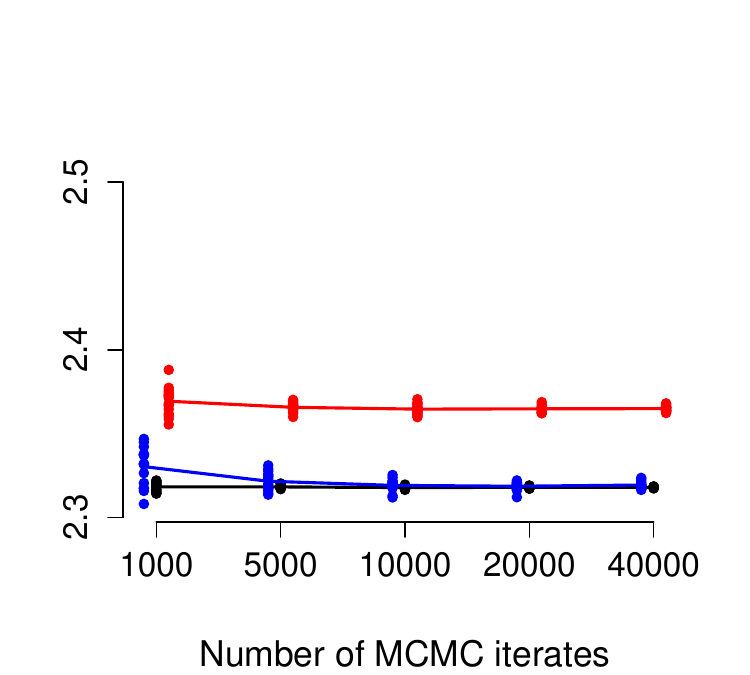} &
\includegraphics[width = 0.45\textwidth]{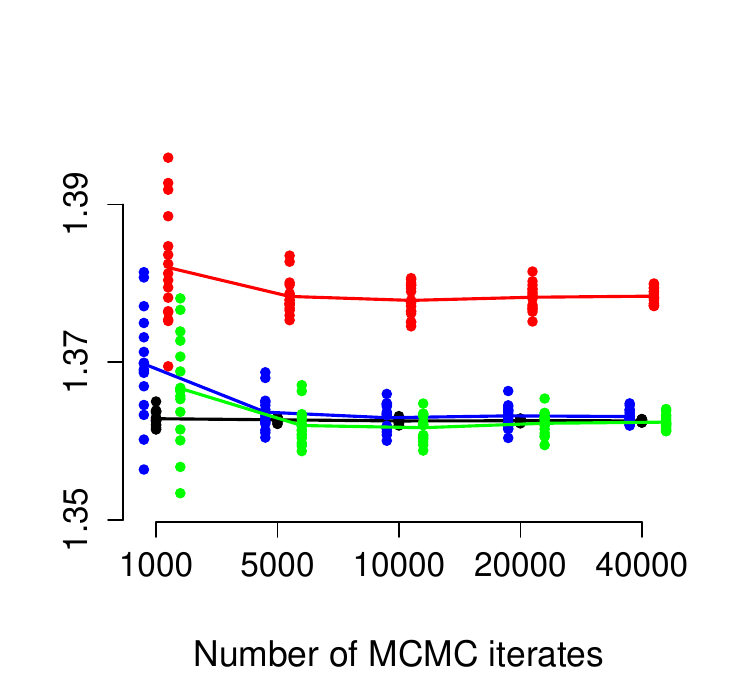} \\ 
\multicolumn{2}{c}{Three quarters ahead}\\
Logarithmic score & CRPS \\ [-15mm]
\includegraphics[width = 0.45\textwidth]{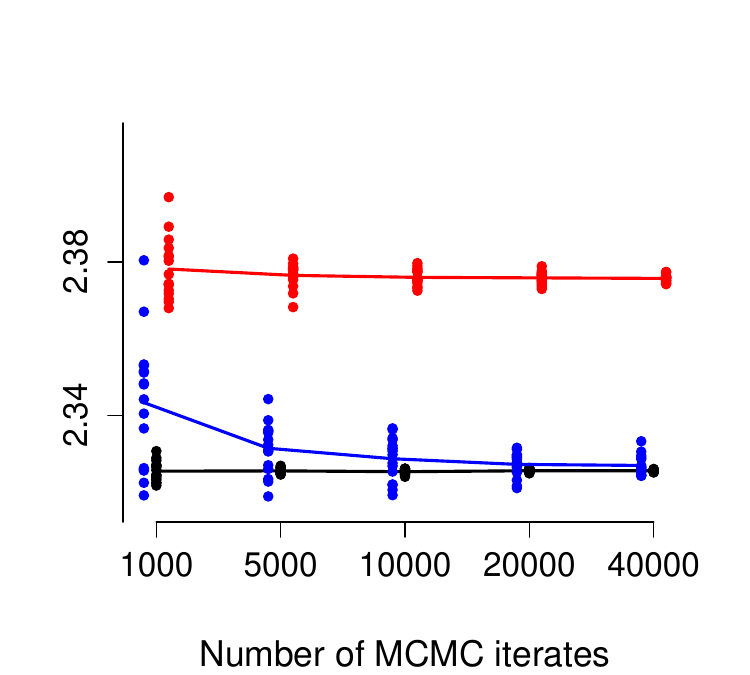} &
\includegraphics[width = 0.45\textwidth]{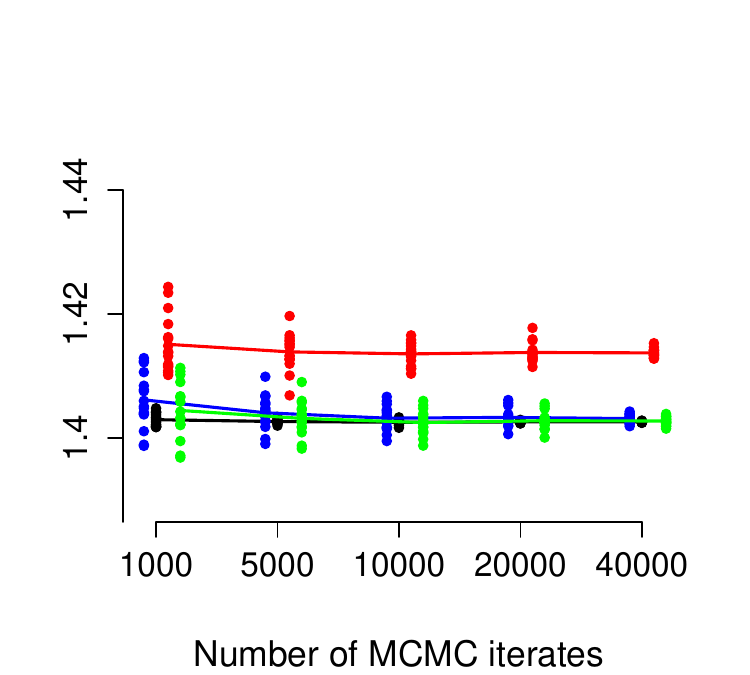} \\ [-1.2cm]
\multicolumn{2}{c}{\includegraphics[width = .8\textwidth]{legend.pdf}} \\ [-1.8cm]
\end{tabular}
	
\caption{Same as Figure 3 in the paper, but at prediction horizons of
two quarters ahead (first row) and  three quarters ahead (second row).  \label{fig:empscores1b}}
	
\end{figure}

\clearpage

\renewcommand\refname{References for the supplemental material}
\bibliographystyle{myims2} 
\setlength{\bibsep}{0mm}
\bibliography{bibliography_all_abbrev_table-update}